\documentclass[12pt, preprint]{aastex}

\usepackage{psfrag}
\usepackage{textcomp}
\usepackage{arydshln}
\usepackage{graphicx, subfigure}
\usepackage{natbib}

\slugcomment{Resubmitted to ApJ (Mar 25, 2013)}

\begin{document}

\title{Transitional disks and their origins: an infrared spectroscopic survey of Orion A}

\author{K. H. Kim\altaffilmark{1},
Dan M. Watson\altaffilmark{1},
P. Manoj\altaffilmark{1},
W. J. Forrest\altaffilmark{1},
Joan Najita\altaffilmark{2},
Elise Furlan\altaffilmark{2,8},
Benjamin Sargent\altaffilmark{3},
Catherine Espaillat\altaffilmark{4,5},
James Muzerolle\altaffilmark{6},
Tom Megeath\altaffilmark{7},
Nuria Calvet\altaffilmark{9},
Joel D. Green\altaffilmark{10},
Laura Arnold\altaffilmark{1,11}
}

\altaffiltext{1}{Department of Physics and Astronomy, University of
                 Rochester, Rochester, NY 14627, USA;
                 {\sf khkim@pas.rochester.edu}}
\altaffiltext{2}{National Optical Astronomy Observatory, 950 North Cherry Avenue, Tucson, AZ 85719, USA}
\altaffiltext{3}{Center for Imaging Science and Laboratory for Multiwavelength Astrophysics, Rochester Institute of Technology, 54 Lomb Memorial Drive, Rochester, NY 14623, USA}
\altaffiltext{4}{Harvard-Smithsonian Center for Astrophysics, 60 Garden St, Cambridge, MA 02138}
\altaffiltext{5}{Sagan Fellow}
\altaffiltext{6}{Space Telescope Science Institute, 3700 San Martin Dr., Baltimore, MD 21218, USA}
\altaffiltext{7}{Department of Physics and Astronomy, University of Toledo, 2801 W. Bancroft St., Toledo, OH 43606, USA}
\altaffiltext{8}{Visitor at the Infrared Processing and Analysis Center, Caltech, 770 S. Wilson Ave., Pasadena, CA, 91125, USA}
\altaffiltext{9}{Department of Astronomy, University of Michigan, 830 Dennison Building, 500 Church St, Ann Arbor, MI 48109}
\altaffiltext{10}{Department of Astronomy, University of Texas, 1 University Station, Austin, TX 78712}
\altaffiltext{11}{The college at Brockport, State University of New York, Brockport, NY 14420}

\begin{abstract}

Transitional disks are protoplanetary disks around young stars, with inner holes or gaps which are surrounded by optically thick outer, and often inner, disks. Here we present observations of 62 new transitional disks in the Orion A star-forming region. These were identified using the \textit{Spitzer Space Telescope}'s Infrared Spectrograph and followed up with determinations of stellar and accretion parameters using the Infrared Telescope Facility's SpeX. We combine these new observations with our previous results on transitional disks in Taurus, Chamaeleon I, Ophiuchus and Perseus, and with archival X-ray observations. This produces a sample of 105 transitional disks of ``cluster" age 3 Myr or less, by far the largest hitherto assembled. We use this sample to search for trends between the radial structure in the disks and many other system properties, in order to place constraints on the possible origins of transitional disks. We see a clear progression of host star accretion rate and the different disk morphologies. We confirm that transitional disks with complete central clearings have median accretion rates an order of magnitude smaller than radially continuous disks of the same population. Pre-transitional disks --- those objects with gaps that separate inner and outer disks --- have median accretion rates intermediate between the two. Our results from the search for statistically significant trends, especially related to $\dot{M}$, strongly support that in both cases the gaps are far more likely to be due to the gravitational influence of Jovian planets or brown dwarfs orbiting within the gaps, than to any of the photoevaporative, turbulent or grain-growth processes that can lead to disk dissipation. We also find that the fraction of Class II YSOs which are transitional disks is large, 0.1-0.2, especially in the youngest associations.
\end{abstract}


\section{INTRODUCTION}

Transitional disks (TDs) are protoplanetary disks around young stars which are optically thick and gas-rich, but which have AU-scale radial gaps or central clearings in their dust distribution. Usually the gap, which is depleted of small dust grains, is revealed as a deficit in the host T Tau star's infrared excess, with respect to its siblings. Such disks are thought to represent an evolutionary stage between Class II and Class III young stellar objects (YSOs).\footnote{As in \citet{diskionary}, we mean by Class II YSO an object with 2-20 \micron\ spectral index between -1.6 and -0.3, which physically corresponds to a (T Tauri) star surrounded by a (radially-continuous) accretion disk which is not viewed close to edge-on. Similarly, by Class III YSO we mean an object with 2-20 \micron\ spectral index less than -1.6, corresponding to a pre-main-sequence star with little or no circumstellar material. } \citet{strom89} first drew attention to these objects, using examples with small --- almost photospheric --- infrared excess at short mid-infrared ($5-50 \micron$) wavelengths but switching sharply to large and strong excess at longer wavelengths.

During the past decade, there have been significant improvements on the study of YSOs and protoplanetary disks based on data from the \textit{Spitzer Space Telescope} \citep{werner04} launched in 2003. Detailed studies of disk structure have been possible with mid-IR spectra taken using from the Infrared Spectrograph \citep[IRS;][]{houck04}. In particular, there has been major progress revealing the detailed disk structures and uncovering a variety of transitional disks.

After the distinctive spectrum of CoKu Tau/4 was discovered by \citet{forrest04}, several transitional disks in the Taurus star-forming region were studied in detail with self-consistent disk modeling by \citet{d'alessio05} (CoKu Tau/4), \citet{calvet05} (DM Tau and GM Aur) and \citet{espaillat07a, espaillat07b, espaillat2008} (LkCa 15 and UX Tau A). \citet{calvet05} and \citet{d'alessio05} suggested that DM Tau and CoKu Tau/4, respectively, can be explained with an empty central cavity in the disk surrounding the central star, but GM Aur requires a disk structure with a gap separating an inner optically thin disk and an outer optically thick disk to match the observed SED \citep{calvet05}. \citet{espaillat2008, Espaillat2010ptd} confirmed that LkCa 15 and UX Tau A have gaps between optically thick inner and outer disks. We will refer in the following to these three types of disks as classical transitional disks (CTDs; e.g. DM Tau), weak-excess transitional disks (WTDs; e.g., GM Aur) and pre-transitional disks (PTDs; e.g., LkCa 15), respectively.

There is generally good agreement with the gap sizes inferred from SED modelling, and direct observations of the outer edges of the gaps. Gaps in some 15 transitional disks have been confirmed by submillimeter interferometry \citep{Pietu+2006,Hughes+2007_TWHya,Hughes_GMAur_2009ApJ...698..131H,Andrews+2009-ophdisk,andrews2011_sma_CTDs}, and one of these, LkCa 15, also in near-infrared light scattered by the outer disk \citep{Thalmann2010_LkCa15_NIRimage}. The resolved inner disk of a pre-transitional disk, T Cha, also confirms its gapped structure \citep{Olofsson+2011_TCha_VLTI}.

The definition and selection criteria of transitional disks have been far from homogeneous considering all objects called `transition' or `transitional' objects from other works. Many authors define transitional disks rather loosely, with diverse nomenclatures\footnote{The definition of diverse nomenclatures are summarized in \citet{diskionary}}. Circumstellar disks with a deficit of dust emission from the inner disk and large 30/13 $\micron$ flux ratio indicating an optically thick outer disk have been designated ``cold disks'' by \citet{brown07} and \citet{Merin2010}. A more relaxed definition includes disks with a much smaller excess for wavelengths beyond $>$ 13 $\micron$ representing a depleted or settled outer disk compared to our definition of transitional disks. Those disks are named ``anemic disks'' \citep{Lada_2006_IC348anemic}, ``homologously depleted disks'' \citep{Currie_2009_homodep}, or ``weak-excess disks''\footnote{The definition of WTD (weak excess transitional disk) in this work and in \citet{Muzerolle2010} are different.} \citep{Muzerolle2010}. Such SEDs can result from dust settling to the disk midplane \citep{Luhman+2010_TauDisks, Espaillat_TDclass_2012}, or outward truncation of a disk by gravitational interaction with a companion \citep{McClure+2008}. Therefore, it is important to clarify by which definition and selection criteria the sample of transitional/transition disks are selected, especially if the sample is to be used for searching for any trends to understand the properties of transitional disks and their origin.

Several physical mechanisms have been suggested to explain inside-out disk dispersal: photoevaporation \citep{cgs01,acp06a,acp06b,Gorti2009b_time_evo_photoevaporation,Owen_photoevaporation2012}; grain growth/coagulation/settling \citep{duldom05,ciesla06}; inside-out clearing by Magneto-Rotational Instability (MRI) \citep{chimur07,Suzuki_MRI_2010,perez-becker_chiang2011MRI-FUV}; dynamical effects by stellar/substellar companions \citep{Lubow_Dangelo_2006}; giant planet formation \citep{Marsh+Mahoney1992,Rice+2003MNRAS.342...79R,quillen04,varni06a,Lubow_Dangelo_2006,Zhu_2011_multiplanets,Zhu+DustFiltration2012ApJ}. To distinguish and understand which mechanisms are dominant for the origin of transitional disks, not only flux density information at all wavelengths for a specific target but also empirical trends from observational data in a large sample are indispensable.

There have been several efforts to use statistical trends among several properties of transitional disks and their host stars to constrain models of the origins of these objects. \citet{kimkh2009} analyzed IRS spectra of 13 transitional disks in Taurus and Chamaeleon (1-3 Myr) and found a strong correlation between stellar mass and outer gap radius, $R_{wall}$. \citet{Merin2010} collected broadband SEDs of some fifteen ``cold" disks from several young associations. Including some objects reported in the literature \citep[e.g.][]{brown07,kimkh2009} to improve the sample size, they found the gap radius to scale linearly with $M_\star$, and to be significantly correlated with disk mass, with more massive disks tending to have larger holes. In the somewhat older (4-12 Myr) clusters, Tr 37 and NGC 7160, \citet{sicilia-aguilar_2010_MdotCepOB2} found the median accretion rates in the transitional-disk systems to be about an order of magnitude smaller than in Taurus and Chamaeleon, but similar to these associations' normal Class II YSOs, in contrast to \citet{najita07} who found that accretion rates for transitional disks in the Taurus-Aurigae association are systematically a factor of 10 smaller than the normal Class IIs.

In this paper, we present IRS spectra of 62 newly selected transitional disks in the Orion A star-forming region and near-IR spectra of 52 of them from SpeX/IRTF followup observation. Orion A consists of the Lynds 1640 dark cloud which includes the Orion Nebula Cluster (henceforth ONC) and the Lynds 1641 (henceforth L1641). This cloud stretches over $\sim$ 30 deg$^2$ on the sky with Declination from $-4^d$$30^m$ to $-9^d$. In this paper, we assume objects to the north of $\delta_{J2000}$ = $-6^d$ belong to ONC while objects to the south belong to L1641.
The distance to the Orion A complex has been estimated to be somewhere between 300 to 600 pc (\citet{Muench_ONC_review_2008} and references therein). Here, we adopt the distance to Orion A as 414 pc based on the study by \citet{Menten_2007_Oriondistance}.
We take the median age of ONC members as less than 1 Myr \citep{Hillenbrand97_ONCpopulation_age} and that of L1641 members as 1 Myr \citep{galfalk_olofsson_L1641N}.

We consider 105 transitional disks not only of Orion A including ONC and L1641, but also of Taurus \citep[henceforth Tau;][]{Furlan2011Taurus}, Chameleon I \citep[henceforth Cha I;][]{manoj2011}, Ophiuchus \citep[henceforth Oph;][]{McClure2010}, and NGC 1333 \citep[henceforth N1333;][]{Arnold2012}. From this very large sample of transitional disks selected homogeneously from disk properties measured with IRS spectra, and comprehensive information on mass accretion rates from SpeX spectra and X-ray luminosity, we are now able to search for trends between stellar parameters and disk parameters with robust statistics. We anticipate this work to help in understanding mechanisms responsible for the origins of transitional disks.

In Section 2, we present IRS and SpeX observations and data reduction procedures. In Section 3, we describe the measurements of stellar properties and mass accretion rates. We describe the extraction of the outer gap radius, $R_{wall}$, after explanation of how we selected transitional disks and their sub-classification in Section 4. We discuss the fraction of transitional disks and age trends in Section 5. In Section 6, we explore correlations and trends of the transitional disks properties and compare them to those of radially-continuous disks where possible. Then, we review and discuss what these findings mean and how they can be used to help our understanding of the origin of transitional disks in Section 7. Summary and conclusions drawn from our finding follows in Section 8.

\section{OBSERVATIONS AND DATA REDUCTION}
We observed the Orion A star-forming region between 2006 November and 2007 October, during \textit{Spitzer}-IRS campaigns 36, 39, 40 and 44. In all, our targets were 555 objects selected on the basis of their \textit{Spitzer}-IRAC and MIPS colors to belong to YSO Classes 0, I, flat-spectrum and II, with flux densities in excess of 2 mJy at 8 \micron\ and 15 mJy at 24 \micron\ \citep{Megeath_OriA+B_survey_2012}. Of these, 303 objects were classified as Class II based on their spectral index between 4.5 \micron\ (IRAC channel 2) and 24 \micron\ (MIPS channel 1) or have the colors of transitional disks seen in other regions; they do not show evidence of envelopes in their IRS spectra, and thus their infrared excess is due to circumstellar disks.. We observed 241 objects (114 in L1641; 127 in ONC) with the IRS with full IRS wavelength coverage of 5-37 $\micron$ in the low resolution mode. We observed 62 additional objects located close to the Trapezium region with only partial wavelength coverage (5-14 $\micron$), as the detectors for the longer wavelengths would have been saturated by the bright background emission. We estimate that our sample is complete for star-and-disk dominated objects for which the host-star spectral type is M4 or earlier. Analysis of the full sample will be presented in more detail by \citet[in preparation]{khkim2013-OriA}. The complementary sample of objects with envelopes -- Class 0, I and flat-spectrum objects -- will be discussed by \citet[in preparetion]{Poteet2013}.

\subsection{IRS/\textit{Spitzer}}
The IRS spectra of the 62 transitional disks were selected from among the 241 objects which were observed with the IRS low resolution modules ($\lambda / \Delta \lambda$ $\sim$ 90; Short-Low (SL):5.3-14 $\micron$; Long-Low (LL): 14-38 $\micron$), based on the selection criteria described in Section 4. Each object was observed at two nod positions separated by one-third the length of the slit in the staring mode. To extract the spectra, we used version S15.3 of the basic calibrated data (BCD) product from the Spitzer Science Center IRS pipeline for both SL and LL.

To fix bad, hot and rogue detector-array pixels before extracting objects from the 2D spectral images, we generated a set of ``grand rogue masks" for Orion A data. A grand rogue mask for the general data reduction process is generated by combining rogue mask files available up to a recent IRS campaign supplied from the Spitzer Science Center, and adding additionally identified rogue pixels from data images. However, we applied additional special treatment to make LL grand rogue masks for these data because (1) the LL array exposure to cosmic rays had increased continuously up to campaign 44 \footnote{In IRS campaign 45 the bias voltage on the Long Low array was reduced from 1.8 to 1.6 volts and the array temperature was reduced from 4.4 to 4.1 K)}, which caused the S/N to decrease and the number of rogue pixels and hot pixels to increase, (2) fluxes from our objects are fainter than similar sources in other nearby star-forming regions due to the greater distance to OriA, (3) the large number of rogue pixels and hot pixels severely degraded the signal-to-noise ratio of our targets, and introduced spectral artifacts when many rogue pixels were grouped in clusters. To avoid rejection of too many pixels which might be perfectly good, we chose only the rogue pixels which appear in the campaign rogue masks more than 10 times through campaign 44. To these we added hot pixels selected by eye from LL 2D images from campaign 39 and campaign 44. The usage of grand mask files including rogue pixels and hot pixels cannot perfectly clean bad pixels. However, we tested several version of grand rogue masks for our LL data and chose a set of grand rogue masks to apply for all of OriA ClassII data. We fixed those rogue pixels in the grand rogue mask (and the permanently bad pixels) by interpolation in the spectral direction from nearest-neighbor good pixels.

Spectral extraction of Class II YSOs in Orion A is generally much more challenging than in other nearby star-forming regions. This is due not only to the source faintness but also to the range of spatial structure in sky emission from the Orion HII regions, and the high stellar density. We used four source extraction methods to derive the final point-source spectra, choosing among them to optimize the rejection of emission from the sky and other nearby point sources.

As a first, basic source extraction step, we used an automated extractor (``auto") based upon the IRS instrument team's Spectral Modeling, Analysis and Reduction Tool \citep[SMART;][]{higdon04}. In auto we extracted sources from the uniformly-weighted signal along a tapered column, 3-5 pixels wide. In this step we used two versions of background-subtracted images: one prepared by subtraction of the two nods (``off-nod"), and the other by subtraction of the sky spectrum in each grating order obtained while the target was being observed in the other order (``off-order"). If there were no serious sky background issues or no additional sources in the 2D images, the spectra taken from images subtracted by off-order or off-nods backgrounds are very similar. We examined each sky-subtracted image and the resulting spectra. If large sky-subtraction artifacts (e.g. spectral lines in the sky, contamination from nearby point sources) remained in both auto products, we re-reduced the data in SMART by using the same tapered column extraction as before, but with sky removed by fitting and subtracting a 0th- or 1st order polynomial to emission outside the span of the target. We designate the tapered column extraction with subtraction of a polynomial sky selected manually as ``man".

Usually One of three tapered column extractions gave artifact-free results even when the sky emission had complicated structure, but issues remained for targets with a neighbor closer than four pixels away along the slit. In those cases, we resorted to optimal point-source extraction, using both the AdOpt script in SMART \citep{Lebouteiller_2010_AdOpt} and another routine, OPSE, developed by our group \citep[in preparation]{opse}. They are complementary: AdOpt employs an empirical point response function (PRF) and can fit multiple objects along the slit; OPSE uses an analytical PRF and can make corrections to the spectrum for pointing errors, and thus was useful for extracting our target sources, as it can account other sources in a image.

For the flux calibration of the spectra, we multiplied relative spectral response functions (RSRFs), which were generated by dividing a template spectrum (M. Cohen, private communication; J. Marshall, private communication) by a calibrator's spectrum, extracted in the same way as the target spectrum. Our photometric standards were $\alpha$ Lac (A1V) for SL, $\xi$ Dra (K2III) for LL, and Mrk 231 (assumed to radiate as a power law) for LL wavelength greater than 32 $\micron$.

We compared the spectra from all different versions of source extraction for each object, then we selected the final spectrum based upon freedom from artifacts. During the process, we combined spectra from different versions of methods if necessary to get the best spectra. For example, we use the combined spectrum for OriA-19: SL from AdOpt and LL from OPSE. Reduction choices for the final selection of spectra are noted in Table~\ref{table:IRSlog}.

\subsection{SpeX/IRTF}
52 out of 62 transitional objects in Orion A were observed at Near-IR (0.8-2.4 $\mu$m) wavelengths with the medium resolution spectrograph SpeX \citep{SpeX_Rayner}, on the NASA Infrared Telescope Facility (IRTF) on Mauna Kea during the 2010A, 2011A , and 2011B semesters.

We observed our Orion A transitional disks with the Short-wavelength Cross-dispersed mode (SXD) on the nights of 15-17 Feb. 2010 and 25-27 Feb. and 6-10 Nov. 2011. We obtained spectra with various slit widths of 0.3", 0.5" and 0.8" for observations during the 2010A semester depending on the seeing conditions of each night. We used only the 0.8" slit width for the observations 2011 February because the weather and seeing were generally poor. Among the objects observed during the 2010A semester, 8 lacked reliable spectral types in the literature, so we used their spectra for the determination of spectral type as well as accretion rate. Of the 8 objects, OriA-34 was only observed with the 0.3" slit to get the highest resolution (R = 2000) available at SpeX in order to determine the spectral type as well as measure the mass accretion rate under good conditions (seeing $\sim$ 0.3"). The spectra of the other 7 objects were obtained with the 0.8" slit (R $\sim$ 800-1200), sufficient to narrow the spectral type range down to 2-3 sub-types by comparing absorption features of the Na I, Al I, Mg I, Ca I and CO to those of template spectra \citep{IRTF_spectral_Library_2009,IRTF_SpeX_MLTdwarf_Cushing_2005}. The details on spectral type determination are described in \S 2.2.1. For targets with very close neighbors, we oriented the slit so as to observe both simultaneously. For airmass greater than 2 we kept the slit orientation fixed to the parallactic angle. The SpeX/IRTF observation log is given in Table~\ref{table:SpeXlog}. We reduced our spectra with the Spextool package \citep{SpeXtool}, and the flux calibration and telluric absorption correction \citep{SpeX_tellcor_Vacca2003} were done with a spectrum of an A0V star, HD 37887, observed near in time and close in air mass to each object.

\subsubsection{Spectral Type Determination}
\begin{figure}[t]
\epsscale{1.0}
\plotone{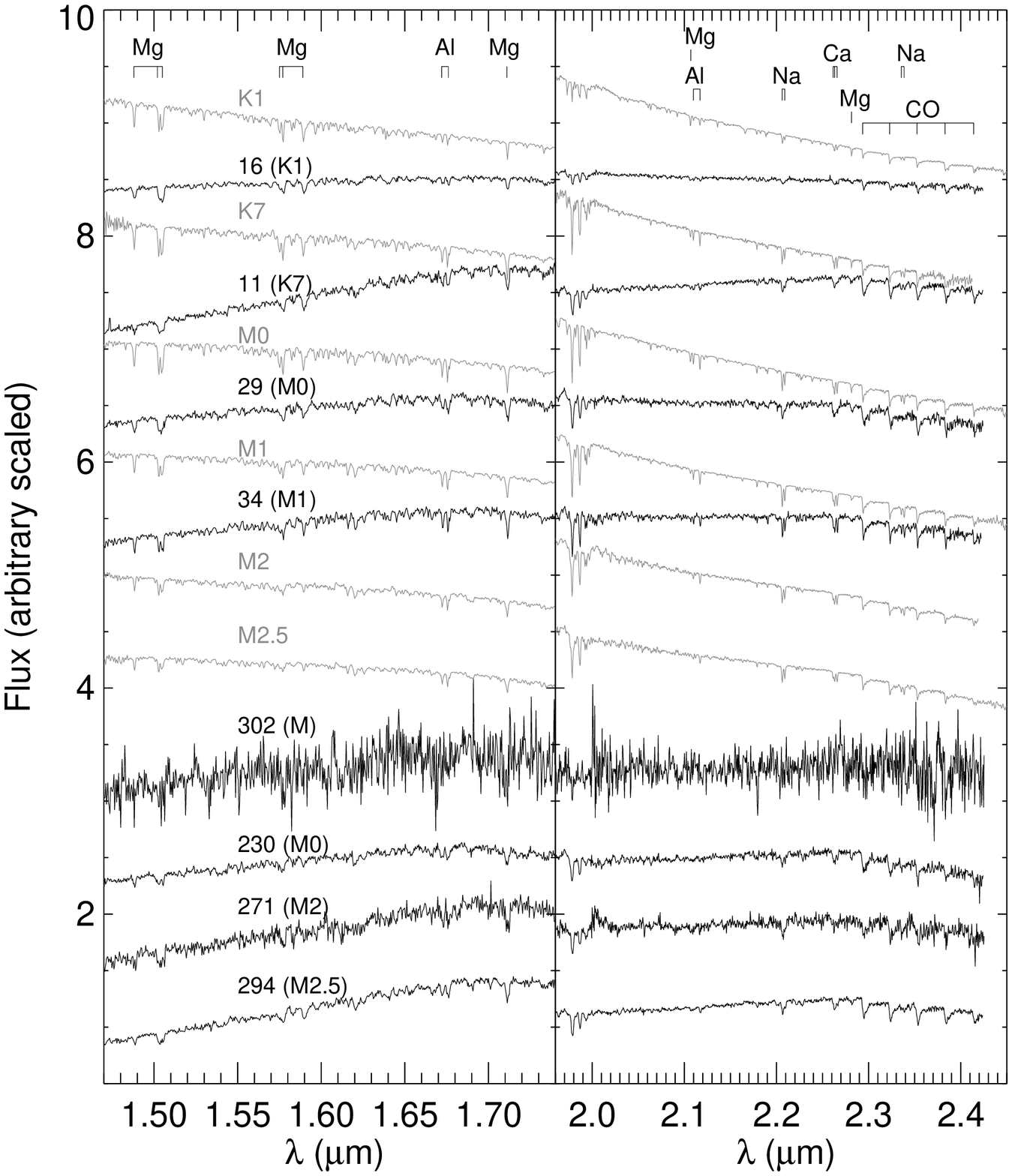}
\caption{Spectral typing from SpeX spectra. The gray spectra with spectral type are the standard spectra of the spectral types \citep{Rayner09IRTFspectralLib}. The black spectra with number and spectra type are the SpeX spectra of the objects with the ID number. We estimated their spectral type as the spectral type next to the object's number. \label{fig-spttyping}}
\end{figure}

We list in Table~\ref{table:SpTypeAv} the spectral type of the host star of each of our transitional disks. Most of our spectral-type information is gathered from the literature \citep{H97b,RHS2000,Allen_thesis_L1641,Dario2009_ubvri,Fang2009,Parihar2009_ONC_variables_SpT4V1498Ori}, or from unpublished results kindly provided by Lori Allen, John Tobin and Jes\'{u}s Hern\'{a}ndez. In addition, we determined several new or improved spectral types from our SpeX spectra. In Figure~\ref{fig-spttyping}, we show the SpeX spectra of these objects and illustrate our spectral-typing procedure. We used Mg I, Al I, Na I, Ca I, and CO features in the H and K band as the SpeX sensitivity is best in those bands and as those absorption features are very sensitive to spectral types in the G-M range.

It would have been best if the spectral resolution of our target spectra was the same as that of the standard spectra in order to distinguish adjacent lines and spectral depth. However, most of our SpeX spectra were taken at R $\sim$ 800-1200. Only OriA-34 was observed with 0.3 arcsec slit giving R $\sim$ 2000. Even though some lines are blended due to the modest spectral resolution, we were able to narrow down spectral types to about $+/-$ 2-3 subtypes. We adopt the spectral type of the first five objects from top of Figure~\ref{fig-spttyping} from the spectral typing using our SpeX spectra. The S/N on OriA-302 is not good enough to determine a sub-class of its spectral type, but OriA-302 is thought to be an M-type star based on the broad spectral features attributable to Mg I, Na I, and CO overtone bands. We found that the uncertain spectral type from HECTOSPEC data (Lori Allen, private communication) of OriA-230, OriA-271, and OriA-294 are in reasonably good agreement with those from our spectra.

\subsection{Ancillary Data: Photometry}
We also compiled broadband photometry from ultraviolet to mid-infrared wavelengths from the literature. From the Naval Observatory Merged Astrometric Dataset (NOMAD) catalog \citep{NOMAD_2005} we collect B (0.44 $\micron$), V(0.55 $\micron$), and R (0.64 $\micron$).
There is a rather recent photometry data set for ONC objects from \citet{Dario2009_ubvri}. From their tables, we gathered optical photometry at U (0.347 $\micron$), B (0.454 $\micron$), V (0.538 $\micron$), TiO (0.6217 $\micron$), and I (0.862 $\micron$).
Photometry from one more optical band (I at 0.8 $\micron$) was taken from the DENIS database as well as two near-IR bands, J (1.25 $\micron$) and K (2.16 $\micron$).
Most of our targets (except OriA-302) have 2MASS photometry in the J (1.25 $\micron$), H (1.65 $\micron$), and K (2.17 $\micron$) bands. We collected 2MASS information from \citet{2MASS_catalog_2006}. The photometry in the JHK bands for OriA-302 in Figure 2 are the averaged fluxes from our SpeX spectra.
All of our targets were observed with IRAC (3.6, 4.5, 5.8, and 8.0 $\micron$) and MIPS (24 $\micron$) prior to the IRS observations \citep{Megeath_OriA+B_survey_2012}.
The SEDs (Spectral Energy Distributions) constructed with the broadband photometry are shown in Figure 2.

\section{STELLAR PROPERTIES AND ACCRETION PROPERTIES}
\subsection{Extinction Correction}
Extinction toward protoplanetary disks can lead to misclassification of evolutionary stages and misinterpretation of disk spectra. To minimize the effects of extinction toward our targets, we de-reddened our data based on the estimates of visual extinction ($A_{V}$) obtained using the following relationship between $A_{V}$ and the color excess $E({\lambda_1}-{\lambda_2})={([{\lambda_1}]-[{\lambda_2}])_{obs}}-{([{\lambda_1}]-[{\lambda_2}])_{int}}$:
\begin{equation}
{A_V}={\frac{\frac{A_V}{A_{\lambda_2}}}{\frac{A_{\lambda_1}}{A_{\lambda_2}}-1}}\times{E({\lambda_1}-{\lambda_2})}
\end{equation}
To get a color from two wavelengths, $\lambda_1$ and $\lambda_2$, we used 2MASS JHK photometry for most of the sample and DENIS IJH photometry for the rest, when available. We measured $A_{V}$ in several ways for each object. We use either $(J-H)_{2MASS}$, $(H-K)_{2MASS}$, or $(I-J)_{DENIS}$ as an observed color $([\lambda_{1}]-[\lambda_{2}])_{obs}$. We adopt $I-J$, $J-H$, and $H-K$ photospheric colors from \citet{kh95} or $J-H$ and $H-K$ of the Classical T Tauri Star (CTTS) locus of colors from \citet{cttslocus97} as the intrinsic color, $([\lambda_{1}]-[\lambda_{2}])_{int}$.
To get $A_\lambda$, for each set of $\lambda_{1}$ and $\lambda_{2}$, we used the Mathis (1990) extinction curve for $R_V$ = 5.0 if the resulting $A_V$ $<$ 3. In case of $A_V$ $>$ 3, we followed the empirical extinction curves from \citet{McClure2009}: two composite extinction curves, one for $3 < A_{V} < 8 $ and one for $A_{V} > 8$.
After calculation of extinction corrections with each intrinsic color choice, we examined the extinction-corrected SEDs and selected a final result based on freedom from artifacts of the correction (e.g. artificial CO$_2$ ice features or structure in the silicate features) and good agreement with the photospheric spectrum of the star's type, at short wavelengths. ($<$ 1 $\micron$). In Table~\ref{table:SpTypeAv}, we list the $A_V$ method selected for each object: $I-J$, $J-H$, and $H-K$.

\subsection{Stellar Properties}
We list the adopted spectral types in Table~\ref{table:SpTypeAv}. Among 62 Orion A TDs, we have 55 objects with well-determined spectral types, one with spectral type constrained to a broad range, and 6 with unknown spectral types. In Figure~\ref{fig-OriAX-SpT} we show the spectral type distribution of objects with known spectral types in ONC and L1641. The spectral types of ONC objects range from G5 to M5, while that of the L1641 objects are more concentrated around the M1 type. The SpT distribution of 27 TDs in ONC and 28 TDs in L1641 generally agree with the SpT distribution of the general stellar populations of ONC \citep{RHS2000} and L1641 \citep{Hsu-L1641-2012}, respectively.

The effective temperatures, $T_{eff}$, are adopted from \citet{kh95} (see Table~\ref{table:property}), corresponding to the spectral type of each object. For the objects of unknown spectral types, we used 3850K, the mean $T_{eff}$ of Class II sources with known SpT in Orion A. The stellar luminosity ($L_\star$) of each object was derived from the stellar effective temperature and the stellar radius ($R_{\star}$). $R_{\star}$ was calculated from the scaling factor ($(R_{\star}/d)^{2}$) applied to the photosphere model, where $d = 414$ pc was assumed to be the distance to Orion A. The photosphere was derived from the intrinsic colors from \citet{kh95} at temperature $T_{eff}$, scaled to match the de-reddened the 2MASS J band photometry.
The mass of star ($M_{\star}$) was inferred from the Siess PMS evolutionary tracks \citep{siess2000} using the assumed luminosities and effective temperatures. TDs with known spectral types in Orion A are shown on an H-R diagram along with Z=0.02 evolutionary tracks in Figure~\ref{fig-HRD}. We also plot H-R diagrams for our TD sample from the Tau, ChaI, Oph, and N1333 associations, which we also use in the present analysis, in Figure~\ref{fig-HRD}. We see that the host stars of transitional disks in Orion A lie furthest above the main sequence. This agrees with the generally-accepted age sequence in which Orion A is younger than that of Tau, ChaI, or Oph.

We compiled X-ray observations for our TD sample from a variety of sources. We searched for X-ray data in HEASARC, in the published literature (\citet{Guedl_XEST_tauXray} for Tau; \citet{Winston_N1333_X-ray} for NGC 1333), and in the Chandra Source Catalog \citep{CSC1.1_chandra} for objects in ONC. We were permitted pre-publication access to data from the XMM-Newton survey of L1641 \citep[SOXS,][]{Pillitteri-2013arXiv1303.3996P}. We have also used data from the Second XMM-Newton Serendipitous Source Catalog \citep{XMMSSC-catalogue}. The X-ray luminosity $L_{X}$ we adopt in this work is that within 0.2-12 keV, the total band of XMM-Newton.

These properties, $T_{eff}$, $L_\star$, $M_{\star}$, $R_{\star}$, and $L_{X}$ are listed in Table~\ref{table:property}.

\subsection{Mass Accretion Rates: $\dot{M}$}
We observed 52 of Orion A transitional disks with SpeX/IRTF in SXD mode from 0.8-2.4 $\micron$ to measure mass accretion rates from their hydrogen recombination lines: Pa$\gamma$ (1.094 $\micron$), Pa$\beta$ (1.282 $\micron$), and Br$\gamma$ (2.166 $\micron$). From the de-reddened SpeX spectra with the $A_{V}$ determined as described above, we obtained mass accretion rates of all the objects except the G5 star, OriA-88, which shows strong Hydrogen absorption lines.

The method of mass accretion rate measurement is as follows. We fit each hydrogen recombination line with a gaussian function plus a local continuum. We measure the line luminosity of each line from the fit. Then we use the following relations to convert line luminosity to accretion luminosity $L_{acc}$ \citep{muzerolle98Brgamma,gatti2008pagamma}:
\begin{equation}
log(L_{acc}/L_{\odot}) = 1.36 \times log(L_{Pa \gamma}/L_{\odot}) + 4.1
\end{equation}
\begin{equation}
log(L_{acc}/L_{\odot}) = 1.14 \times log(L_{Pa \beta}/L_{\odot}) + 3.15
\end{equation}
\begin{equation}
log(L_{acc}/L_{\odot}) = 1.26 \times log(L_{Br \gamma}/L_{\odot}) + 4.43
\end{equation}

whence we obtain the disk-star accretion rate:
\begin{equation}
{\dot{M}}={\frac{L_{acc}R_{\star}}{GM_{\star}}}
\end{equation}

In Figure~\ref{fig-estMdot} we show as an example the results for OriA-59.

In general, the three recombination lines in our spectra yield similar results for accretion rate within a factor of 2-3, so we report the average in Table~\ref{table:property}. When fewer than three lines were detected we adopt the resulting average $\dot{M}$ as an upper limit and indicate them as such in Table~\ref{table:property}.

\section{DISK PROPERTIES}

\subsection{Transitional Disks: Selection Criteria and Their Variety}

Several spectral indices derived from IRS spectra of Class II YSOs have been used as a first step to separate transitional disks from the radially-continuous bulk of the population \citep{Furlan2009TauChaOph,dmw_2009,manoj2011,McClure2010,Arnold2012}, which we also use to identify the transitional disks in Orion A. The continuum spectral indices are defined as
\begin{equation}
n_{\lambda_{1} - \lambda_{2}} = \frac{\log(\lambda_{2} F_{\lambda_{2}})-\log(\lambda_{1} F_{\lambda_{1}})}{\log(\lambda_{2})-\log(\lambda_{1})}~~.
\end{equation}
The wavelengths, $\lambda_{1}$ and $\lambda_{2}$, are selected to avoid emission features, and thus to represent the spectral shape of the optically-thick disk continuum emission. This in turn reveals disk structure, both radial (central clearings and gaps) and vertical (degree of flaring). The equivalent width of the 10 $\micron$ silicate emission feature (EW(10$\micron$)),
\begin{equation}
EW(10\micron)=\ \int_{8\micron}^{13\micron} {\frac{F_{\lambda}-F_{\lambda,con}}{F_{\lambda,con}}} \,d\lambda\,~~,
\end{equation}
is a measure of the amount of optically thin dust per unit area of optically thick disk.\footnote{In our analysis EW(10$\micron$) does not depend much on temperature and composition. As is evident in the ubiquity of 10 \micron\ excesses and silicate emission features, disks around young stars or brown dwarfs always have a distribution of temperatures which exceed that required for efficient excitation of 10 \micron\ continuum (T $>$ 300 K), and therefore are well above that required for the silicate features, which are dominated by emission from cooler material at larger radii. We also know from the shape of the silicate emission profiles that essentially all the dust grains we see are optically thin (internally, that is), and composed of amorphous and crystalline silicates (see, e.g., \citet{Sargent+2009-GrainGrowth}). In this case EW(10$\micron$) does not depend upon mass fractions of amorphous and crystalline material. One way to see this is to note that a grain with a given number N of silicate monomers has a fixed number of oscillators, with possibly a small range of oscillation frequency, in any given vibrational mode. Absorption or emission integrated over that mode is, to first order, proportional to N. (This is essentially the Thomas-Kuhn sum rule.) Thus two grains with the same N but different mineral fractions have the same equivalent width in the same vibrational mode -- such as that which produces the 10 \micron\ silicate feature -- though the one with larger mineral fraction will have a larger number of oscillators at a small set of fixed frequencies, and identifiable sub-structure to the silicate feature. } Thus, a large EW(10$\micron$) is a sign of a large amount of optically thin dust and/or an indication of a reduction in disk continuum due to the absence of optically thick disk in the region where the 10 micron emission feature is formed. Here we use the continuum spectral indices along with $EW(10\micron)$ to identify disks with central clearing or radial gap. The principles are that small values of $n_{K-6}$ --- down to the color of photospheres --- and large values of $n_{13-31}$ indicate the spectral ``transition" that signifies a gap with outer gap radius in the few- to few-tens-of AU range; and that large values of $EW(10\micron)$ connote warm optically thin dust in gaps.

To find out break points for TDs among the Class II YSOs distribution in these observed parameter spaces, we utilized all ($\sim$600) IRS spectra of Class II YSOs observed in Orion A, Tau, ChaI, Oph, and N1333. Using the properties of TDs already well identified in Tau \citep{Furlan2011Taurus}, Cha I \citep{manoj2011}, Oph \citep{McClure2010}, and N1333 \citep{Arnold2012} as a guide, we found the breaks occur to at

\begin{tabbing}
 \= $\bullet$ $n_{K-6}$ $\leq$ -2.1 (the lowest octile for the distribution of $n_{K-6}$)\\
 \= $\bullet$ $n_{13-31}$ $\geq$ 0.5 (the highest octile for the distribution of $n_{13-31}$)\\
 \= $\bullet$ $EW(10\micron)$ $\geq$ 4.3 (the highest octile for the distribution of $EW(10\micron)$)
\end{tabbing}
In Figure~\ref{fig-OriA-tdselection} we plot $n_{13-31}$ vs. $n_{K-6}$ and $n_{13-31}$ vs. $EW(10\micron)$ for objects in Orion A. We also added objects in Tau \citep{Furlan2011Taurus} in the plots for comparison.

To identify TDs in Orion A, we first selected objects satisfying one of the above conditions as possible candidates. Second, we rejected objects with spectral types earlier than G because they have significantly higher masses ($>$ 2 $M_\odot$) and their circumstellar disks evolve much faster by possibly different disk clearing mechanisms than the case of transitional disks around the low-mass T Tauri stars. However, we keep objects with unknown spectral type as all appear to lie in the same luminosity range as the low-mass T Tauri stars. Third, we examined the detailed shape of the SEDs, and selected objects with deficits of infrared excess in the 2-8 micron range, compared to the appropriate median IRS spectrum of Class II disks in Tau \citep{Furlan2011Taurus}. We used the median of Taurus K5-M2 for objects with the spectral type earlier than M3, and the median of Taurus M3-M5 for objects with the spectral type of M3 or later. We adopt the Taurus median spectra instead of Orion A median because transitional disks in Tau and the median SED of Class II disks in Tau have been well studied and used widely for comparison in studies of other star-forming regions. As shown in Figure~\ref{fig-other-tdselection}, the previously characterized TDs in Tau, Cha I, Oph, and N1333 pass these filters clearly.

In Table~\ref{table:TDtype}, we indicate how the TDs in this paper were selected based on the selection criteria. Three objects, OriA-44, OriA-88, and OriA-172, are classified as TDs based upon examination of their SEDs which are indicative of gap/central hole, despite the fact that they do not pass any of the three criteria. The spectra of 62 transitional disks in Orion A appear in Figure 2. We use these 62 TDs in Orion A and 43 TDs from Tau (13), Cha I (11), Oph (10), and NGC 1333 (9) to unveil properties of transitional disks in the following.

\subsubsection{Subclassification of Transitional Disks: CTD, WTD, and PTD}
The mid-infrared spectrum of a CTD like DM Tau and CoKu Tau/4, having a few-AU to few-tens of AU central clearing, is distinctive compared to a radially-continuous Class II disk: it shows very little continuum excess over the photosphere from near-IR (1-2 $\micron$) to wavelengths in the mid-infrared (around 8-13 \micron), at which point the excess increases exponentially with increasing wavelength until it matches or exceeds the median spectrum. The other two types of TDs have excesses from near-IR to the shorter mid-infrared wavelengths ($\sim$5-8 \micron) that are smaller than or similar to the median. In some, like LkCa 15 and UX Tau A, veiling in the near IR spectra points to an underlying optically thick inner disk \citep{espaillat07b,espaillat2008}, and their distinctive spectrum corresponds to optically thick inner and outer disks separated by a gap. These are PTDs. Intermediate between CTDs and PTDs are WTDs, in which the weaker near- and mid-infrared excess is best explained by an optically-thin inner disk separated by a gap from an optically-thick outer disk, as in GM Aur \citep{calvet05,Espaillat2010ptd}.
Therefore, the distinction between CTDs, WTDs, and PTDs is whether an (optically thick/thin) inner disk exists or not.

It is useful to define the Inner Disk Excess Fraction (\textit{IDEF}) to characterize the near-infrared and shorter-wavelength mid-infrared excess fraction relative to the K5-M2 median spectrum of Class II sources in Tau. Since our sample is complete in H band data from the 2MASS catalog and we have IRS spectra starting at $\sim$5.2 $\micron$, we interpolate H band to 6 $\micron$ to acquire the assumed spectrum covering 1.65-6 $\micron$:
\begin{equation}
\textit{IDEF} = \frac{\ \int_{1.65\micron}^{6\micron} {F_{\lambda,obj}-F_{\lambda,photosphere} \,d\lambda\,}}{\ \int_{1.65\micron}^{6\micron} {F_{\lambda,median}-F_{\lambda,photosphere} \,d\lambda\,}}
\end{equation}
In the case of OriA-88 for which we do not have an SL2 spectrum, the flux at IRAC channel 3 (5.8 \micron) is used instead of IRS fluxes.

By taking the already well studied Taurus TDs as references, we adopt a set of infrared-excess ranges to subclassify TDs:
\begin{tabbing}
 \= CTD: \textit{IDEF} $<$ 0.25\\
 \= WTD: 0.25 $\leq$ \textit{IDEF} $<$ 0.5\\
 \= PTD: \textit{IDEF} $\geq$ 0.5
\end{tabbing}

The \textit{IDEF} values for the TD subclassification criteria are derived by adopting the K5-M2 median spectrum. The M3-M5 median spectrum is fainter than the K5-M2 median spectrum due to an effect of lower stellar luminosity and lower disk emission\citep{Furlan2011Taurus}. Therefore the excess emission over photosphere at 2-8 $\micron$ is weaker. It may lead to misclassification to compare objects of M3 or later spectral type to the K5-M2 median spectrum: an \textit{IDEF} value of an object with M3 or later spectral type derived from the K5-M2 median spectrum is generally lower than that derived from the M3-M5 median spectrum. Furthermore, there are no WTDs and few PTDs with M3 or later spectral type among the already well studied TDs in Tau, Cha I, and Oph.
Thus, for M3-M5 types, we rely on confirmation by other TD selection criteria. Considering the additional contributions such as the disk inclination and scattered light to the degeneracy of the interpretation of the excess emission in the near-infrared, we consider the subclassification of TDs based on \textit{IDEF} as preliminary. We note the subclassification based on \textit{IDEF} values and some cases of exceptions of using this criteria in Table~\ref{table:TDtype}.

With these criteria we obtain 34 CTDs, 15 WTDs and 13 PTDs in Orion A. Adding the four other nearby associations of Tau, ChaI, Oph, and N1333 brings the totals to 47 CTDs, 17 WTDs and 41 PTDs. Spectral indices of these objects are plotted in Figures~\ref{fig-OriA-tdselection} and \ref{fig-other-tdselection}.

Most CTDs are placed below the lower octile of $n_{K-6}$, which reflect their negligible short-wavelength infrared excess from the inner disk, while PTDs are distributed over the whole range greater than the lower octile of $n_{K-6}$. The WTDs are mostly located between the distribution of CTDs and PTDs in $n_{K-6}$.

In the plot of $n_{13-31}$ vs. $EW(10\micron)$ in Figure~\ref{fig-OriA-tdselection} and Figure~\ref{fig-other-tdselection}, we also indicate the ranges occupied by radially-continuous disk models with a range of inclination angles, stellar masses, degrees of dust settling, and mass accretion rates \citep{d'alessio06,dmw_2009,espaillat_2009_phDthesis}. Most of TDs in Orion A lie outside the model polygon. However, in contrast to the positions of TDs in other star-forming regions, some of CTDs, WTDs, and PTDs in Orion A are located inside the polygon, indicating that they could be modeled as radially-continuous disks, with respect \textit{only} to these two parameters. With the exception of IRS-18 (PTD) and IRS-154 (WTD) in ONC, the transitional disks inside of the polygon are located in the region for vertically well-mixed disks (the area toward the upper right side in the polygon). Few objects lie in the domain of well-mixed disks for Tau, ChaI, Oph and N1333. \citet{Furlan2009TauChaOph} concluded from this that substantial disk structural evolution, especially settling of dust to the disk midplane, has occurred in 1-2 Myr. \citet{Arnold2012} showed that the disks in N1333 are statistically indistinguishable from those in Tau, Cha I and Oph in this regard.

In the ternary plot of Figure~\ref{fig-OriA-ternary} we see the distribution of TDs are separated from the radially-continuous disks in the three dimensional parameter spaces of $n_{K-6}$, $n_{13-31}$, and $EW(10 \micron)$: especially CTDs and WTDs are nicely located in the different region from the region occupied by the radially-continuous disks.

We plot $n_{13-31}$ and $EW(10 \micron)$ of disks in Orion A along IDEF in Figure~\ref{fig-OriA-IDEF}, as well as TDs in Tau as the references of criteria along IDEF. We find that $n_{13-31}$ of TDs decreases as IDEF increase, i.e., CTDs tend to have larger $n_{13-31}$ than PTDs. We note that CTDs with low IDEF span through the ranges of $EW(10 \micron)$, whereas most PTDs have high values ($>$4) of $EW(10 \micron)$.

\subsubsection{TD Selection Criteria in the Literature}
We have compared our selection criteria for TDs to other selection criteria used in the literature. We plot TDs as well as Class II sources in OriA on the color-color diagrams used for the selection criteria by \citet{Merin2010} (left panel), \citet{cieza_2010_TDsOph} (middle panel), and \citet{Muzerolle2010} (right panel) in Figure~\ref{fig-c2dselection}. These authors used \textit{Spitzer} IRAC and MIPS broadband photometry, not IRS spectra, for their selection criteria.

\citet{Merin2010} identified and characterized disks with inner holes from the \textit{Spitzer} c2d Legacy program. They used [3.6]-[8.0] vs. [8.0]-[24] color-color diagram to identify TDs that fell in two separate regions:
\begin{tabbing}
    \= Region A: 0.0 $<$ [3.6]-[8.0] $<$ 1.1 and 3.2 $<$ [8.0]-[24] $<$ 5.3 \\
    \= Region B: 1.1 $<$ [3.6]-[8.0] $<$ 1.8 and 3.2 $<$ [8.0]-[24] $<$ 5.3
\end{tabbing}

Region A selects TDs with central clearings like CTDs in our sample. Region B corresponds to the disks with some excess flux in the IRAC bands like WTDs and PTDs in our sample. Their selection criteria of region A agrees with our criteria for CTDs in L1641, but some number of CTDs in Orion A fall in Region B or outside the Region A and B (the left panel of Figure~\ref{fig-c2dselection}).

\citet{cieza_2010_TDsOph} set selection criteria based on the [3.6]-[24] vs. [3.6]-[4.5] color-color diagram, in which TDs are located in the region where [3.6]-[4.5] $<$ 0.25 and [3.6]-[24] $>$ 2.
We see that most of our CTDs fall in that region, but PTDs have [3.6]-[4.5] $>$ 0.25 (the middle panel of Figure~\ref{fig-c2dselection}).

\citet{Muzerolle2010} used spectral slopes (i.e. spectral indices) at two different wavelength intervals, 3.6-5.8 $\micron$ and 8-24 $\micron$. The criteria for the classical transitional disks with central holes based on a red $\alpha_{8-24}$ in \citet{Muzerolle2010} agrees with our criteria for CTDs. However, similar to \citet{cieza_2010_TDsOph}, most of our WTDs and all of our PTDs will be missed by this criterion (the right panel of Figure~\ref{fig-c2dselection}).

From the comparisons of selection criteria for TDs, we find that (1) CTDs can be commonly identified by a variety of criteria; (2) a criterion utilizing the largest wavelength interval in the IRAC channels (between ch1 (3.6 $\micron$) and ch4 (8 $\micron$) in \citet{Merin2010}) can select WTDs and PTDs, but a color (i.e. the spectral slope) based on other broadband channels, such as [3.6]-[4.5] in \citet{cieza_2010_TDsOph} or 3.6-5.8 $\micron$ in \citet{Muzerolle2010}, cannot find PTDs.

\subsection{Radial Properties of Transitional Disks: $R_{wall}$}
One of the most important properties of a transitional disk is how large the gap or hole is; that is, the radius at which the inner wall of the optically thick outer disk lies. Bearing in mind the limitations of inference of structure from the SED, we derive the radius, $R_{wall}$, from the shape of the spectrum at the transition from small to large infrared excess within the IRS spectrum. Our procedure, illustrated in Figure~\ref{fig-estRw}, is the same as in our previous work \citep{kimkh2009}: we model the inner edge of the outer disk as an optically thick insulating wall, with dust temperature $T$, as follows. We first subtract a power-law fit to the IRS spectrum in the 5-8 $\micron$ region to remove the flux from the photosphere or an excess from the inner disk. To the residuals, we fit a model with two components tightly constrained at 13-16 $\micron$ and 30-33 $\micron$. One component is emission from optically-thin astronomical-silicate dust \citep{dl84} with 0.1 $\micron$ radius, to represent the inner disk and the bulk of the optically thin atmosphere of the outer disk. The other component is a single-temperature ($T$) blackbody that represents the insulated inner edge of the optically thick outer disk. As we described in \cite{kimkh2009}, we do not aim to fit the details of the spectrum perfectly, but merely to separate the optically-thick continuum --- for which the SED shape is the signature of the wall --- from the silicate emission features centered at 10 and 20 $\micron$. With the temperature $T$, we calculate $R_{wall}$ using radiative equilibrium at the inner wall:
\begin{equation}
R_{wall} = \sqrt{\frac{L_{\star}(1-A_{\star})}{4 \pi \sigma T^{4} \epsilon_{IR}}} \left( \equiv \sqrt{\frac{L_{\star}}{4 \pi \sigma T^{4}}} \right)
\end{equation}
where $A_{\star}$ is the effective albedo at stellar wavelengths and $\epsilon_{IR}$ is the effective emissivity at mid-IR wavelengths. We adopt $(1-A_{\star})/\epsilon_{IR} = 1$ : a perfectly black wall.

We compared the resulting $R_{wall}$ of several transitional disks in Tau (CoKu Tau/4; DM Tau; GM Aur; LkCa 15; UX TauA), Cha I (CR Cha; SZ Cha; T11; T25; T35; T56), and Oph (Rox 44, 16126-2235AB) with the same quantity determined from detailed self-consistent models \citep{Espaillat2010ptd,Espaillat2011TDvar}. We found the derived $R_{wall}$ from Equation (9) are within about $33\%$ of those obtained from detailed models. We also compared our $R_{wall}$ estimates to the cavity radius ($R_{cav}$) derived from the Submillimeter Array (SMA) observation for DM Tau, GM Aur, LkCa 15, UX Tau A, and ROX 44 \citep{andrews2011_sma_CTDs}, and these results lie within $34 \%$ of each other. Comparison of $R_{wall}$ from the detailed model and the $R_{cav}$ observed by SMA also gives a similar range of difference, about $34\%$. Therefore, we adopt 33 $\%$ as the uncertainty of $R_{wall}$ estimated from Equation (9). The $R_{wall}$ results are listed in Table ~\ref{table:property}.

\section{FREQUENCY AND AGE DISTRIBUTION}
Most previous studies of transitional disks \citep[e.g.][]{kimkh2009,Muzerolle2010,Merin2010,Currie-Sicilia_Aguilar2011_TD} and protoplanetary disks \citep{Furlan2009TauChaOph,Luhman+2010_TauDisks,Manoj2010-ASPC} have estimated the fractions of transitional disks to infer when the transition occurs and how long the transition stage persists. We take the fraction of CTDs+WTDs to compare to the Muzerolle's CTDs because the qualitative definition of the classical transitional disks in \citet{Muzerolle2010} is most nearly equivalent to the definition of the combined set of CTDs and WTDs in our sample. We note that the quantitative selection criteria used by \citet{Muzerolle2010} missed some WTDs as discussed in \S4.1.2

The fraction of TDs in this work is defined as n(TD type)/n(disks) using the number of TDs and number of Class II sources identified from the IRS survey of each region (see, Table~\ref{table:agefreq}). We exclude samples in Ophiuchus for the frequency and age trend search even though it is listed in Table~\ref{table:agefreq} because the selected TDs are from several sub-regions of Ophiuchus with differing estimated ages, and the sample sizes are too small to separate by subregion with high statistical significance.

The TD fraction is plotted as a function of age in Figure~\ref{fig-xdiskfreq}. We note that (1) the fraction of CTDs+WTDs can be high over the broad age ranges (1-10 Myr) and (2) the fraction of TDs varies from region to region at young ages ($<$ 3 Myr), with some having a fraction of only a few $\%$ while others have fractions of $>$20 $\%$ even at age $\leq$ 1 Myr.

Also shown in Figure~\ref{fig-xdiskfreq} are the fractions of transitional disks of each type from this work only: CTD (square), WTD (circle), PTD (star), and the total of them (cross). If we consider the fraction of TDs including all CTD, WTD, and PTD, the fractions of all TDs in each region are significantly high even at very young ages of less than 1 Myr: 17$\pm$6$\%$ for N1333, 25$\pm$4$\%$ for ONC, and 26$\pm$5$\%$ for L1641 versus the TD fractions for the 2-3 Myr old associations: 8$\pm$2$\%$ for Tau and 16$\pm$5$\%$ for Cha I.

\section{DISTRIBUTION AND TRENDS OF PROPERTIES}

\subsection{Transitional Disk Types {\it vs.} Spectral Type, $R_{wall}$, and $\dot{M}$}
In this section, we examine how the distributions of host star spectral types, $R_{wall}$, and $\dot{M}$ compare for transitional disks with different inner disk structures: CTD, WTD and PTD types. In this analysis we exclude the TDs for which we do not have reliable host-star spectral types.

In Figure~\ref{fig-histSpT}, we show the spectral type distribution for all TDs and for each subtype.
The median spectral type differs slightly among the TD subtypes: M2 for CTDs; M1 for WTDs; K7 for PTDs; M0 for WTD+PTD; M1 for TDs. The spectral type distributions of CTDs and PTDs differ noticeably, even if we consider the general spectral type uncertainty of one to two subtypes.
However, a Kolmogorov-Smirnov (K-S) test shows that the spectral type difference between CTDs and PTDs is of marginal statistical significant: for CTDs vs. PTDs, $D=0.31$ and $p=0.037$; where $D$ is the maximum deviation between the cumulative distribution of two groups and $p$ indicates the probability that there is no significant difference between the distributions.

In Figure~\ref{fig-histRw}, we show the frequency distribution of $R_{wall}$ for each TD subtype. For almost 90$\%$ of TDs $R_{wall}$ is less than 30 AU. This pattern is similar whether the inner disk exists or there is an empty inner cavity. However, we see that the median $R_{wall}$ is smaller for CTDs than WTDs and PTDs.
Most CTDs have $R_{wall} <$ 10 AU (about 52 $\%$), whereas 71 $\%$ of WTDs and PTDs with inner disks have $R_{wall} > 10$ AU. A K-S test that compares the $R_{wall}$ distributions of CTDs and PTDs yields values similar to those for the comparison of the spectral type distributions: $D=0.29$ and $p=0.05$, which is not very significant. The $p$ value decreases to $0.02$ when comparing CTDs and WTDs+PTDs.
The difference in the $R_{wall}$ distributions of the TD subtypes could also be due to the impossibility of distinguishing PTDs with very small gaps from radially continuous disks based on the IR spectrum alone. Overall, Figure~\ref{fig-histRw} shows that the IRS spectra are most sensitive to disk holes with $R_{wall}$ $<$ 30 AU.

In Figure~\ref{fig-histMdot} we plot the $\dot{M}$ distribution for all TDs and broken down by subtype. TDs with central clearings (CTDs) and gapped disks (WTDs and PTDs) differ significantly in $\dot{M}$. Both groups also have $\dot{M}$ substantially smaller than the typical $\dot{M}$ of radially-continuous disks. The median $\dot{M}$ of the gapped disks is $10^{-8.25}M_{\odot}/yr$ and that of CTDs is $10^{-8.7}$ $ M_{\odot}/yr$. This visible difference is confirmed by statistical analysis with the K-S tests. When we include the upper limits, the K-S test results $D=0.46$ and $p=0.001$. Even when we do not include the upper limits, the statistical significant difference between the two groups is still valid with $D=0.5$ and $p=0.002$. Thus we confirm that the $\dot{M}$s of CTDs are smaller on average than those of WTDs and PTDs.

For fair comparison of the $\dot{M}$ of radially-continuous disks and TDs, we need to be sure that these two groups have similar ages and stellar masses because $\dot{M}$ decreases with age \citep{hartmann98} and increases with $M_{\odot}$ \citep{muzerolle03mdot,calvet04mdot}. Because systematic differences of stellar masses and ages can arise from a choice of different evolutionary tracks \citep[e.g.][]{SiomDutreyGuilloteau2000-evocomp,najita07}, the best would be to have $\dot{M}$ estimated with the same method and assumptions used for $\dot{M}$ estimation of TDs. However, there is currently no such $\dot{M}$ survey of low-mass T Tauri stars. A comparison between $\dot{M}$ of Orion A TDs and other radially-continuous Class II disks measured in an homogeneous method will be discussed by \citet[in preparation]{khkim2013-OriA}.

Therefore, for the next best comparison, we use the $\dot{M}$ measurements of T Tauri stars in Taurus by \citet{gullbring98} for the following reasons. \citet{gullbring98} adopted the evolutionary tracks of \citet{D'Antona+mazzitelli-1997}, which tends to result in younger stellar ages and lower stellar masses. \citet{SiomDutreyGuilloteau2000-evocomp} found both the \citet{Baraffe+1998} model and \citet{siess2000} model agree with dynamical masses in the 0.7-1 $M_\odot$ range, while \citet{D'Antona+mazzitelli-1997} do not agree as precisely. Even though \citet{White+Ghez-2001} measured $\dot{M}$ of T Tauri stars in Taurus as adopting an evolutionary track combined \citet{Baraffe+1998} and \citet{Palla+Stahler-1999}, the samples are binary systems and $\dot{M}$ of binary system is not comparable to $\dot{M}$ of TDs.
Hence, we adopt the median $\dot{M}$ from \citet{gullbring98} as the median $\dot{M}$ of radially-continuous disks in Taurus, $\dot{M}=10^{-7.8} M_{\odot}/yr$.\footnote{We excluded three TDs, GK Tau, GM Aur, and IP Tau, in the sample of \citet{gullbring98} to measure the median $\dot{M}$ of radially-continuous disks in Taurus. If we include those three TDs, the median Mdot decreases to $\dot{M}=10^{8} M_{\odot}/yr$.}
While bearing in mind the possible uncertainties ($\sim$30$\%$) of stellar masses between two different evolutionary tracks, we confirm that the ages and stellar masses of the \citet{gullbring98} sample are similar to those of TDs studied in this work. Thus our results confirm that TDs in general have substantially smaller $\dot{M}$ than radially-continuous disks, in accord with the findings by \citet{najita07}, \citet{kimkh2009}, and \citet{Espaillat_TDclass_2012}.

\subsection{Trends among TD Properties}

Understanding the correlations between disk and stellar properties of TDs is an essential and important key to understand how protoplanetary disks evolve from radially-continuous optically thick disks to a final planetary system.
We have gathered the largest sample of TDs which are identified by homogeneous criteria. From this large sample, we are able to find not only trends of TDs in general but also detailed trends of the three different types of TDs.

Except for $\dot{M}$, we estimate the stellar/disk properties of TDs in Tau, ChaI, Oph and N1333 in the same ways as for Orion A TDs. The $\dot{M}$ of TDs of other star-forming regions are from literature or personal communication. Adopting $\dot{M}$ measured by different techniques from that for the TDs of OriA should not affect significantly these trends, based on the demonstration of the tight correlation between the luminosities of IR hydrogen recombination lines and $L_{acc}$ measured from the blue excess spectrometry and/or U-band photometry by \citet{muzerolle98Brgamma}. Therefore, we expect that any discrepancies in luminosity between the two (IR and UV) data sets on average will be small \citep{muzerolle98Brgamma}.
These stellar/disk properties and other properties obtained from the literature are also listed in Table~\ref{table:property}.

Among trends from many possible combinations of properties, we present the trends of interesting pairs of properties showing somewhat different behaviors from non-transitional disks in Figure~\ref{fig-trend-MstarMdotLx} through Figure~\ref{fig-subtrend_Mdot-Lx}.
The correlation parameters are calculated by using $linmix\_err.pro$ which was developed for the Bayesian approach to linear regression with errors in both X and Y, by \cite{kelly07linmixerr}, and the trends line in each plot can be read in the manner of
$\log_{10}(Y)$~$=$~$(\alpha \pm e\_\alpha) + (\beta \pm e\_\beta) \log_{10}(X)$.
We indicate the correlation parameters ($corr$) and probabilities ($P$) of pairs of properties in Table~\ref{table:correlation} for TDs without separation by subtype and in Table~\ref{table:correlation_subTDtype} for two subgroups of TDs separated by their radial disk structures, i.e., CTDs and WTPs+PTDs. In general, a 5$\%$ or lower $P$ value is considered to be statistically significant. Sometimes $P$~$\lesssim$~2$\%$ is considered as a conservative threshold of statistically significant. Therefore, we interpret a correlation to be a statistically significant with $P$~$\lesssim$~2$\%$ and to be a marginally significant if $P$ is 2-5$\%$.

\subsubsection{Trends of TDs}
Our search for trends related to $\dot{M}$ utilized weighted linear regression to account properly for upper limits of $\dot{M}$ in our samples. The resulting trend for $\dot{M}$-$M_\star$ in Figure~\ref{fig-trend-MstarMdotLx}-(a) is $\dot{M}$ $\propto$ ${M_{\star}}^{1.6\pm0.3}$. To compare this to the $\dot{M}$-$M_\star$ relations from the previous studies ($\dot{M}$ $\propto$ ${M_{\star}}^{1.95}$ \citep{calvet04mdot}; $\dot{M}$ $\propto$ ${M_{\star}}^{2.0}$ \citep{muzerolle03mdot}; $\dot{M}$ $\propto$ ${M_{\star}}^{2.1}$ \citep{muzerolle05}), we include in Figure~\ref{fig-trend-MstarMdotLx}-(a) a plot of the result, $\log \dot{M} \approx 2 \log M_{\star}-7.5$ \citep{muzerolle03mdot,Telleschi07}.
The slope of the $\dot{M}$-$M_\star$ relation of TD host stars is roughly consistent with previous studies among T Tauri stars, but the trend line for $\dot{M}$ is shifted downward by factor of about 10 with respect to the thick dashed line representing T Tauri stars in Taurus, consistent with the results discussed in Section 6.1.

In Figure~\ref{fig-trend-MstarMdotLx}-(b), we plot $L_{X}$ as a function of $M_{\star}$. Compared to the results in Taurus $\log L_{X} = 1.69\log M_{\star} + 30.33$ \citep{Telleschi07}, the regression of our TD data is $\log L_{X} = 1.1\log M_{\star} + 30.1$: slightly smaller slope than that of T Tauri stars.

Figure~\ref{fig-trend-MstarMdotLx}-(c) shows $\dot{M}$ as a function of $L_X$ compared to the Taurus results derived from Taurus's $\dot{M}$-$M_\star$ and $L_{X}$-$M_\star$, $\dot{M}$-$L_X$ relation for T Tauri stars $\log \dot{M} = 1.2 \log L_{X} - 43$. The $\dot{M}$-$L_X$ trend line for our TD sample, $\log \dot{M} = 0.2\log L_{X} - 14.6$. This shows that TD's $\dot{M}$ is not related to $L_{X}$, unlike CTTSs.

The most interesting relations are the correlation of $R_{wall}$ to stellar properties. In Figure~\ref{fig-trend-Rw}-(a), we find a very strong correlation between $M_\star$ and $R_{wall}$.
For both cases of $\dot{M}$ and $L_X$, we also see the increasing trends of $\dot{M}$-$R_{wall}$ in Figure~\ref{fig-trend-Rw}-(b) and $L_X$-$R_{wall}$ in Figure~\ref{fig-trend-Rw}-(c). However, we should recall that there is a tight correlation of $M_\star$-$R_{wall}$ and a statistically significant correlation of $\dot{M}$-$M_\star$. Therefore, the correlation showing in the plots of $\dot{M}$-$R_{wall}$ may merely reflect the combination of correlations between $M_\star$-$R_{wall}$ and $\dot{M}$-$M_\star$. To test this hypothesis, we looked for a trend between $\dot{M}$ and $R_{wall}$ in more restricted mass bins (e.g., 0.2-0.4 $M_\odot$, 0.4-0.7 $M_\odot$, and 0.7-2.3 $M_\odot$) and did not find significant trends. Similarly, the increasing trend of $L_X$-$R_{wall}$ may be the result of combination of the correlations between $L_X$-$M_\star$ and $M_\star$-$R_{wall}$.

For further examination of the difference between $\dot{M}$-$R_{wall}$ relationship of TDs and that of CTTSs under the strong dependence on $M_{\star}$, we derived an expected $\dot{M}(M_{\star})$-$R_{wall}$ of CTTSs by combining the $\dot{M}$-$M_{\star}$ of CTTSs ($\log \dot{M} = 2 \log M_{\star}-7.5$) and a strong $M_{\star}$-$R_{wall}$ correlation of TDs ($\log M_\star = 0.7 \log R_{wall}-1.0$): $\log \dot{M}(M_{\star}) = 1.4 \log R_{wall}-9.5$ (Figure~\ref{fig-trend-Rw}-(b)).
The expected $L_{X}(M_{\star})$-$R_{wall}$ of CTTSs considering the $M_\star$ dependence is also derived by combining $L_X$-$M_\star$ of CTTSs and $M_\star$-$R_{wall}$: $\log L_{X}(M_{\star}) = 1.2 \log R_{wall}-28.6$ (Figure~\ref{fig-trend-Rw}-(c)).
We find that the trends of TDs in $\dot{M}$-$R_{wall}$ and $L_{X}$-$R_{wall}$ generally agree with the trends of CTTSs.

Several other interesting correlations for TDs are listed in Table~\ref{table:correlation}. The very strong correlation of $L_{\star}$~$\sim$~${M_\star}^{1.6}$ which is similar to that ($L_{\star}$~$\sim$~${M_\star}^{1.5}$) found from T Tauri stars in Tau by \citet{Telleschi07} supports that the basic stellar properties, $M_\star$ and $L_{\star}$, of TDs are not very different from the stars hosting radially-continuous flared disks and/or homologously evolved disks.

\subsubsection{Trends of $\dot{M}$ and $L_{X}$ for Transitional Disk Subtypes}
In this section we explore characteristics of $\dot{M}$ and $L_{X}$, which are sensitive to the different inner disk structures and to the different disk evolution mechanisms, by searching for detailed trends of sub-samples grouped by different TD types. The correlation parameters of the detailed trend analysis for the case of inner clearings (CTDs) and the case of radial gaps (WTDs+PTDs) are listed in Table~\ref{table:correlation_subTDtype}.

It is clear that the trend between $\dot{M}$ and $M_\star$ of CTDs is different from that of WTDs+PTDs from the left panels ((a) and (b)) of Figure~\ref{fig-subtrend_Mdot-Lstar}. When the inner disk is essentially empty as for the CTDs, no correlation exists between $\dot{M}$ and $M_\star$. However, for the case of WTDs+PTDs, $\dot{M}$ and $M_\star$ are significantly correlated each other, and the relation ($\dot{M}$ $\propto$ ${M_\star}^{1.9}$) is very close to that of CTTS ($\dot{M}$ $\propto$ ${M_\star}^{2}$). Therefore, the increasing tendency shown in Figure~\ref{fig-trend-MstarMdotLx} probably leads to a strong effect from the significant correlation of $\dot{M}$ and $M_\star$ for WTDs+PTDs which still have an inner disk in their disk.
In contrast to $\dot{M}$, we find no significant difference between the two subgroups of TDs for the $L_{X}$-$M_{\star}$ correlation (the right panels ((c) and (d)) of Figure~\ref{fig-subtrend_Mdot-Lstar}).

Similar to the left panels of Figure~\ref{fig-subtrend_Mdot-Lstar}, in the left panels of Figure~\ref{fig-subtrend_Mdot-Rwall}, we also see no correlation of $\dot{M}$-$R_{wall}$ for CTDs but a significant and strong correlation for WTDs+PTDs.
To find the genuine behavior of TDs between $\dot{M}$ and $R_{wall}$ which is different from $\dot{M}$ of CTTSs, we divided $\dot{M}$ of our sample by the $\dot{M}(M_{\star})$ (the thick long-dashed lines).
The right panels of Figure~\ref{fig-subtrend_Mdot-Rwall} show the residual relation between $\dot{M}$ and $R_{wall}$ of CTDs ((c); anti-correlation) and WTDs+PTDs ((d); no correlation) after taking the dominant effect of $M_{\star}$ out. This supports the ideas that (1) the mass accretion behaves similarly to the case of radially-continuous disks while an inner disk exists in a TD, and (2) the mass accretion rate decreases as the size of inner cavity increases.

In the left panels of Figure~\ref{fig-subtrend_Lx-Rwall}, CTDs have a very strong and significant correlation between $L_{X}$ and $R_{wall}$, while WTDs+PTDs have a weak relationship with large uncertainty in the trend of $L_{X}$ and $R_{wall}$ comparing to that for all TDs combined. Therefore, we infer that the general increasing tendency of $L_{X}$ along $R_{wall}$ in Figure~\ref{fig-trend-Rw} is due to the dominant effect of the strong correlation for CTDs.
After removing the underlying contribution from $M_\star$ to $L_{X}$-$R_{wall}$ relation by dividing $L_{X}$-$R_{wall}$ by $L_{X}(M_{\star})$-$R_{wall}$, we see no correlation for CTDs and insignificant anti-correlation with large uncertainty for WTDs+PTDs from the right panel of Figure~\ref{fig-subtrend_Lx-Rwall}.

From Figure~\ref{fig-subtrend_Mdot-Lx}, we confirm that $\dot{M}$ and $L_{X}$ are not correlated with each other regardless of subtypes of TDs not like the strong correlation between them in case of T Tauri stars (the thick dashed line in the (a) and (b) panels). The residual $\dot{M}$ after removing the strong $\dot{M}$-$L_{X}$ correlation of T Tauri stars in the right panels are anti-correlated to $L_{X}$ for both CTDs and WTDs+PTDs.
We note that this is a similar to the trend found for CTTSs by \citet{Telleschi07}.
\citet{Drake+2009} also shows the similar result that the objects with higher $L_X$ have lower $\dot{M}$ and find a strong anti-correlation between $L_{X}$ normalized to $M_\star$ and $\dot{M}$ with CTTS/WTTS in ONC. Therefore, the anti-correlation shown in the right panels of Figure~\ref{fig-subtrend_Mdot-Lx} for both CTDs (c) and WTDs+PTDs (d) may be a common characteristic of protoplanetary disks, not a unique characteristic of TDs.

\subsection{Summary of Significant Trends}

Here, we summarize the correlations and trends of TD properties we find to help our insights on the origin of TDs, which will be discussed through the next section.

$L_{\star}$ and $M_{\star}$ are independent of the subtypes of transitional disks. The significantly strong correlation between $M_{\star}$ and $R_{wall}$ is consistent regardless of the subtype of TDs as $M_{\star}$ $\propto$ ${R_{wall}}$.

$\dot{M}$ and $L_{X}$ vary/evolve with time. We have seen that some trends differ by subtypes of TDs and some trends are comparable to those of the larger T Tauri star population overall. The residual (or normalized) properties after removing the underlying effect of $M_\star$ which is a parameter correlated with $\dot{M}$ and $L_{X}$ may show the effect of the transition process from a radially continuous disk to a transitional disk.

Some properties have very different trends according to whether the inner disk exists (WTD/PTD) or not (CTD).

(1) Trends related to $\dot{M}$ show a strong correlation with WTDs+PTDs, but no correlation with CTDs: \\
$\bullet$ $\dot{M}$ vs. $M_\star$: WTDs+PTDs show a very similar correlation to that of T Tauri stars with $\dot{M}$ $\propto$ ${M_\star}^{1.9}$. In stark contrast, no correlation shows for CTDs. \\
$\bullet$ $\dot{M}$ vs. $R_{wall}$: WTDs+PTDs show a very tight correlation between the two properties, $\dot{M}$ $\propto$ ${R_{wall}}^{1.9}$, but the two properties are not correlated for CTDs.\\

(2) Trends of some properties are strongly correlated for CTDs, but not for WTDs+PTDs:\\
$\bullet$ $L_{X}$ vs. $M_\star$: The trend of CTDs ($L_{X}$ $\propto$ ${M_\star}^{1.8}$) alone is very close to the trend of T Tauri stars ($L_{X}$ $\propto$ ${M_\star}^{1.7}$), but the slope ($\beta$) of the trends of WTDs+PTDs is smaller by about a factor of two ($L_{X}$ $\propto$ ${M_\star}$).\\
$\bullet$ $L_{X}$ vs. $R_{wall}$: A significant correlation between $L_{X}$ and $R_{wall}$ for CTDs, but the trend for WTDs+PTDs is very uncertain.\\
$\bullet$ \emph{residual} $\dot{M}$ vs. $R_{wall}$: The residual trend of $\dot{M}$ and $R_{wall}$ is opposite to the trend before removing $M_\star$ effect from $\dot{M}$ vs. $R_{wall}$: CTDs show an anti-correlation, but WTDs+PTDs do not show any correlation.\\

On the other hand, $\dot{M}$ vs. $L_{X}$ show no correlations from any subtype of TDs, in contrast to the strong correlation for T Tauri stars with full disks.\\

\section{MECHANISMS FOR THE ORIGINS OF TRANSITIONAL DISKS AND CONSTRAINTS}
In this section, we briefly review proposed mechanisms for disk dispersal and compare how our findings are consistent, or not, with these predictions from the mechanisms: dust coagulation and settling; photoevaporation; inside-out disk clearing by MRI; gravitational effects of one or more low-mass companions.

\textit{Grain growth and settling.} Dust grains should grow and settle to the disk midplane during the protoplanetary disk evolution process. As grains grow larger, the opacity of the grains becomes smaller, leading to weaker continuum emission in IR range. On this basis, it has been proposed that the flux deficit shown in SEDs of TDs may be due to the existence of an opacity hole caused by grain growth and settling in the inner disk rather than a real material deficit in the inner disk \citep[e.g.][]{duldom05}. \citet{Tanaka_2005_DustGrowthSettling} suggested that the SEDs of TDs could be due to differing opacity as a function of radius; e.g., smaller disk optical depth in the inner disk than in the outer disk at 10 $\micron$. Some authors \citep{Garaud2007,Brauer+2007,Brauer+2008a} have demonstrated the short time scale of grain growth and settling. However, considering grain growth/settling only as the main mechanism of TDs cannot explain the sharp transition represented by the sharp edge of the inner wall of the outer disk which is implied from the distinctive characterstic of TD's SEDs and is supported by many resolved submillimeter images of TDs \citep{Hughes_GMAur_2009ApJ...698..131H,andrews2011_sma_CTDs}.In any case, grain coagulation and settling is a process that takes monotonically longer at increasing radius within the disk, so it cannot possibly explain the gaps in WTDs and PTDs.

\textit{MRI.} A mechanism important in the large scale mixing and turbulence in a disk is the Magnetorotational Instability (MRI). \citet{chimur07} showed how MRI accelerates mass accretion and leads to inside-out disk clearing. \citet{perez-becker_chiang2011MRI-FUV} considered the ionization, necessary for the MRI mechanism, by stellar FUV radiation and demonstrated that the surface layer accretion driven by this could reproduce the trend of increasing accretion rate with increasing hole size seen in TDs. \citet{Suzuki_MRI_2010}, using MHD simulations with X-rays as the ionization source, showed the disk winds driven by MRI leads to a decrease in surface density in the manner of inside-out dispersal irrespective of the existence of a deadzone.

These models based on the inside-out disk clearing by MRI draining predict correlations between $R_{wall}$, $\dot{M}$, and $L_{X}$: (1) a positive correlation between $\dot{M}$ and $R_{wall}$ at a given $M_{\star}$ because a larger ionized area leads to more mass accretion; (2) a positive correlation between $\dot{M}$ and $L_{X}$ at a constant $R_{wall}$ because more ionization of disk surface as exposed by stronger energy sources leads to more mass accretion. However, our findings on the trends of $\dot{M}$, $L_{X}$, and $R_{wall}$ do not support these predicted correlations between $R_{wall}$, $\dot{M}$, and $L_{X}$. First, a strong correlation of $\dot{M}$ with $R_{wall}$ is shown only for the case of TDs with radial gaps (WTDs+PTDs) not for the TDs with inner cavities (CTDs). Furthermore, that correlation disappears when differing stellar masses are accounted for. Second, we found no correlation between $\dot{M}$ and $L_{X}$ for TDs regardless of their subtypes. To test for correlation between $\dot{M}$ and $L_{X}$ at a given $R_{wall}$, we examine trends for several different $R_{wall}$ ranges in Figure~\ref{fig-Mdot-Lx-constRwall}. We find no correlation between $\dot{M}$ and $L_{X}$ for any ranges of $R_{wall}$.

\textit{Photoevaporation.} Photoevaporation has long been thought to be an important and major disk dispersal mechanism \citep{Shu_Najita_Ostriker_Wilkin_1994,Hollenbach_1994,cgs01,Font_2004,acp06a,acp06b}. High energy radiation from the central star ionizes and heats the disk surface, which can be unbound and leaves as a wind beyond a certain radius. A gap can be opened when the accretion rate is small enough that material from the outer disk beyond the photoevaporation radius cannot replenish the inner disk. The inner disk, decoupled from the outer disk, drains on to the central star while the outer disk is evaporated into space. These models encounter contradictions when only the EUV ionizing radiation is considered: (1) the mass accretion rates of TDs is higher than the photoevaporation rate (or mass loss rate), (2) the ages of many TDs are less than the dispersal time scale of photoevaporation, and (3) the existence of the large radial gaps of PTDs.
Recently, the photoevaporation model has been modified to address those issues by taking much higher energy radiation sources (FUV and X-ray) as sources \citep{ercolano_2009_XrayPhoEva,Gorti2009a_photoevaporation,owen_2010_X-EUV,owen_2011b_Xray,owen_2011a_edgeon,Owen_photoevaporation2012}: (1) much higher mass loss rate can be driven by X-rays \citep{ercolano_2009_XrayPhoEva}; (2) 1-10 AU gap creation at relatively early epochs (3-4 Myr) \citep{Gorti2009b_time_evo_photoevaporation}. Considering disk dispersal by photoevaporation through EUV, FUV, and X-ray radiation, \citet{Owen_photoevaporation2012} concluded that the stars' intrinsic X-ray luminosity should have a decisive role in disk's life times and evolution.

We compare our data to the theoretical predictions from the X-ray photoevaporation (XPE) model by \citet{Owen_photoevaporation2012}.
Both the TD data and the XPE model have tendencies of increasing $R_{wall}$ and $\dot{M}$ as $L_{X}$ and $M_\odot$ increase. However, most TDs of all three types in our sample do not fall into the model domains, as indicated in Figure~\ref{fig-owen-XPE-fig18}\footnote{\citet{Owen_photoevaporation2012} note the difficulty of determining a correlation between any two of the three properties because $L_{X}$, $R_{wall}$, and $M_\star$ are interdependent. Therefore, the predictions in the Figures are the results from numerical simulation.} and Figure~\ref{fig-mdot-Rw-model}. In particular, the XPE model fails to explain the objects with either large $R_{wall}$ or the large $\dot{M}$. More than half of all TDs fall outside the region allowed by the XPE in Figure~\ref{fig-mdot-Rw-model}. Therefore, we conclude that X-ray photoevaporation is not the dominant mechanism for creating transitional disks.

\citet{morishima_2012_deadzone_photoevaporation} developed a gas disk model taking into account layered accretion driven by MRI and X-ray photoevaporative winds. With a central star of 1 $M_\odot$ and an initial disk mass of 0.1 $M_\odot$, they found their gas dispersal model can open a gap at large radii while the mass accretion rate is still similar to that of CTTS when dead zones are considered. They show that their model with a dead zone can reconstruct the distribution of the observed transitional disks with high $\dot{M}$ and large $R_{wall}$, which is the range X-ray photoevaporation cannot reconstruct.

The average disk mass of Class II objects measured by observations of the submm-mm continuum of disks (\citet{aw05}; \citet{aw_2007_Oph}) in Tau and Oph is about 5 $M_{J}$. This is much less than the minimum mass solar nebular (MMSN) and the requirement of disk mass to form giant gas planets or multiple planets which are observationally confirmed. It seems likely that submillimeter continuum observations systematically underestimate disk masses (e.g. \citet{D'Alessio2001}, \citet{hartmann06}). Alternatively, systems could have more substantial disk mass at the Class 0/I stage \citep{greaves_rice_2010}.

\textit{Gravitational effects of companions.} If the disks start with larger masses, photoevaporative gas dispersal models may encounter another challenge to deplete disk material because more massive disks requires stronger energy sources to evaporate disk material. On the other hand, for a more massive disk, the more favorable mechanism to open and clear a gap in a protoplanetary disk is giant planet formation by gravitational instability. Recent theoretical results on planet formation by gravitational instability suggest the possibility of gas giant planet formation at even much closer distances to the central star (10 AU by \citet{Inutsuka_2010_gaseousplanet_GI}; R $<$ 25 AU by \citet{Meru_Bate_2010_giantplanet_GI}). Even giant planet formation \textit{via} core accretion model, which requires longer time scales, is possible in $\sim$1 Myr with disks substantially larger than the MMSN \citep{Dodson-Robinson2009-MMSN,greaves_rice_2010}.

Giant planet formation in protoplanetary disks has been proposed as the origin of gaps/holes of TDs and debris disks \citep[e.g.][]{Holland_1998Natur.392..788H,Jura+Turner_1998Natur.395..144J,Zuckerman+Song_DebrisPlanets2004}. Several hydrodynamical simulations show the companion's presence can reproduce the sharp inner disk truncation and gap formation in a short time scale \citep{quillen04,Rice_Armitage+2006MNRAS.373.1619R,varni06a,Zhu_2011_multiplanets,Zhu+DustFiltration2012ApJ}.

Whether the companion is a planetary object or a sub-stellar companion, the observational features can be explained by the dynamical effects of the companion on the central star. The distribution of $\dot{M}$ shown in Figure~\ref{fig-histMdot} strongly supports the idea of gap opening and disk dispersal by planet/companion formation. The displacements of $\dot{M}$ of the gapped disks and the CTDs from the median $\dot{M}$ of CTTS is almost a factor of 10, which is consistent with the estimated decrements of the mass flow across a gap created by a low-mass companion \citep{Lubow_Dangelo_2006}. As we discussed in \citet{kimkh2009}, the strong correlation between $L_\star$ or $M_\star$ vs. $R_{wall}$ is reminiscent of the observed dependence between binary separation and the system's stellar mass. While this similarity might be taken to suggest that the stellar mass vs. separation dependence is imprinted when stars form, this argument has an important caveat. The strong trend of $M_\star$-$R_{wall}$ of TDs is possibly driven by strong correlations of $L_\star$-$M_\star$ and $L_\star$-$R_{wall}$. More detailed studies are needed to investigate this possibility.

Giant planet formation can also explain the trends shown in $\dot{M}/\dot{M}(M_{\star})$-$R_{wall}$ (Figure~\ref{fig-subtrend_Mdot-Rwall}). Assuming (1) $M_{disk}$ $\propto$ $M_\star$ at a given age which implies (2) normalization by $M_\star$ is the same as normalization by $M_{disk}$, the mass accretion rates of WTDs+PTDs may not be strongly related to $R_{wall}$ since material accreting from an inner disk may not be dependent on a planet formation location or $R_{wall}$ and the mass accretion from an inner disk may be dominant than mass accretion through a gap from an outer disk.
In the case of CTDs when the inner disk is depleted, however, the outer disk mass will decrease as the size of an inner cavity increases. That will result less material to accrete through the larger inner hole, therefore the anti-correlation between $\dot{M}/\dot{M}(M_{\star})$ and $R_{wall}$ of CTDs is naturally explained.

The disk-clearing companion can be a stellar companion such as the case of CoKu Tau 4 \citep{ik08} because a large fraction of stars form in multiple systems. If a stellar companions are the dominant origin of TDs, they should be of lower mass than the primary to explain the strong correlations in $\dot{M}$-$M_\star$ of WTD+PTD.
Sensitive searches have been made for 10 AU-scale binaries in Taurus (\citet{Pott_2010}, \citet{Kraus_2011_multiplicities_Tau}), but only one (CoKu Tau/4; \citet{ik08}) of the nine transitional disk systems so studied has a stellar companion been detected with binary separation similar to the gap radius. The limits on companion luminosity for the other eight rule out stars.
In addition, \citet{Kraus_LkCa15_2012AAS...21922803K} have tentatively detected an infant giant planet in the gap of the LkCa 15 disk. Therefore, planet formation could possibly be chiefly responsible for the origin of TDs.

Currently it is not clear that how different paths of disk evolution are followed, and which mechanisms dominate on a track of evolution given different initial conditions. We do not deny the contribution to the dispersal of disk material and disk evolution from other mechanisms such as grain growth, MRI action in the inner disk, and photoevaporation. In other types of evolved disks, such as the anemic, homologously depleted, and weak excess disks, which are not covered in this study, grain growth and photoevaporation could be dominant mechanisms. However, we find that giant planet formation is probably the dominant mechanism in gap formation for our sample of TDs, based on the trends reported here.

\section{Summary and Conclusions}
We presented the SEDs of 62 new TDs identified from IRS spectra in the Orion A star-forming region and discussed selection criteria for TDs. Utilizing the TDs already identified in Taurus, Chamaeleon I, Ophiuchus, and NGC 1333, in a manner similar to TDs in Orion A, we explored statistically robust trends with the largest and most homogeneous set of TDs to date.

We presented a set of TD selection criteria and a quantitative empirical method to classify three subtypes of TDs: CTD, WTD, and PTD.

We found the TD fraction of Class II YSOs is very high ($\sim$20$\%$) even at the youngest ages ($\leq$ 1 Myr, Orion A and NGC 1333). This could indicate early disk evolution even in the Class 0/I stages, most likely due to giant planet formation \citep{Lubow_Dangelo_2006}.

We have examined various mechanisms of disk clearing utilizing our observation of TDs. We confirm that disk clearing mechanisms including MRI action and photoevaporation generated by X-ray/FUV photons are consistent with some of the observations, but not all. In particular, there are several observational trends which cannot be explained by these mechanisms, especially those related to $\dot{M}$ and $L_X$: \\
(1) the observed mass accretion rate suppression of CTD's ($10^{-8.7}$ $M_\odot/yr$) and PTD's ($10^{-8.25}$ $M_\odot/yr$) compared to the radially continuous disks ($10^{-7.8}$ $M_\odot/yr$); \\
(2) the lack (or negative) correlation between $\dot{M}$ and $R_{wall}$ after accounting for $M_\star$'s effects \\
(3) no correlation of $L_X$ and $R_{wall}$ after correcting for $M_\star$'s effects\\
(4) no correlation of $L_{X}$ and $\dot{M}$ at constant $R_{wall}$. \\
However, these properties/trends are naturally explained by substellar companions, formed within the disks in the early stages of disk evolution. Infant jovian-mass planets would nicely explain the size and structure observed for the gaps (\citet{quillen04}; \citet{edgar+quillen+park2007}).

Based on our results from the largest sample of transitional disks to date, as summarized above, we conclude that giant planet formation plays the dominant role in opening gaps and creating transitional disks presented in this work.

\acknowledgments
We thank Lori Allen, Gregory Mosby, and Jes\'{u}s Hern\'{a}ndez for supplying the spectral types of objects, and John Tobin for the information of spectroscopic binaries and spectral types. We also thank Ignazio Pillitteri and Scott Wolk for supplying the X-ray luminosity of L1641 objects from their SOXS project.

Several of the authors were visiting astronomers at the Infrared Telescope Facility, which is operated by the University of Hawaii under Cooperative Agreement no. NNX-08AE38A with the National Aeronautics and Space Administration, Science Mission Directorate, Planetary Astronomy Program.

This publication makes use of data products from the Two Micron All Sky Survey, which is a joint project of the University of Massachusetts and the Infrared Processing and Analysis Center/California Institute of Technology, funded by the National Aeronautics and Space Administration and the National Science Foundation.

This work is based on observations made with the \textit{Spitzer Space Telescope}, which is operated by the Jet Propulsion Laboratory, California Institute of Technology under NASA contract 1407. Support for this work was provided by NASA through Contract Number 1257184 issued by JPL/Caltech, and Cornell subcontracts 31419-5714 to the University of Rochester. N. C. acknowledges support from NASA Origins grant NNX08AH94G. C. E. was supported by a Sagan Exoplanet Fellowship from the National Aeronautics and Space Administration and administered by the NASA Exoplanet Science Institute (NExScI).

\clearpage
\appendix{}

\section{Notes on Individual Objects}

\textit{OriA-5}: It is a single line spectroscopic binary (SB1), the average radial velocity is 18$\pm$4.8 $km/s$, and the maximum velocity difference between two components is 3.8$\pm$4.8 $km/s$ \citep{Tobin_2009_ONC_SB}. Its SED shows steeply decreasing fluxes after 20 $\micron$ and this may be the reflection of the effect of a second companion. It also shows prominent crystalline silicate features.\\

\textit{OriA-8}: This one satisfies the criteria on both $n_{K-6}$ and $n_{13-31}$. There are no obviously resolved sources within 30 arcsec around the target, but the target image in the Digitized Sky Survey (DSS) looks elongated, and it is suspected to have contributions from two sources. \\

\textit{OriA-18}: This object lies inside of the radially continuous disk model region in $n_{13-31}$-$EW(10\micron)$ space; however, its $n_{K-6}$ is much less than the lower octile of $n_{K-6}$, which is one of the selection criteria. The reason for low $n_{13-31}$ is due to the decreasing flux after 20 $\micron$, which could be the effect of external strong radiation evaporating the outer disk. There are 3-4 other sources near this target. This object's spectral type is not known. \\

\textit{OriA-39}: It satisfies only one criterion of $n_{K-6}$. Its IRS spectrum is very noisy (Figure~\ref{fig-SED-2}). It is faint at IRS wavelengths, and the sky emission around it is very complicated because it lies in a fringe area of the bright nebula NGC 1977, even though there are no point sources within 40 arcsec. \\

\textit{OriA-47}: It lies in a very crowded and complicated region. There are many HH objects about 2 arcmin away, and there is much complicated background emission from bright sources in the center of ONC, the Trapezium. There is also an additional point source about 1.5 arcsec away. Recently, OriA-47 has also been identified as a variable star, [PMD2009] 185, as well as the nearby star [PMD2009] 183 \citep{Parihar2009_ONC_variables_SpT4V1498Ori}. The IRS didn't resolve the signals from the two sources, and the IRS fluxes are the composite fluxes from both.\\

\textit{OriA-88}: Its SL2 data is not available because the array of SL2 is saturated due to the bright radiation entering in the IRS Peak-up cameras. This target is located just outside of the HII region, about 9 arcmin away from the center of ONC (M42). The PAH features are real and come from the disk. \\

\textit{OriA-149}: This target lies in a complicated region of OMC2/3. It is reported to be an X-ray source ([TKK] 780) \citep{TKK2003_OMC_Chandra}, and there is a point source identified as an IR-source ([TKK] 774) about 5 arcsec away from the target. IRS SL and LL slits were placed to avoid [TKK] 774 as much as possible, but the source's IR radiation seems to affect LL1: there is a kind of extended emission entering in LL1 as background, very close to our target source. We tried to remove the effect from the extended emission as much as possible by blocking the pixels corresponding the emission and using multiple source extraction in AdOpt. \\

\textit{OriA-154}: Its spectral indices and $EW(10\micron)$ are similar to OriA-18, and it is located in a similar position on $n_{13-31}$-$EW(10\micron)$ space. Source extraction for LL spectra was performed by AdOpt multiple source extraction because an additional source appeared in the LL slit. Its outer disk could be affected by strong radiation from nearby bright sources. \\

\textit{OriA-164}: It lies in a dense dark core region, and it is very faint at optical wavelengths. This supports the large value of $A_{V}$ found here, even though its spectral type is unknown so that there are large uncertainties in our $A_{V}$ estimate. It is barely passed the criteria of $EW(10\micron)$. Its SED, which is restricted from J band to 35 $\micron$ in IRS data, with strong silicate features at 10 $\micron$ and 20 $\micron$, lets us infer that this object is a possible PTD. \\

\textit{OriA-174}: The prominent emission lines in its IRS spectra are molecular hydrogen $\nu=0 \rightarrow 0$ S(1), S(2), S(3) and S(5), which arise in the foreground and background cloud material rather than our target, and did not subtract away precisely. The object is an embedded source which is not shown up at optical wavelength ranges. \\

\textit{OriA-188}: The spectral type of this object is not known. There were no issues in the IRS data reduction, and its SED is like an object with a central clearing or gap. About in 30 arcsec radius from the target there are three faint 2MASS objects, but they don't enter in the IRS slits. A significant environmental concern is that there are two OB type stars about 6.7 arcmin away from the target: iot Ori (O9) and V2451 Ori (B7). The disk surface of the target may be affected by the strong radiation from the bright sources. \\

\textit{OriA-198}: This target also barely passed the criteria on $EW(10\micron)$ while its $n_{K-6}$ and $n_{13-31}$ fail to pass the criteria. There is a bright nebula, [B77] 122, about 7 arcmin away, and no nearby sources within 10 arcsec of the target. OriA-198 has strong silicate features at 10 $\micron$ and 20 $\micron$. These could be due to radial gaps in a more settled disk than a typical ClassII disk.\\

\textit{OriA-221} Its $EW(10)\micron$ satisfies the TD selection criteria, and it lies in the area of the well studied TDs of Tau, the region of outliers in $n_{13-31}$ vs. $EW(10)\micron$ space. After extinction correction, this object's flux at optical wavelengths exceeds considerably that expected for photospheric spectra that fit well at longer wavelengths. This could be overcorrection of extinction toward the star, but we consider it more likely to be due to near-edge-on view and the extinction overcorrection of scattered visible light. There are several 2MASS objects within 40 arcsec, but they do not affect the source extraction at all. There is a bright B type star 7.4 arcmin away, HR 1911 (B1; double or multiple system). This target is also a PTD candidate.\\

\section{Consideration of $\dot{M}$ and $\dot{M_{W}}$ from XPE}
To ponder the mass dissipation through $\dot{M}$ and/or the expected mass loss rate ($\dot{M_{W}}$), we overlay ($\dot{M_{W}}$) by X-ray photoevaporation \citep{Owen_photoevaporation2012} on the observed mass accretion rate of TDs in Figure~\ref{fig-mdot-Lx-owen}. The plus and cross data points in Figure~\ref{fig-mdot-Lx-owen} is the estimated $\dot{M_{W}}$ of each TD with its $M_{\star}$ and $L_X$ as the input condition of $\dot{M_{W}}$. The estimated $\dot{M_{W}}$ have higher values than the mass accretion rates for most targets with high $L_X$ and high $\dot{M}$. However, $\dot{M_{W}}$ is weaker comparing $\dot{M}$ of CTTS indicated as the gray dashed line.
From this, we may raise the question of when $\dot{M_{W}}$ overcomes $\dot{M}$ to create a gap/hole like in TDs. The $\dot{M}$ of TDs is not the $\dot{M}$ when $\dot{M_{W}}$ is able to open a gap. However, the estimated $\dot{M_{W}}$ are not very different whether for radially continuous disks (plus sign) or disks with a large inner hole (cross sign). If we assume that each TD should have followed the gray dashed line for their initial $\dot{M}$-$L_X$ relation before they became TDs, their initial $\dot{M}$ must have been greater than the theoretical $\dot{M_{W}}$. And the current $\dot{M}$ of TDs are much lower than the theoretical $\dot{M_{W}}$ of TDs. Some possible scenarios might explain these inconsistencies: (1) the TDs in our sample have already evolved much after gap opening, and they have lost much inner material to accrete to the central star; (2) there could be some other contribution (e.g. planets) to make the mass accretion rate decrease quickly, such as intercepting disk material on the way to the central star while $\dot{M_{W}}$ forces material to drift away outward.

\section{Trends of TDs in Orion A}
We examine trends for the sample of TDs in Orion A and the linear correlations of any two properties in $log-log$ scale are listed in Table~\ref{appendix-table:correlation-OriA} and Table~\ref{appendix-table:correlation_subTDtype-OriA}. Comparing trends in Table~\ref{appendix-table:correlation-OriA} and Table~\ref{table:correlation} before separating by subtypes of TDs, the trends of TDs in Orion A generally agree within 1 $\sigma$ uncertainties with the trends of the full sample of TDs not only from Orion A but also from Tau, ChaI, Oph, and N1333, despite $L_{X}$ and $M_\star$ of TDs in Orion A tend to distribute toward lower values than that of other region. The trends of subtypes of TDs of Orion A in Table~\ref{appendix-table:correlation_subTDtype-OriA} are also not very different from that of full samples in Table~\ref{table:correlation_subTDtype}, even though uncertainties of the results from linear regression are larger due to smaller sample sizes for some correlations.
We will discuss difference/similarity of properties of TDs and Class II objects in Orion A from that in other star-forming regions in a separate paper \citep{khkim2013-OriA} because the discussion is outside the scope of this paper.

\bibliography{references}

\clearpage
\clearpage

\newcounter{subfig}
\renewcommand{\thefigure}{\arabic{figure}}
\setcounter{subfig}{1}

\clearpage
\setcounter{subfig}{1}
\begin{figure}[t]
\epsscale{0.9}
\plotone{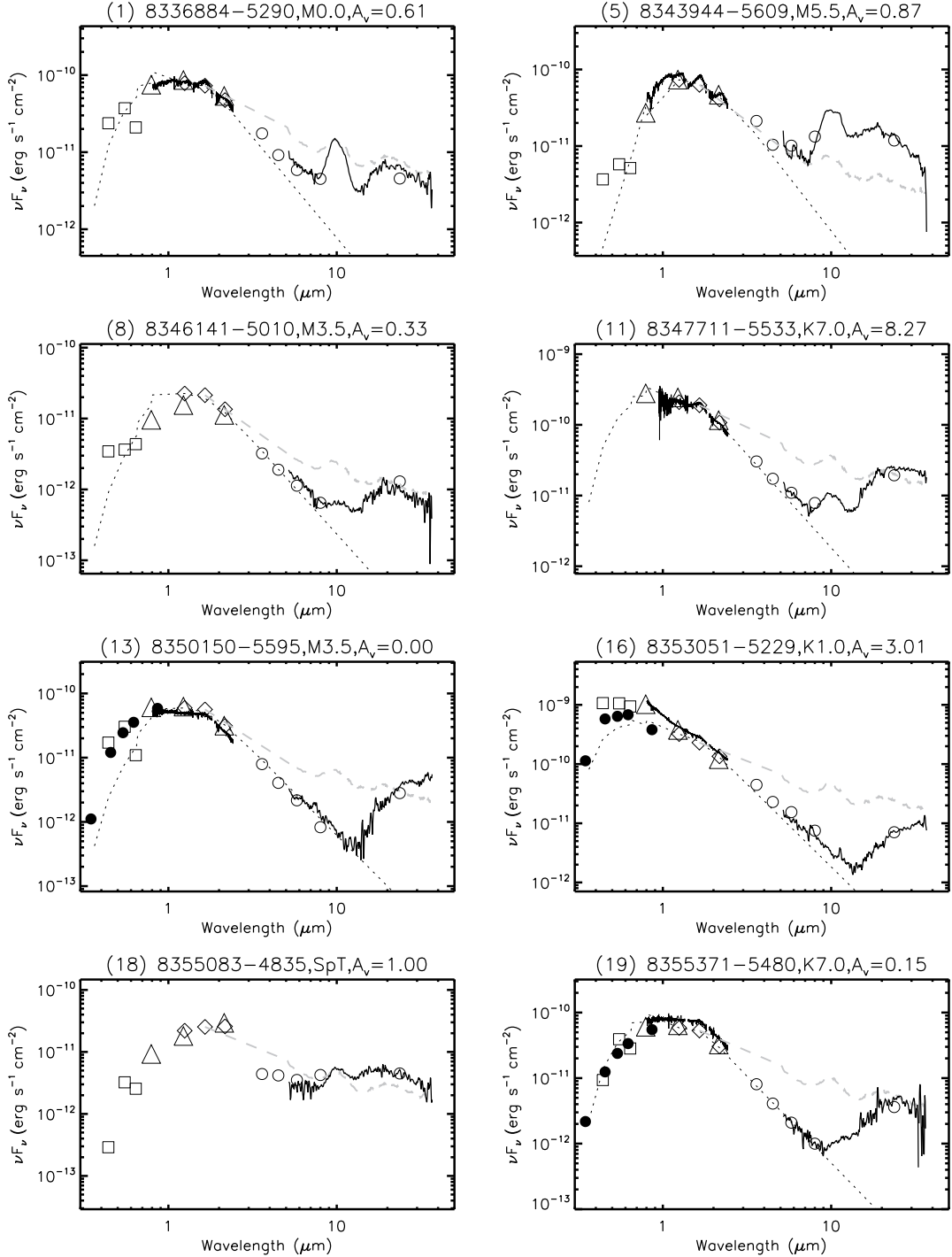}
\caption{De-reddened SEDs of transitional disks in Orion A. The SEDs are composed of the following components: IRS (black line in the wavelength range of 5.2-35 \micron); SpeX (black line in the wavelength range of 0.8-2.4 \micron); IRAC and MIPS (open circles); 2MASS JHK (open diamonds); DENIS IJH(open triangles); UBVRI from \cite{Dario2009_ubvri} (filled circles); BVR from NOMAD (open squares); photosphere (black short dashed line); the median spectrum of protoplanetary disks in Taurus region (gray long dashed line). The gray line from 0.8 to 2.4 \micron\ in the plots of OriA-26, 38, 47, and 290 is for spectra of their companion resolved in the SpeX observations. \label{fig-SED-1}}
\end{figure}

\clearpage
\addtocounter{figure}{-1}
\addtocounter{subfig}{1}
\begin{figure}[t]
\epsscale{0.9}
\plotone{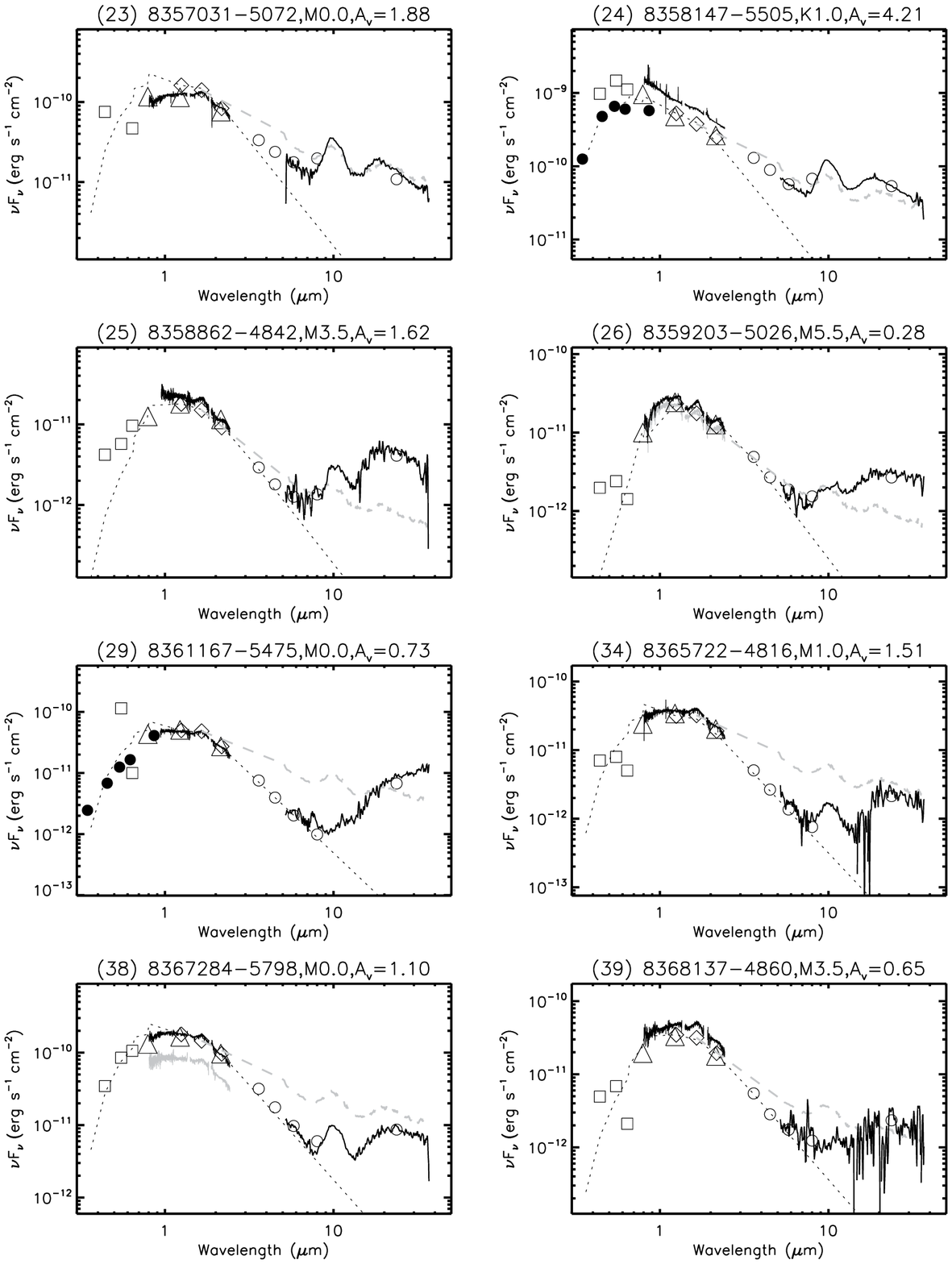}
\caption{Figure 2. continued \label{fig-SED-2}}
\end{figure}

\clearpage
\addtocounter{figure}{-1}
\addtocounter{subfig}{1}
\begin{figure}[t]
\epsscale{0.9}
\plotone{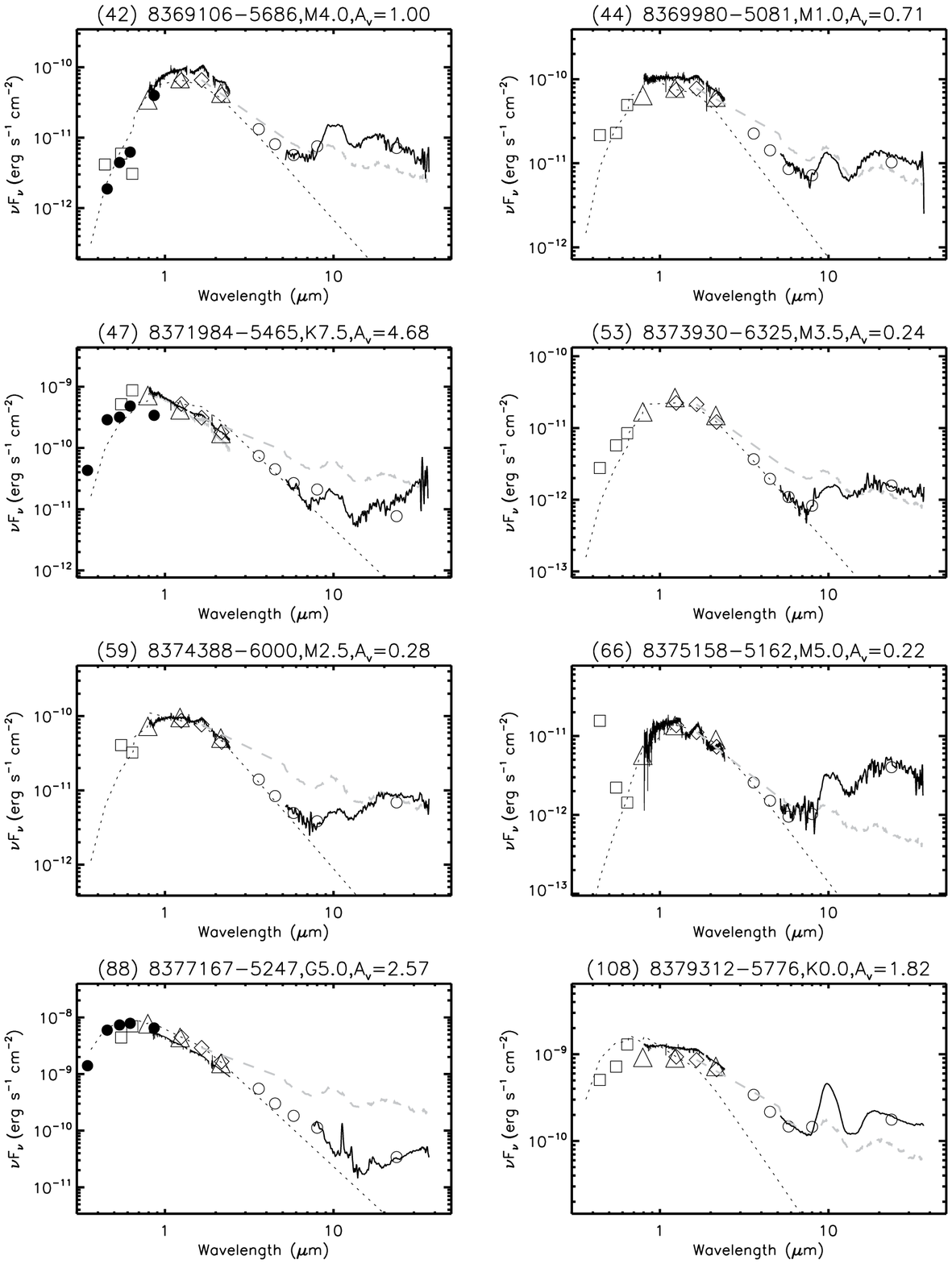}
\caption{Figure 2. continued \label{fig-SED-3}}
\end{figure}

\clearpage
\addtocounter{figure}{-1}
\addtocounter{subfig}{1}
\begin{figure}[t]
\epsscale{0.9}
\plotone{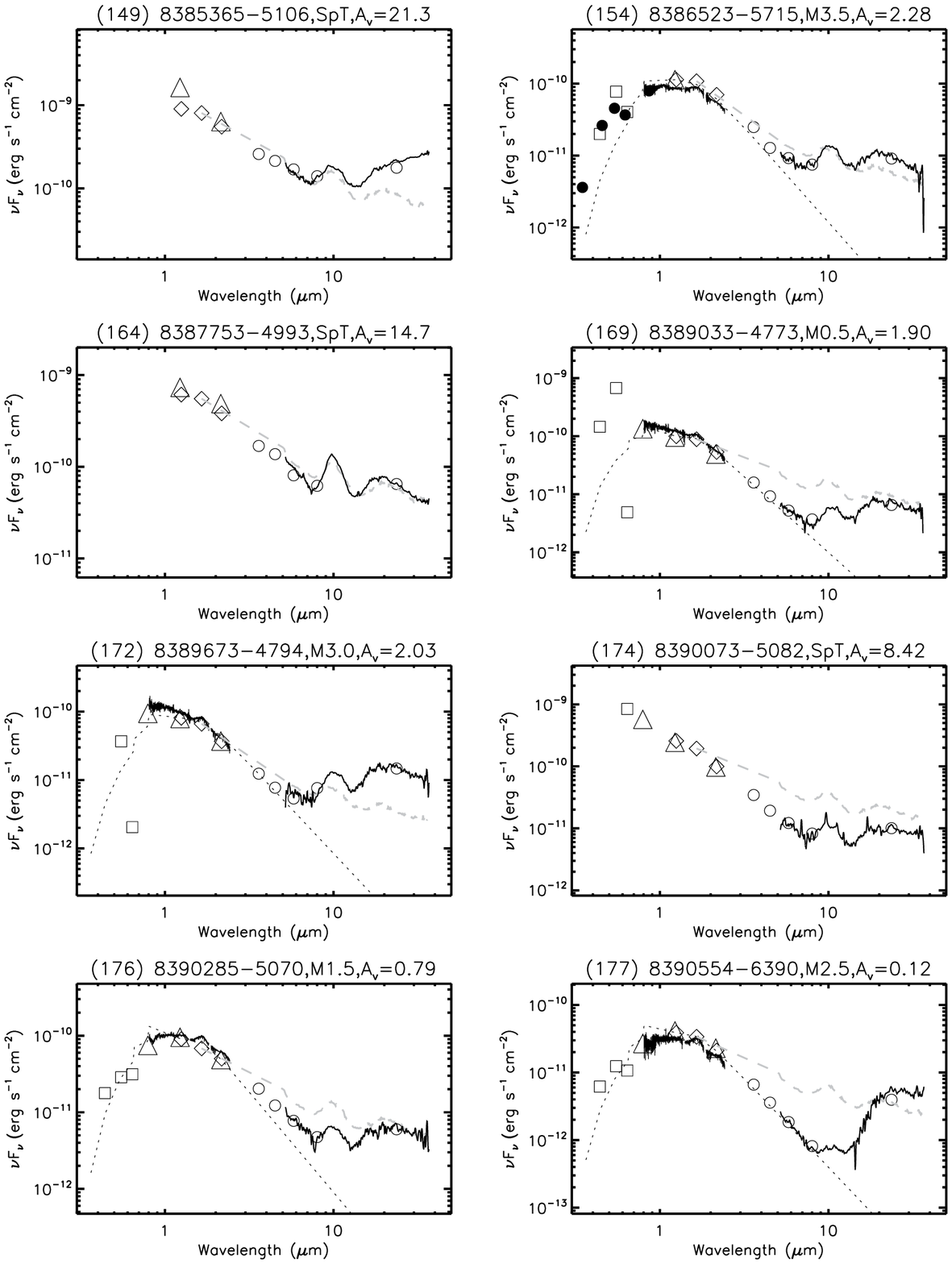}
\caption{Figure 2. continued \label{fig-SED-4}}
\end{figure}

\clearpage
\addtocounter{figure}{-1}
\addtocounter{subfig}{1}
\begin{figure}[t]
\epsscale{0.9}
\plotone{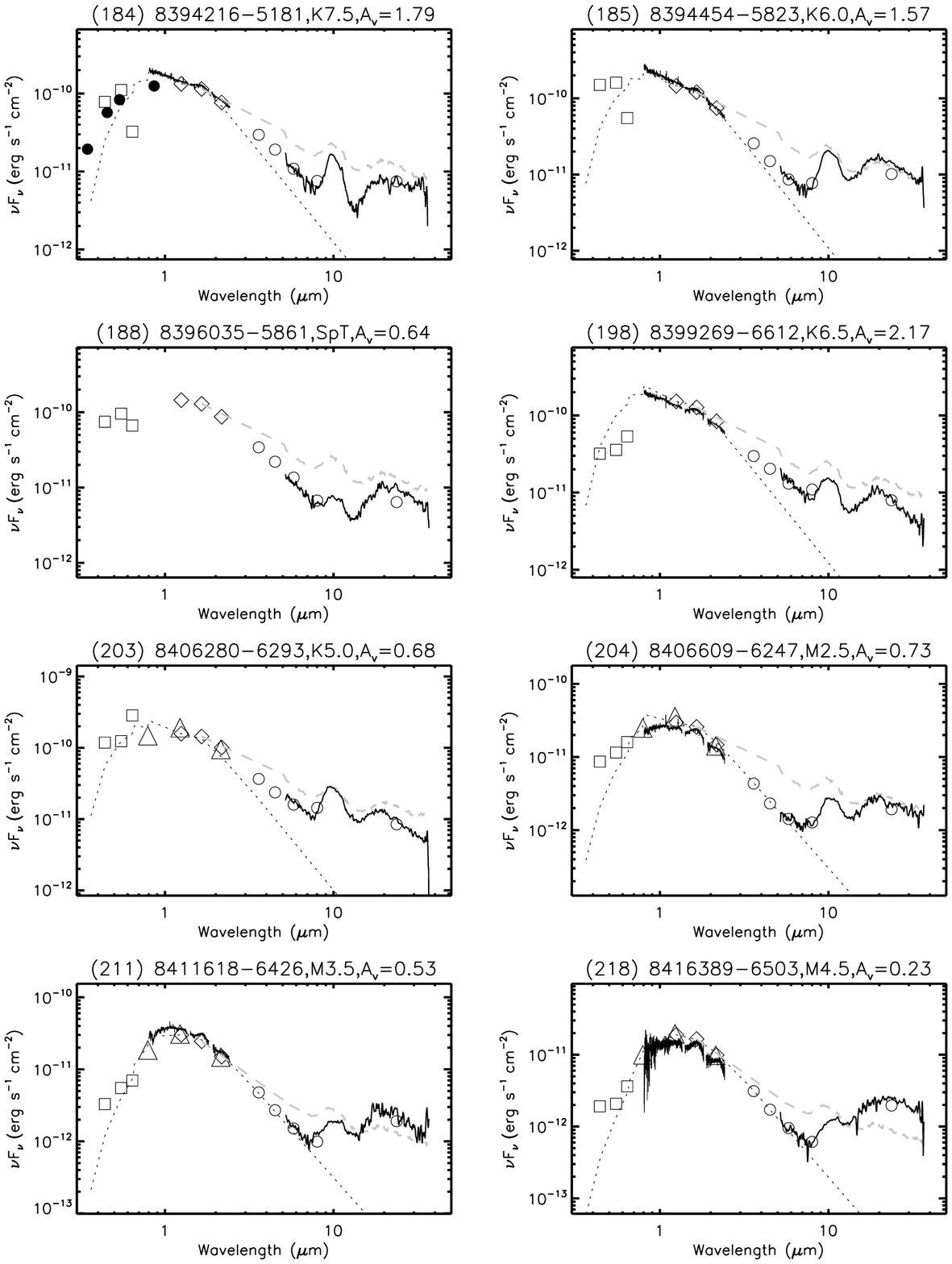}
\caption{Figure 2. continued \label{fig-SED-5}}
\end{figure}

\clearpage
\addtocounter{figure}{-1}
\addtocounter{subfig}{1}
\begin{figure}[t]
\epsscale{0.9}
\plotone{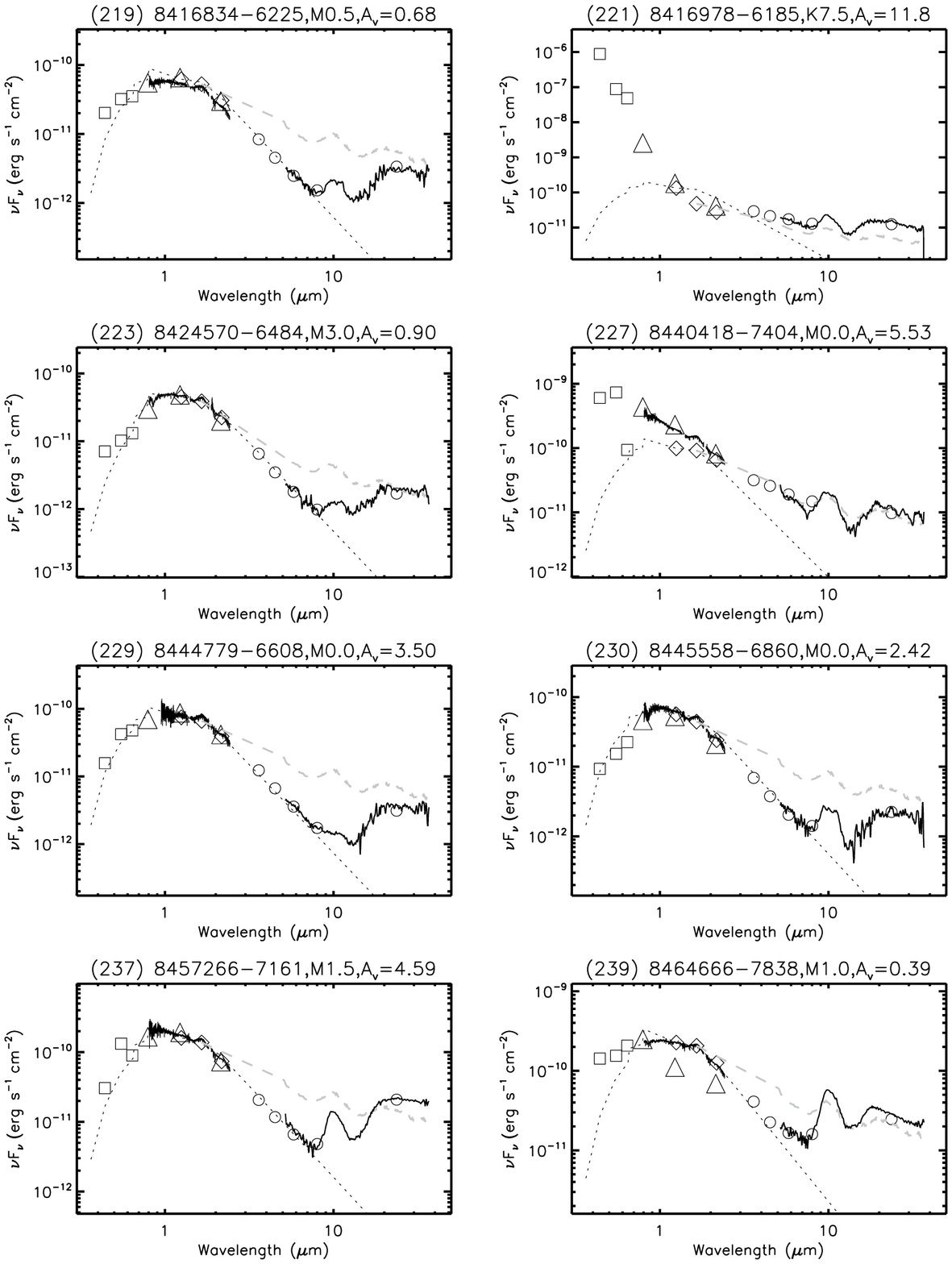}
\caption{Figure 2. continued \label{fig-SED-6}}
\end{figure}

\clearpage
\addtocounter{figure}{-1}
\addtocounter{subfig}{1}
\begin{figure}[t]
\epsscale{0.9}
\plotone{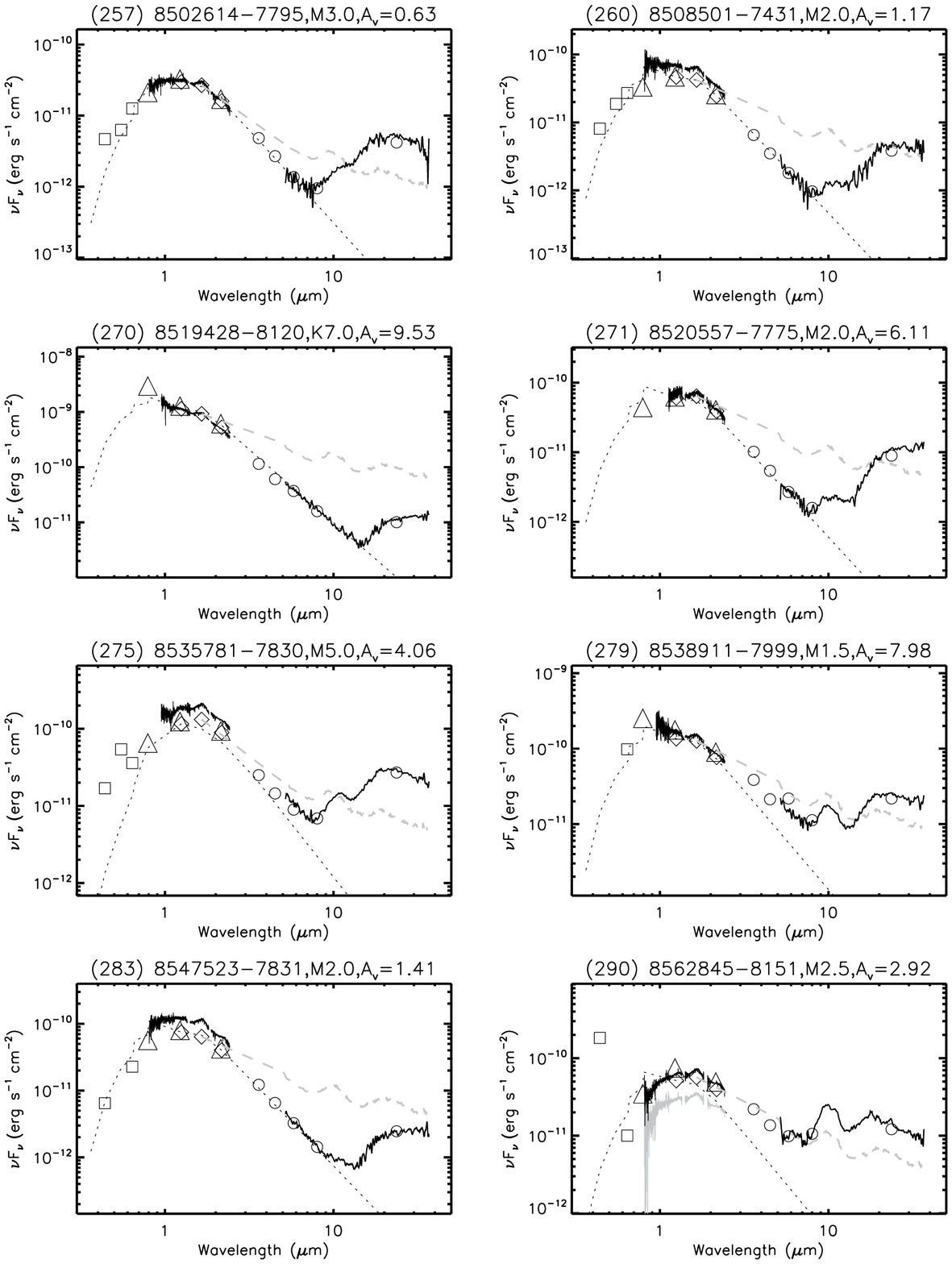}
\caption{Figure 2. continued \label{fig-SED-7}}
\end{figure}

\clearpage
\addtocounter{figure}{-1}
\addtocounter{subfig}{1}
\begin{figure}[t]
\epsscale{0.9}
\plotone{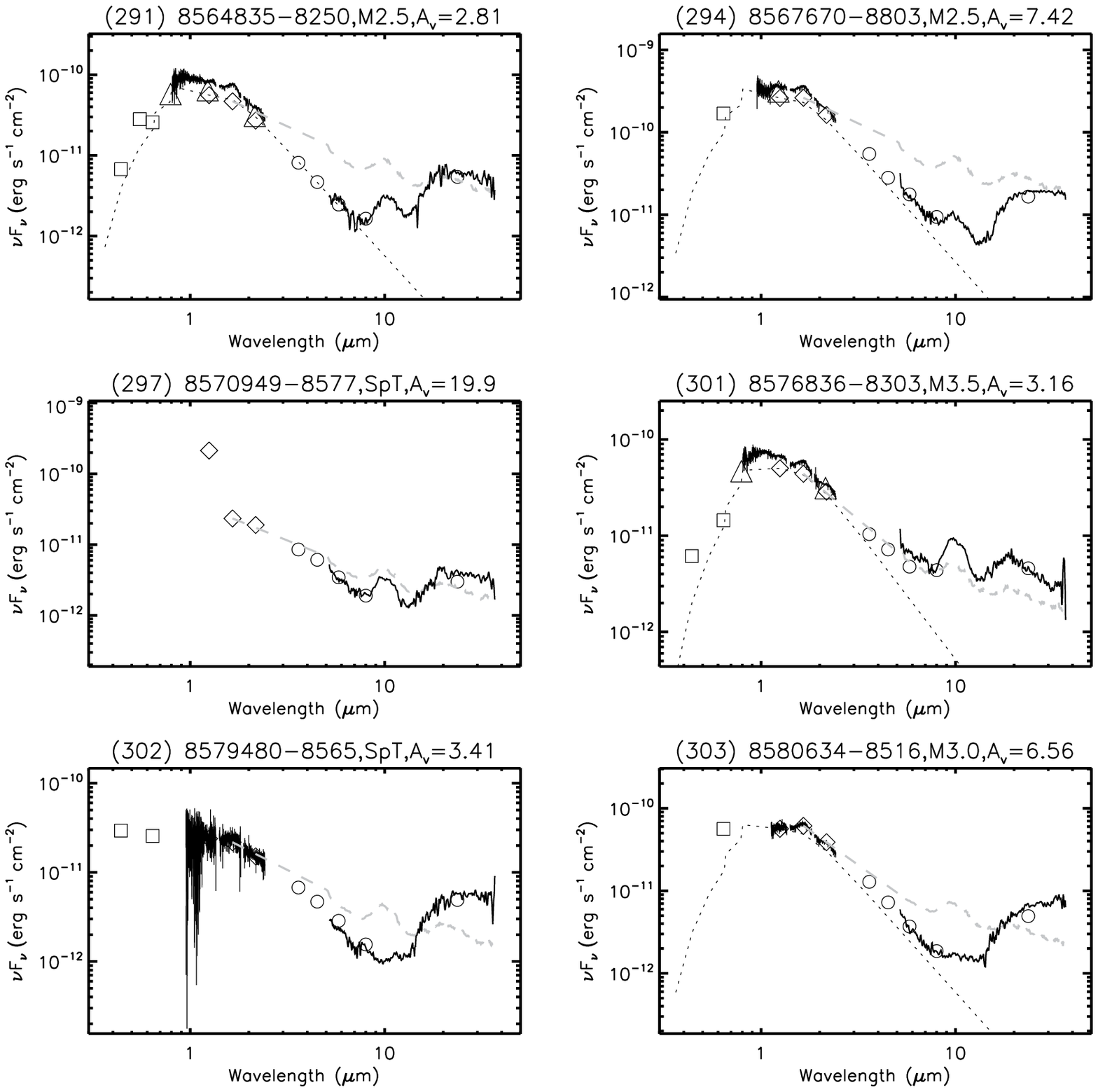}
\caption{Figure 2. continued \label{fig-SED-8}}
\end{figure}

\renewcommand{\thefigure}{\arabic{figure}}

\clearpage
\begin{figure}[t]
\epsscale{1.0}
\plotone{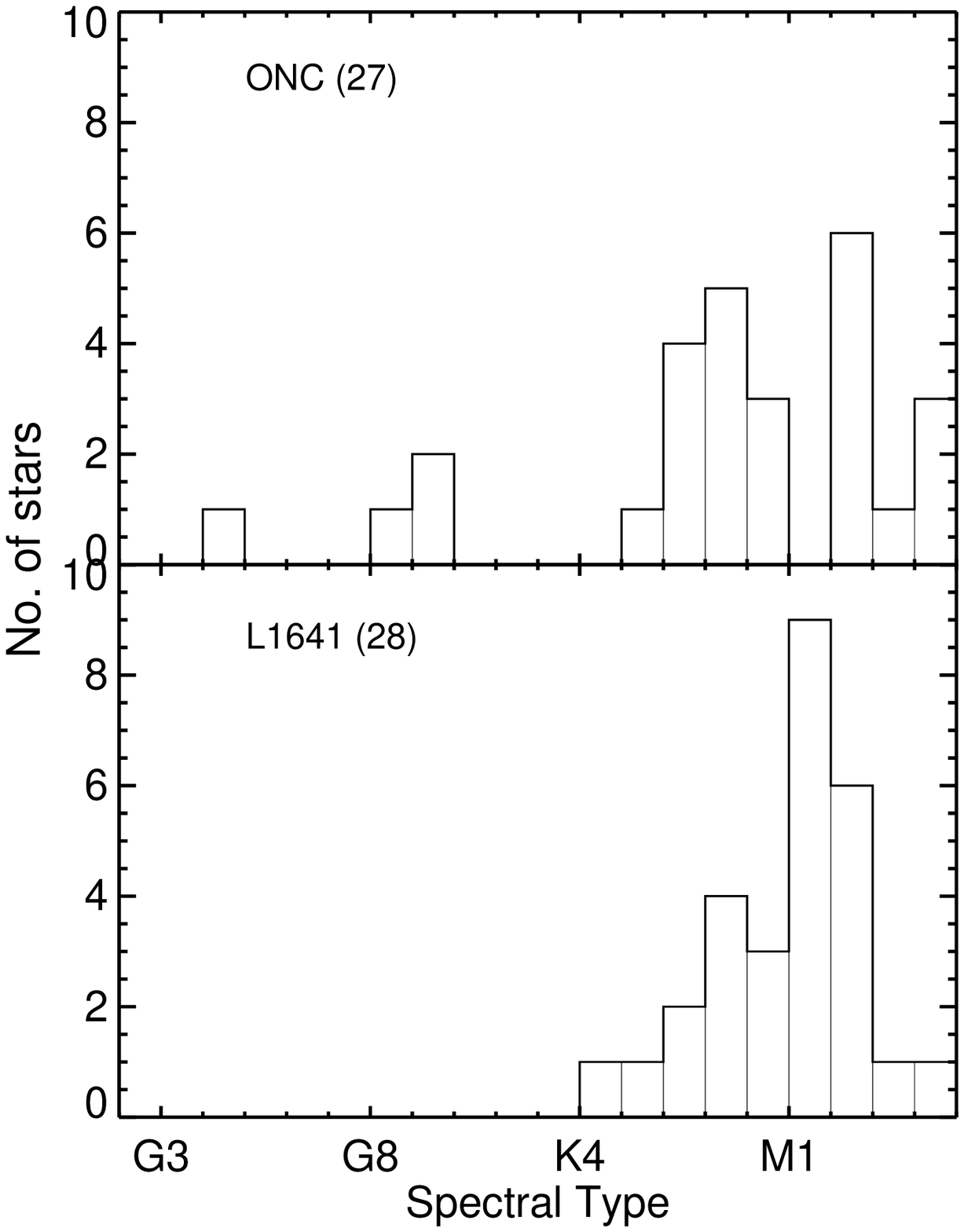}
\caption{Spectral type distribution for host stars of transitional disks in Orion A. \label{fig-OriAX-SpT}}
\end{figure}
\clearpage

\clearpage
\begin{figure}[t]
\epsscale{1.0}
\plotone{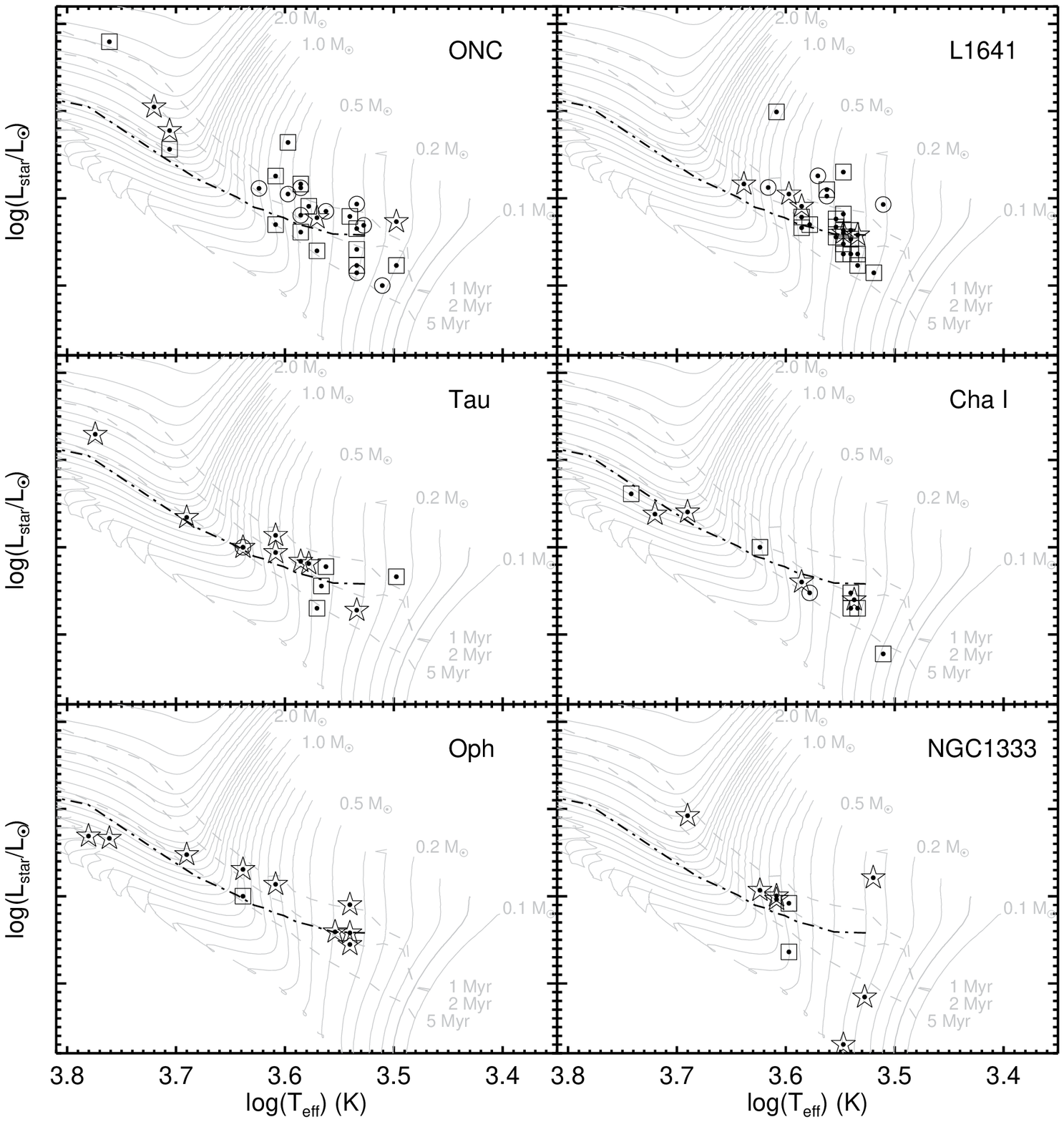}
\caption{HR diagrams for host stars of transitional disks in this paper. Squares are for CTDs; circles are for WTDs; stars are for PTDs. Evolutionary tracks and isochrones are from \cite{siess2000} (Z=0.02). Isochrone ages of various types of transitional-disk systems range from $<$ 1 Myr to $>$ 5 Myr. The average disk life time in Tau-Aur \citep{bertout07disktime} is also shown as a dash-dotted line for reference. \label{fig-HRD}}
\end{figure}
\clearpage

\clearpage
\begin{figure}[t]
\epsscale{1}
\plotone{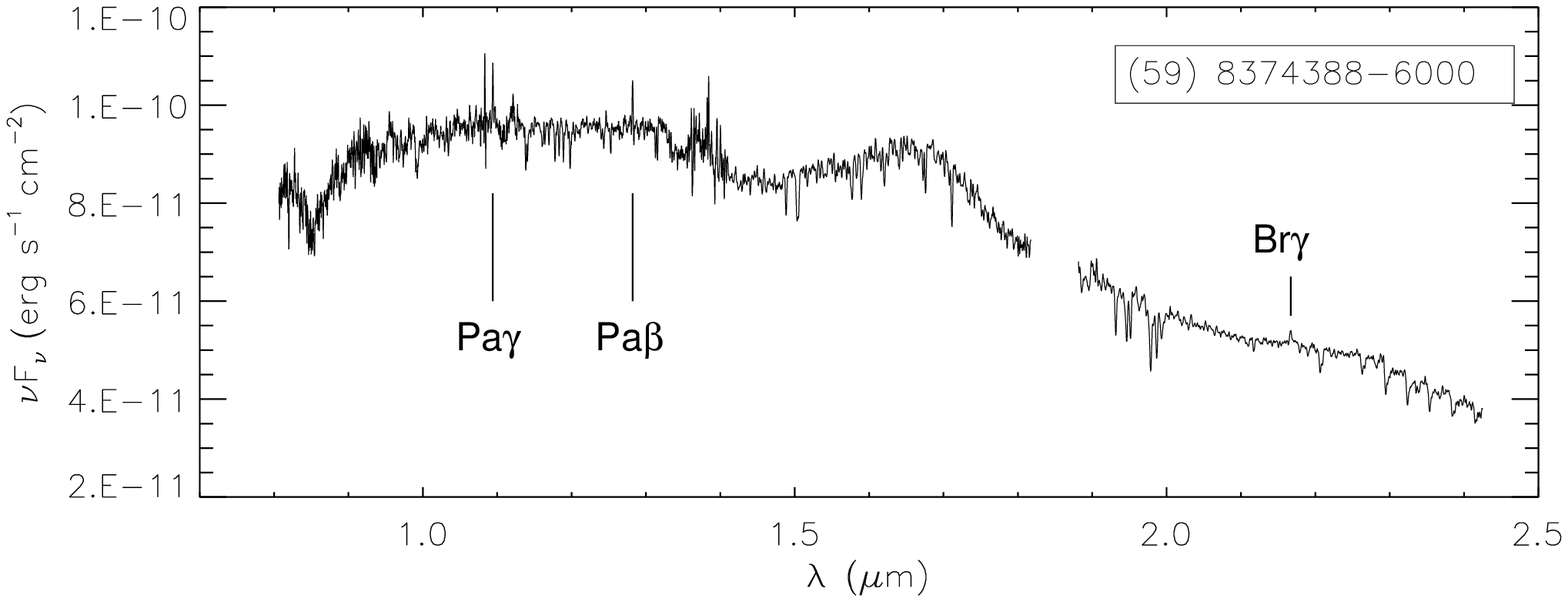}
\plotone{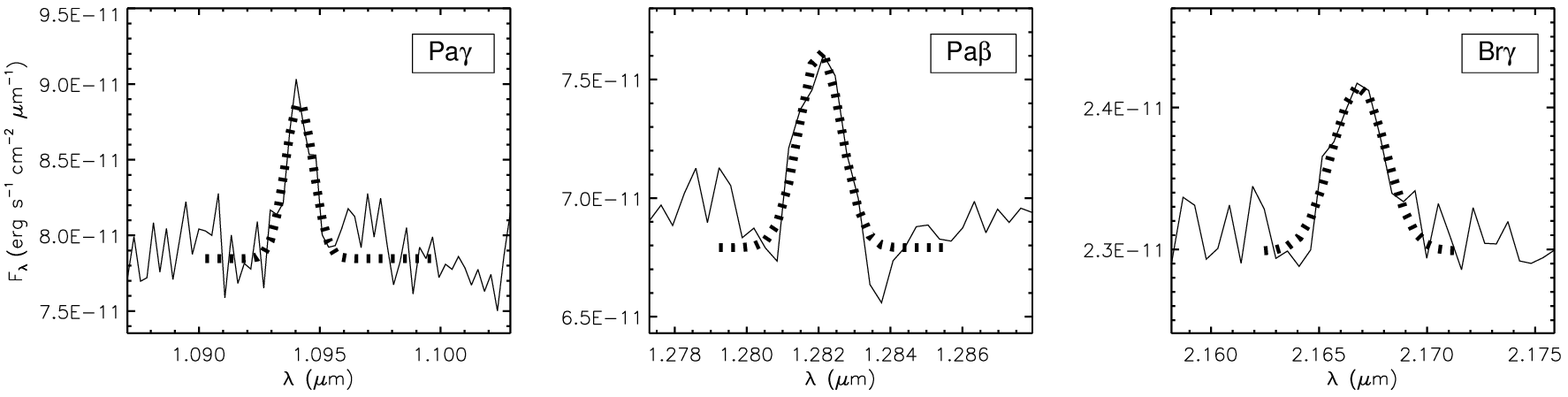}
\caption{Example of mass accretion rate measurement from hydrogen recombination lines. \label{fig-estMdot}}
\end{figure}
\clearpage

\clearpage
\begin{figure}[t]
\epsscale{1}
\plotone{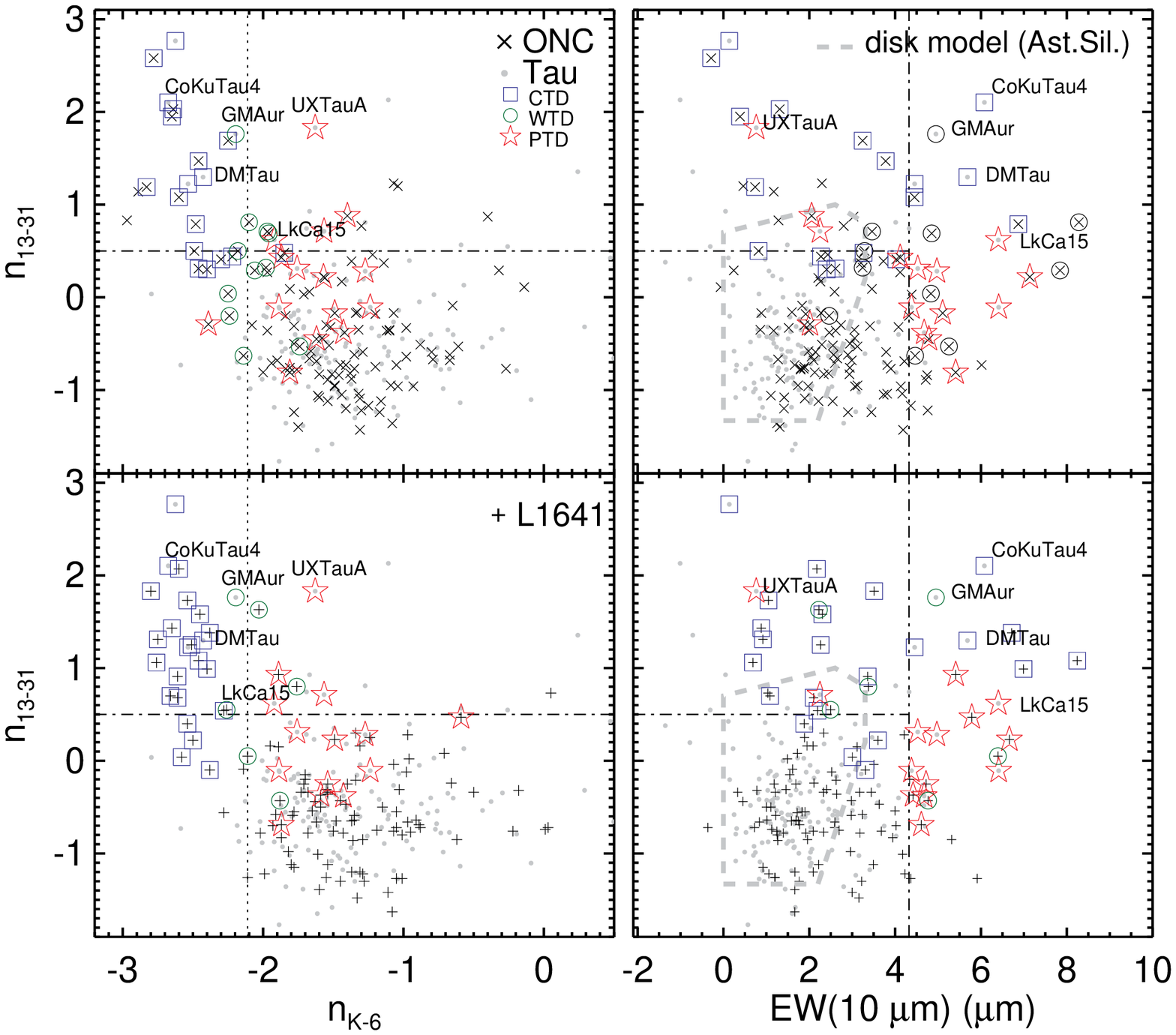}
\caption{Selection of OriA TDs by $n_{13-31}$ vs. $n_{K-6}$ (the left panels) and $n_{13-31}$ vs. $EW(10\micron)$ (the right panels). In the right panels, the polygon with thick dashed line indicates the coverage area by a typical accretion disks model \citep{d'alessio06}. The upper panels are for the TD selection in ONC, and the lower panels are for the TD selection in L1641. In each plot, samples in Tau \citep{Furlan2011Taurus} are also included for comparison. The dash-dotted lines indicate the upper octile; the dotted lines indicate the lower octile. (A color version of this figure is available in the online journal.)
}
\label{fig-OriA-tdselection}
\end{figure}
\clearpage

\clearpage
\begin{figure}[t]
\epsscale{1}
\plotone{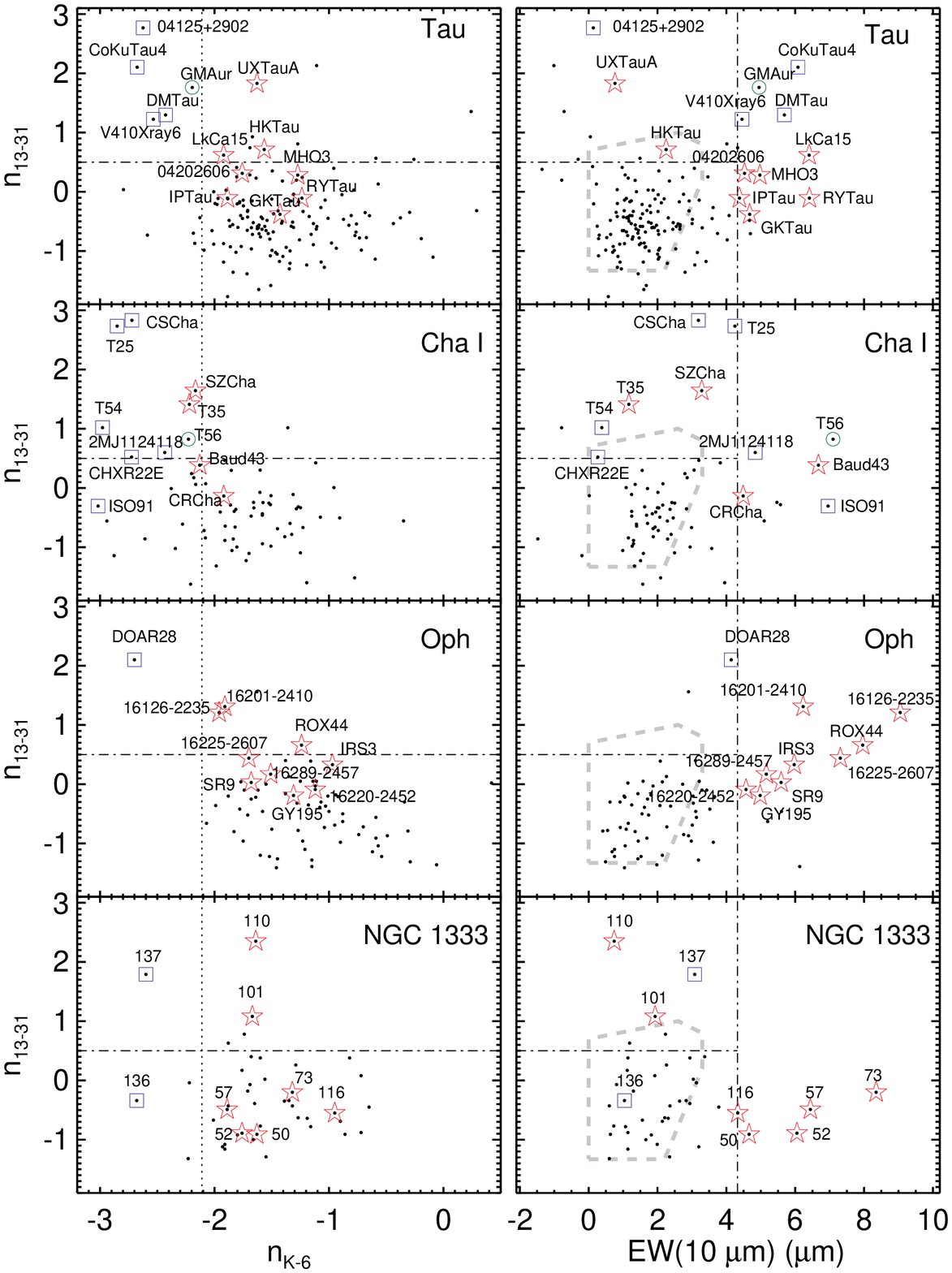}
\caption{Selection of TDs in other star-forming regions, Tau, Cha I, Oph, and NGC 1333, by the criteria in $n_{13-31}$ vs. $n_{K-6}$ (the left panels) and $n_{13-31}$ vs. $EW(10\micron)$ (the right panels). The meanings of different symbols and lines are same as Figure~\ref{fig-OriA-tdselection}. (A color version of this figure is available in the online journal.)
}
\label{fig-other-tdselection}
\end{figure}
\clearpage

\clearpage
\begin{figure}[t]
\epsscale{1}
\plotone{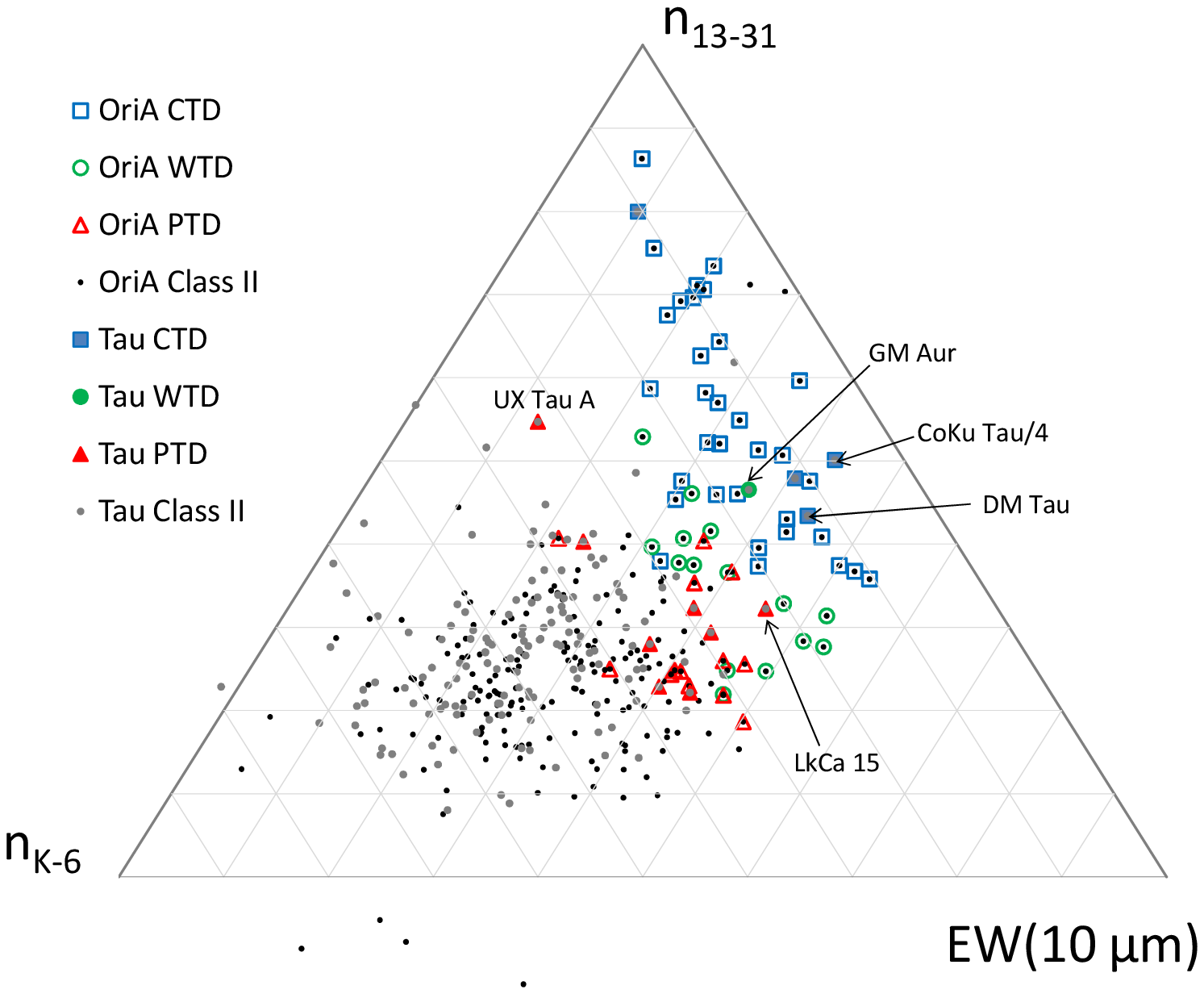}
\caption{Disks samples of Orion A (black dots) and Tau (gray dots) are plotted in the spaces of three parameters. TDs are indicated by subtypes: CTDs (squares), WTDs (circles), and PTDs (triangles). Orion A TDs are open symbols and Tau TDs are filled symbols. The center of this plot is corresponding to the projected origin of the three axes. The three axes values are from 0 to 100, which are linearly transformed from their original values of [-3, 0] for $n_{K-6}$, [-2, 3] for $n_{13-31}$, and [-1, 9] for $EW(10\micron)$. (A color version of this figure is available in the online journal.)
}
\label{fig-OriA-ternary}
\end{figure}
\clearpage

\clearpage
\begin{figure}[t]
\epsscale{1}
\plotone{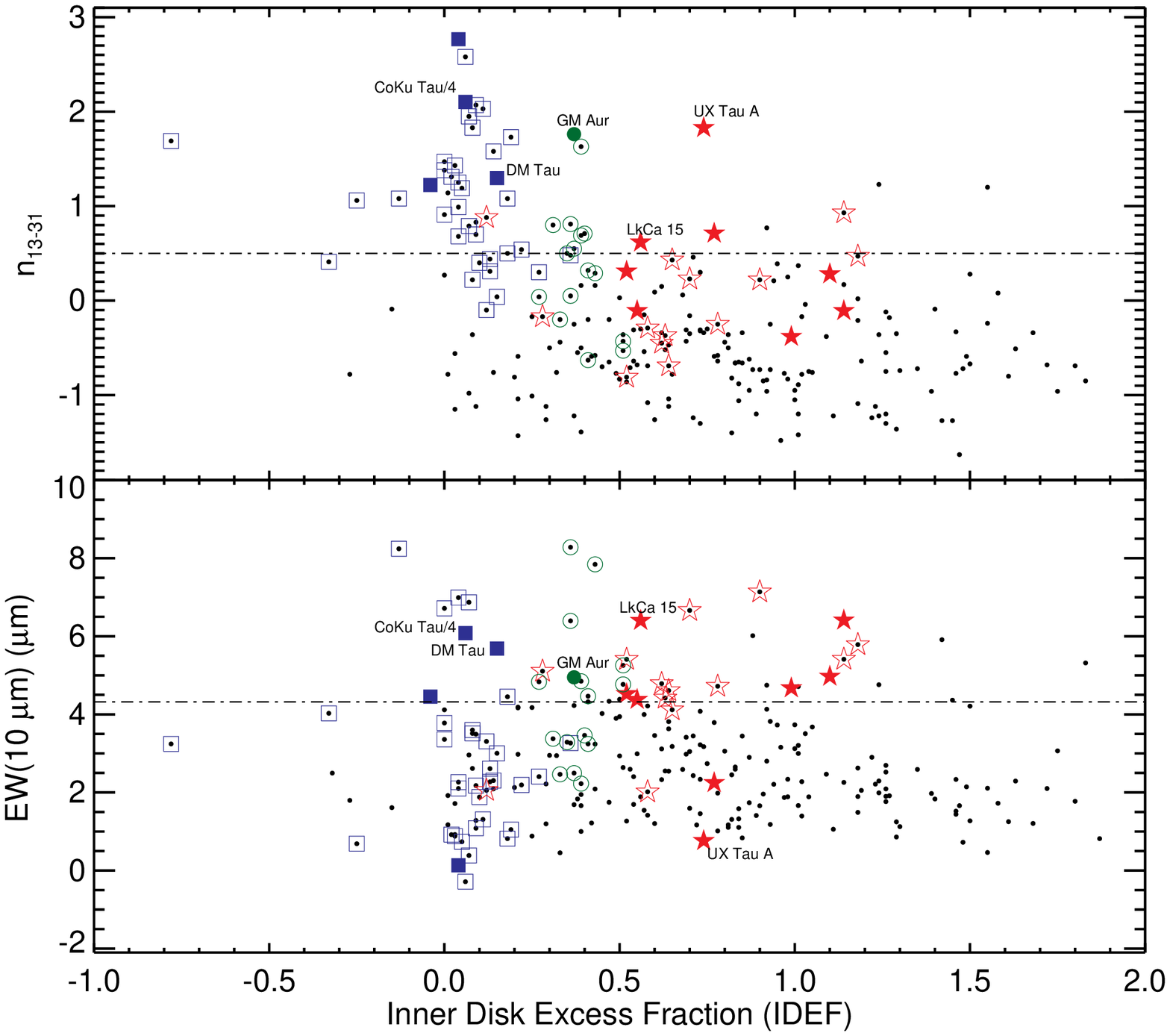}
\caption{$n_{13-31}$ and $EW(10\micron)$ plotted against IDEF (Inner Disk Excess Fraction). TDs of Orion A and Tau are indicated in open and filled symbols: CTDs (squares), WTDs (circles) and PTDs (stars). The dots indicate radially-continuous disks of Orion A. (A color version of this figure is available in the online journal.)
}
\label{fig-OriA-IDEF}
\end{figure}
\clearpage

\clearpage
\begin{figure}[t]
\epsscale{1}
\plotone{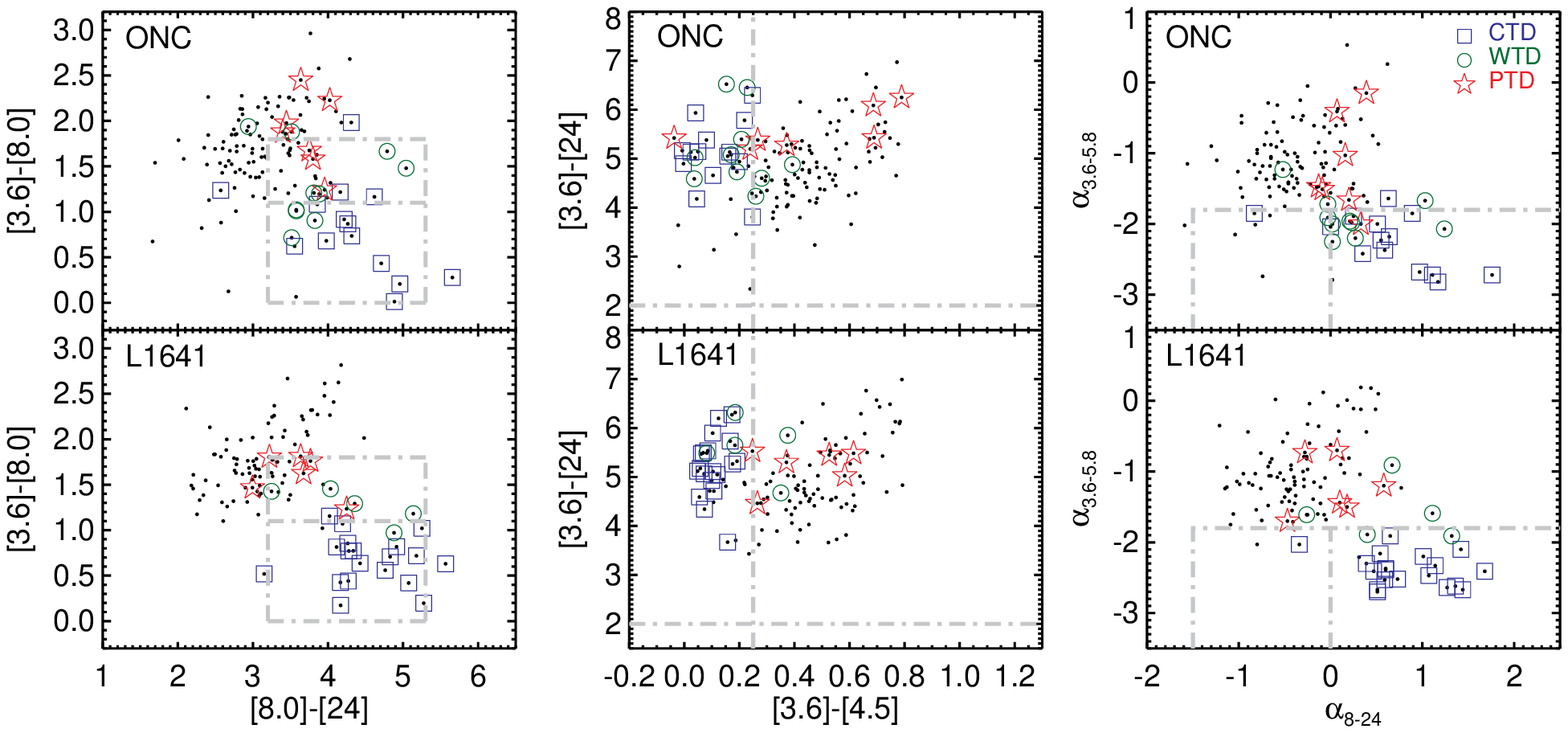}
\caption{TDs of Orion A under other selection criteria. In each panel, the data points are our Orion A samples: the upper panels for ONC and the lower panels for L1641. Each selection criteria is indicated with the dot-dash lines. The left panel is for the selection criteria from \citet{Merin2010}. The middle panel shows the selection criteria of \citet{cieza_2010_TDsOph}. The right panel is the selection criteria from \citet{Muzerolle2010}. (A color version of this figure is available in the online journal.)}
\label{fig-c2dselection}
\end{figure}
\clearpage

\clearpage
\begin{figure}[t]
\epsscale{0.7}
\plotone{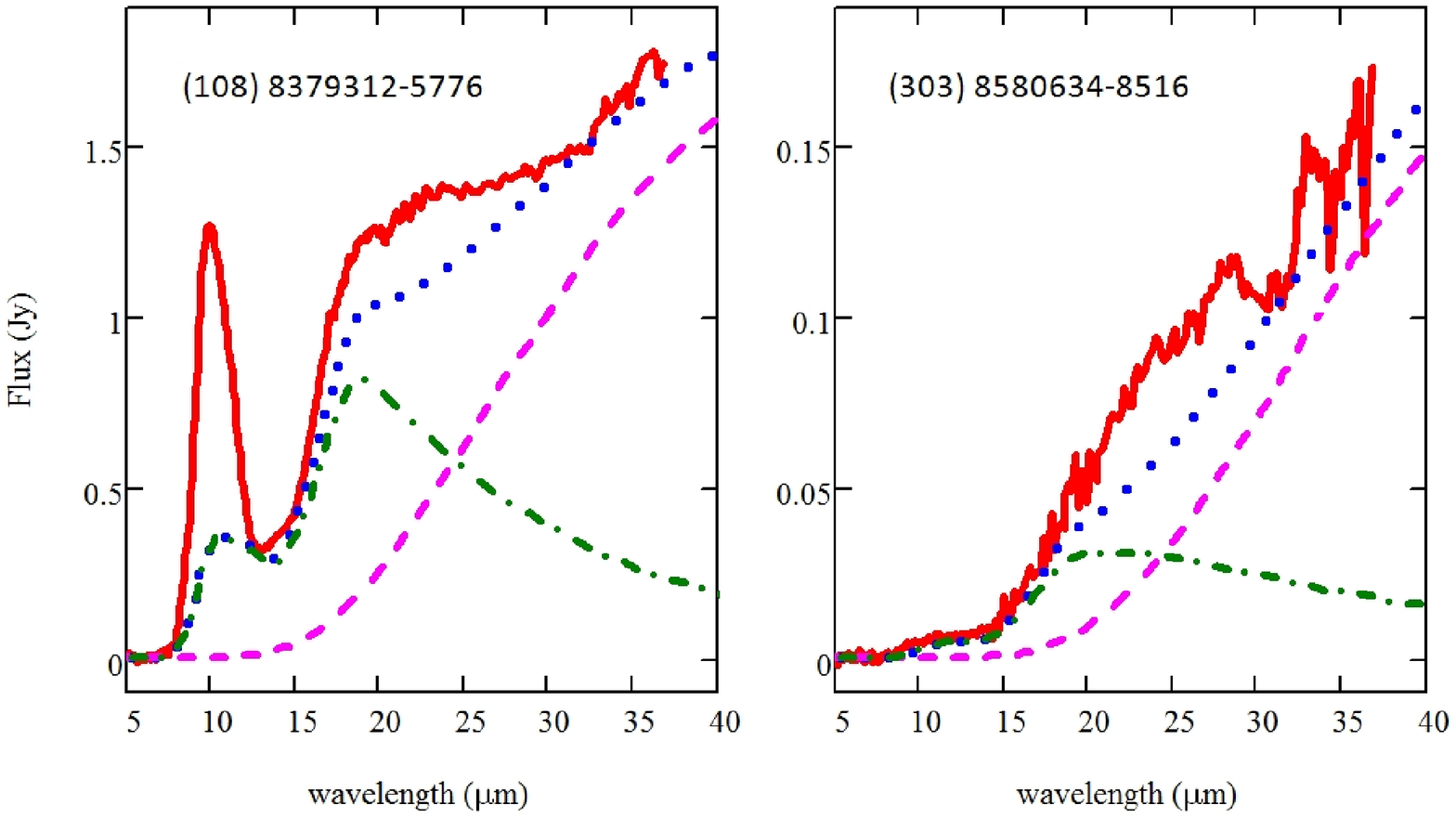}
\caption{Example of fits using the simple model to estimate $R_{wall}$ of transitional disks in Orion A and other star-forming regions included in this paper. The thick solid line: a residuum of IRS spectra after subtracting a power-law fitted to 5-8 $\micron$ of IRS spectra, representing emission from the photosphere or a part of the inner disk. The dash-dotted line: astronomical silicate model to account for some contribution of emission from small dust grains in the atmosphere of wall and disk upper layers. The dashed line: a single blackbody profile with the wall temperature $T$. The dotted line: the continuum fit as the combination of the dash-dotted line and the dashed line. (A color version of this figure is available in the online journal.) \label{fig-estRw}}
\end{figure}
\clearpage

\clearpage
\begin{figure}[t]
\epsscale{1}
\plotone{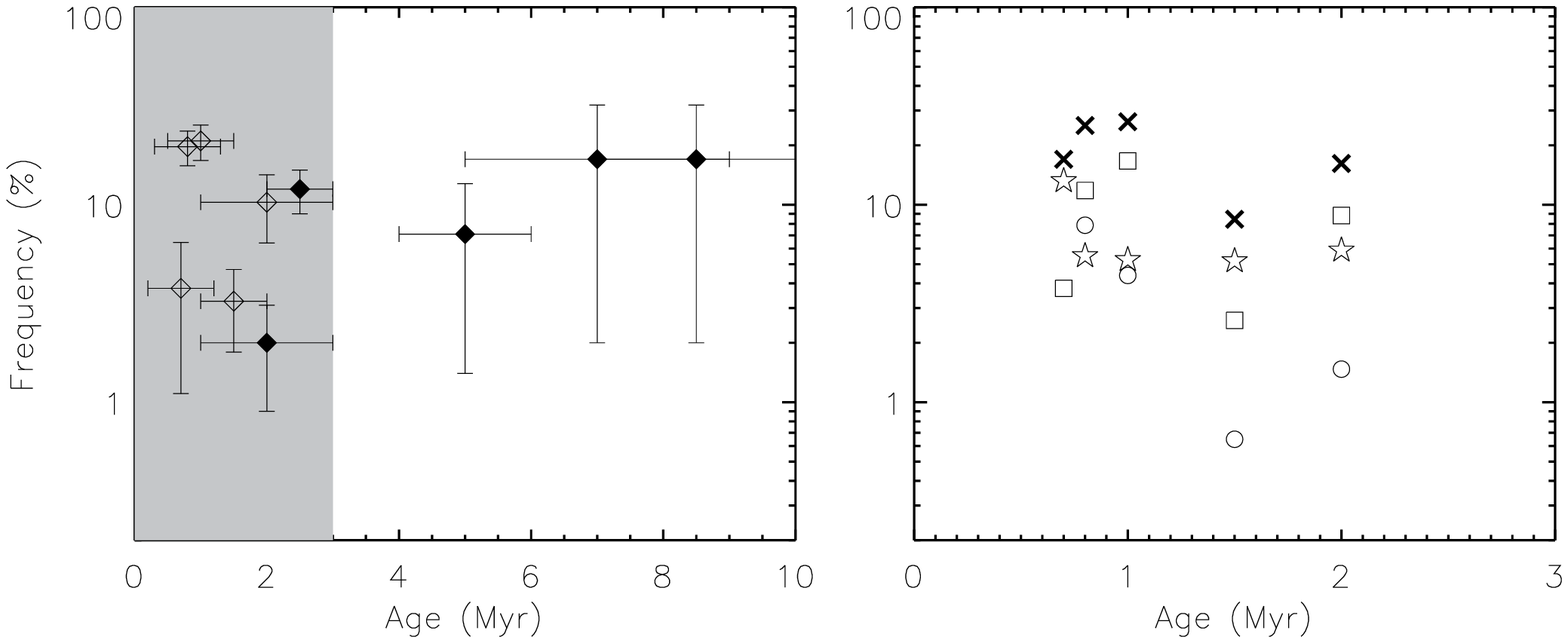}
\caption{The fraction of transitional disks plotted against the estimated age of the association. The solid diamonds in the left panel are from the fraction of classical transitional disks defined by \citet{Muzerolle2010}. The empty diamonds in the left panel are the fraction of CTD plus WTD from this work. The right panel is for the shaded region covering ages of 0-3 Myr in the left panel, and it shows fractions of TD (=CTDs+WTDs+PTDs) types in N1333, ONC, L1641, Tau, and ChaI from younger ages to older ages of star-forming regions. The symbols in the right panel indicate the TD fraction (cross), the CTD fraction (square), the WTD fraction (circle), and the PTD fraction (star).
\label{fig-xdiskfreq}}
\end{figure}
\clearpage

\clearpage
\begin{figure}[t]
\epsscale{0.8}
\plotone{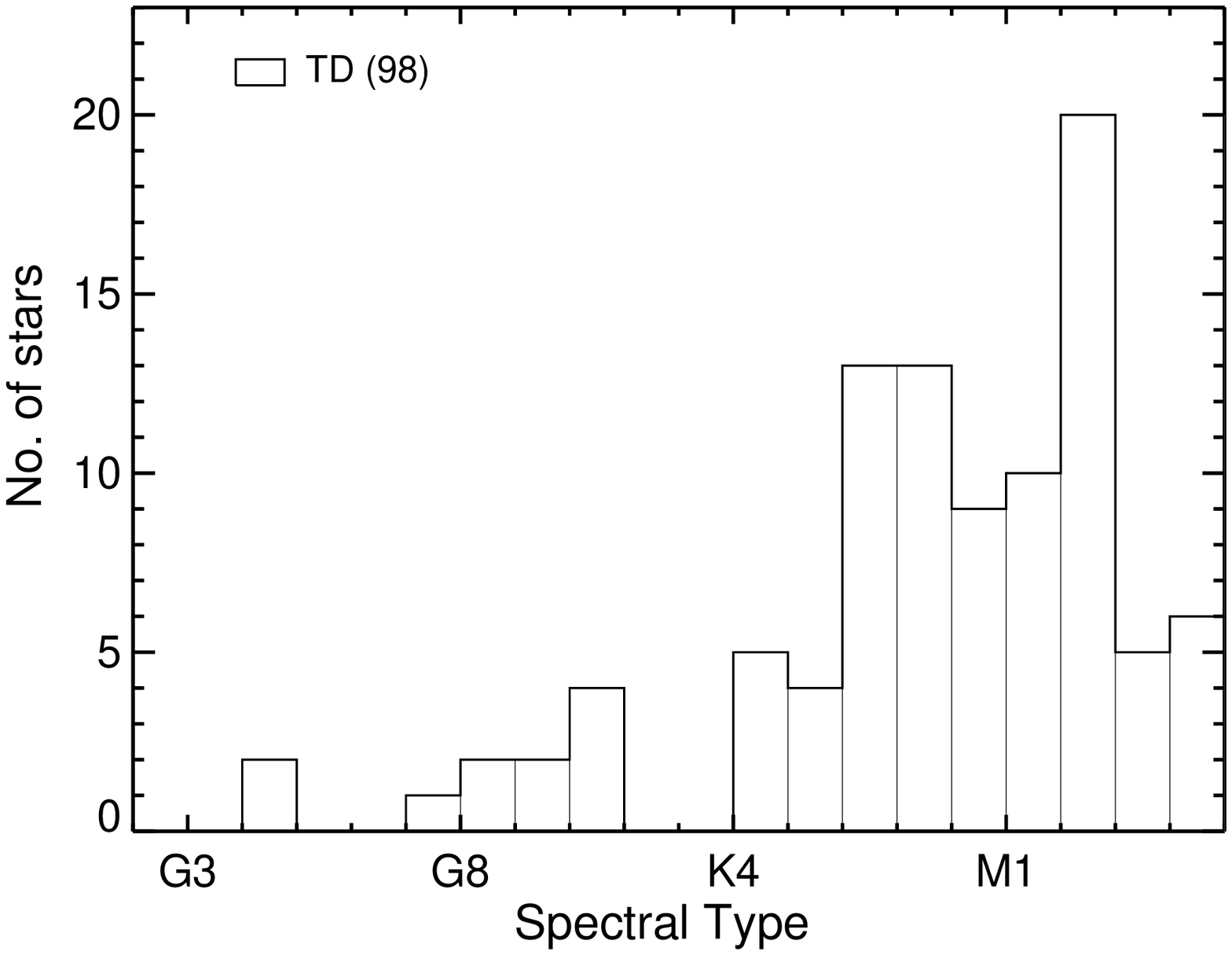}
\plotone{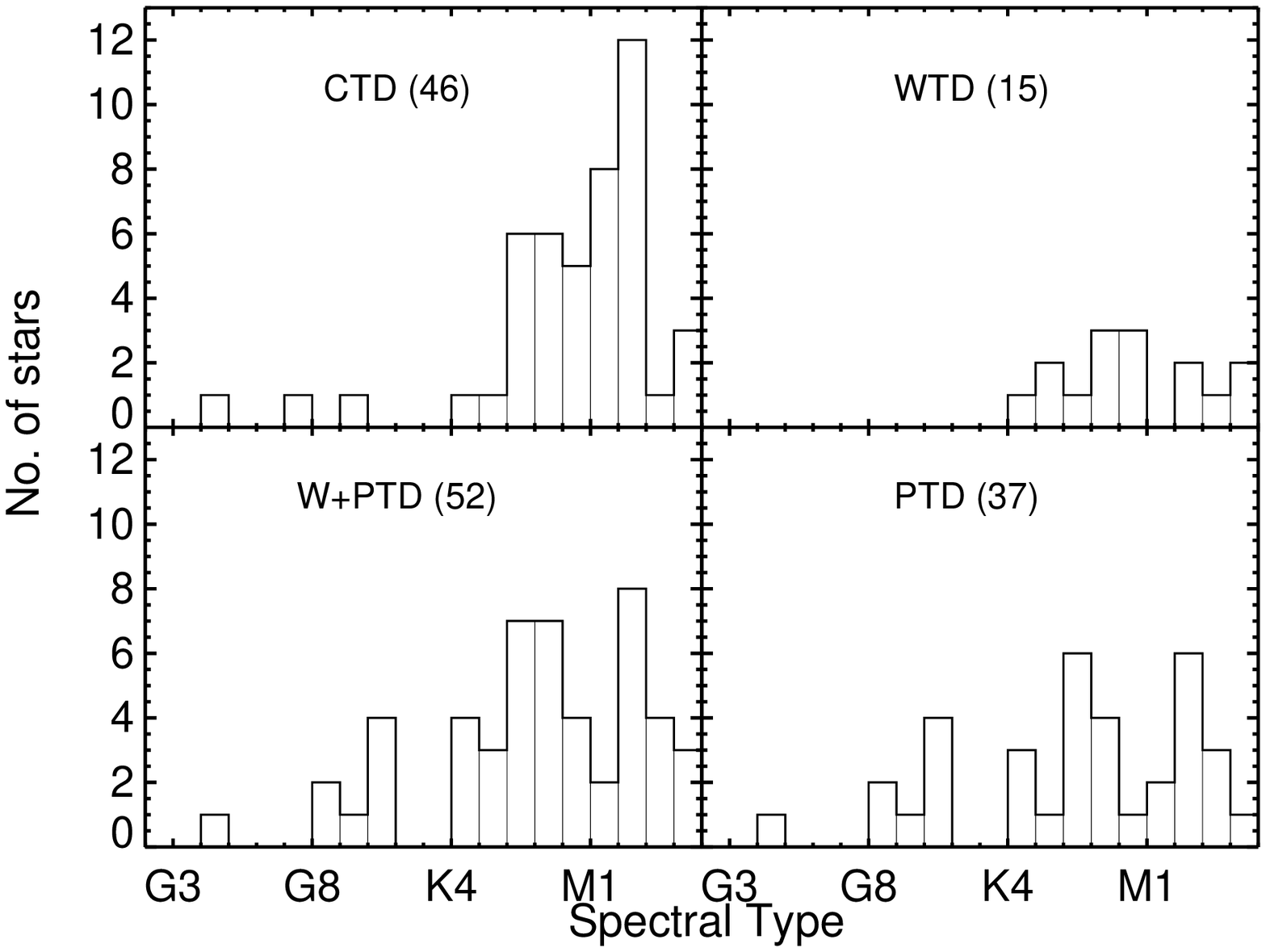}
\caption{Spectral type distribution of transitional disks. The upper bigger panel shows the spectral type distribution of all TDs with known spectral types. The lower multiple panels show the spectral type distributions divided by TD subtypes. The results from K-S tests: (1) CTD vs. WTD: $D=0.20$, $p=0.69$; (2) WTD vs. PTD: $D=0.32$, $p=0.17$; (3) CTD vs. PTD: $D=0.30$, $p=0.04$; (4) CTD vs. WTD+PTD: $D=0.24$, $p=0.10$. \label{fig-histSpT}}
\end{figure}
\clearpage

\clearpage
\begin{figure}[t]
\epsscale{0.8}
\plotone{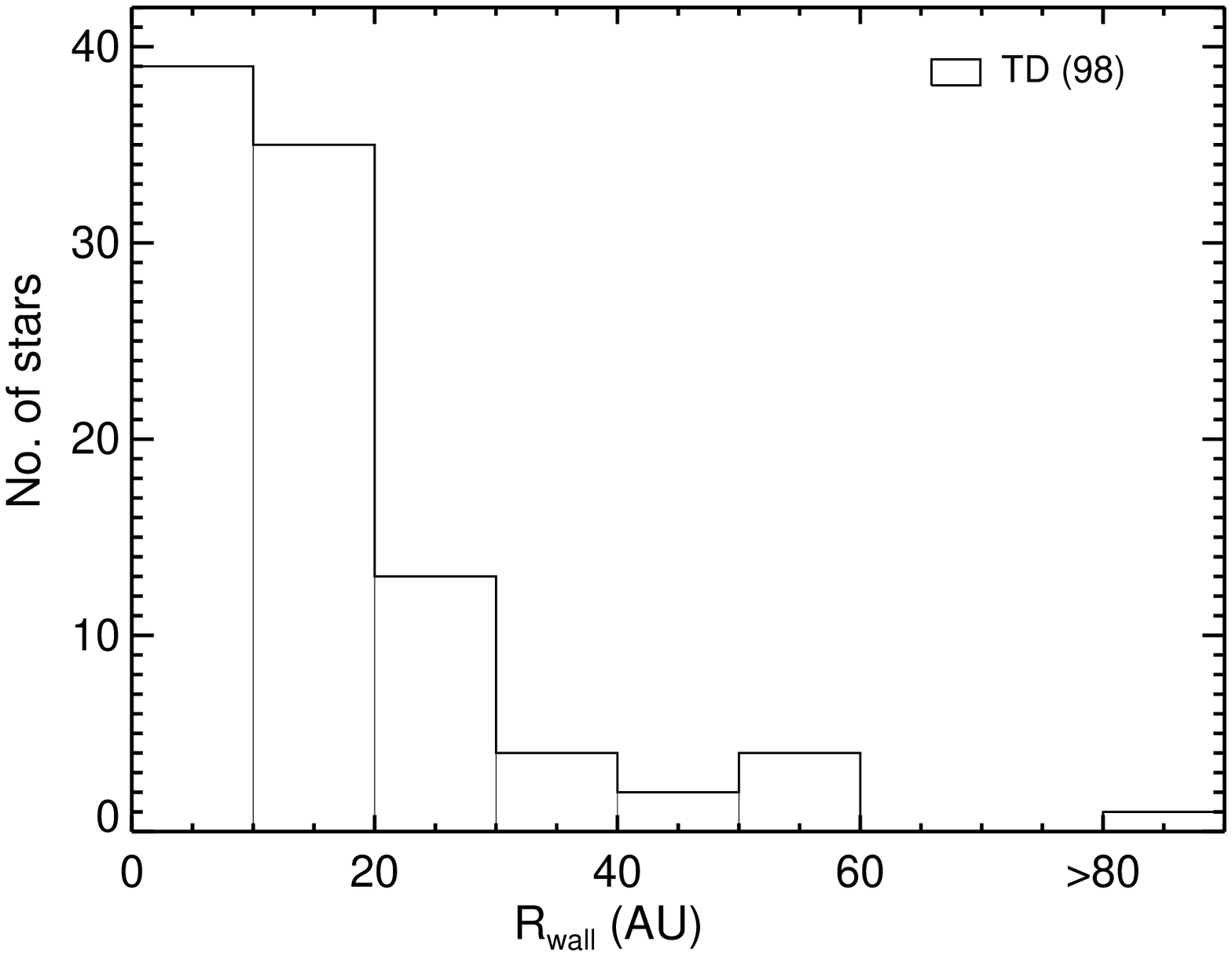}
\plotone{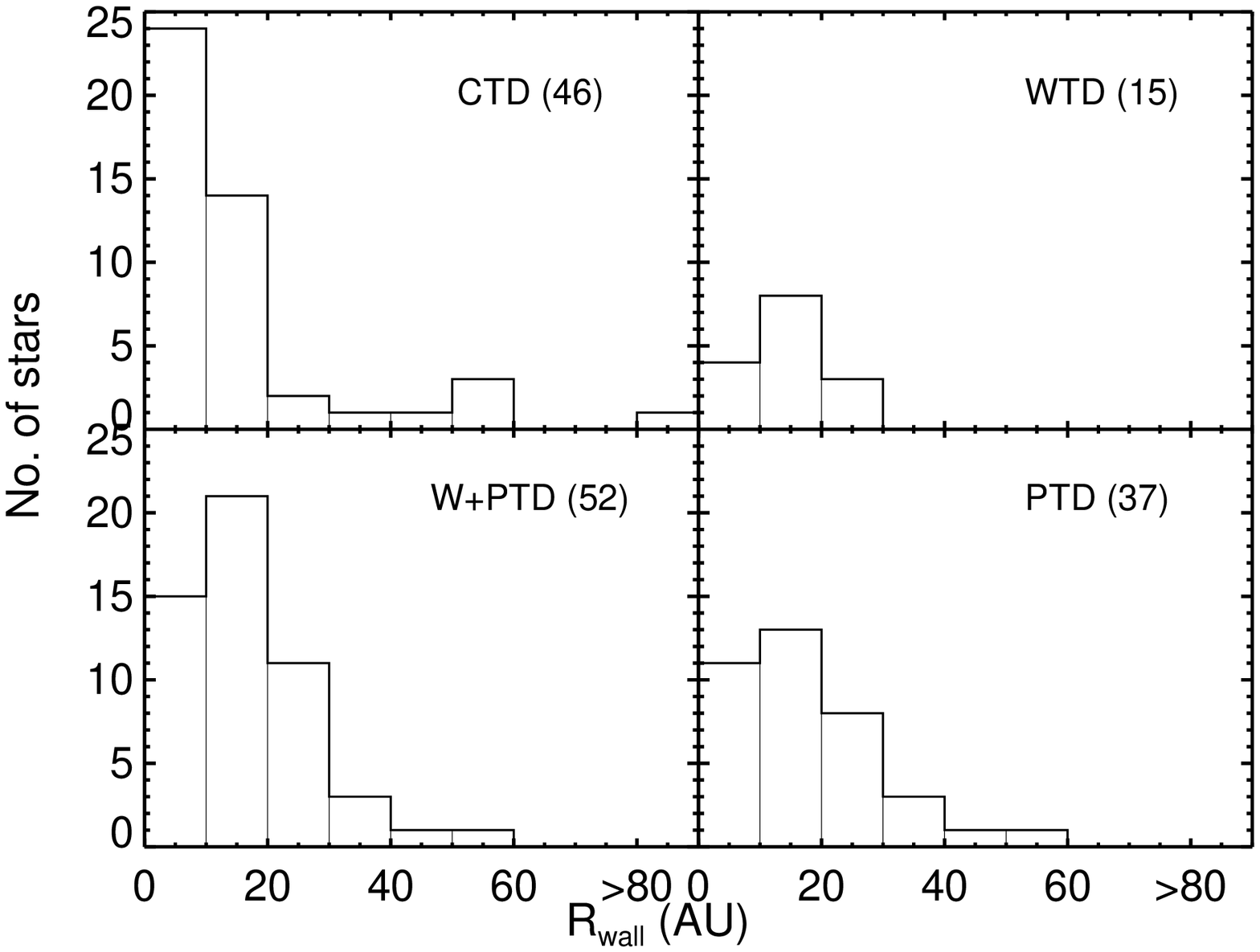}
\caption{$R_{wall}$ distribution of transitional disks. The upper bigger panel shows the $R_{wall}$ distribution of all TDs with known spectral types. The lower multiple panels show the $R_{wall}$ distributions divided by TD subtypes. The results from K-S tests: (1) CTD vs. WTD: $D=0.32$, $p=0.15$; (2) WTD vs. PTD: $D=0.27$, $p=0.36$; (3) CTD vs. PTD: $D=0.29$, $p=0.05$; (4) CTD vs. WTD+PTD: $D=0.29$, $p=0.02$.\label{fig-histRw}}
\end{figure}
\clearpage

\clearpage
\begin{figure}[t]
\epsscale{1}
\plotone{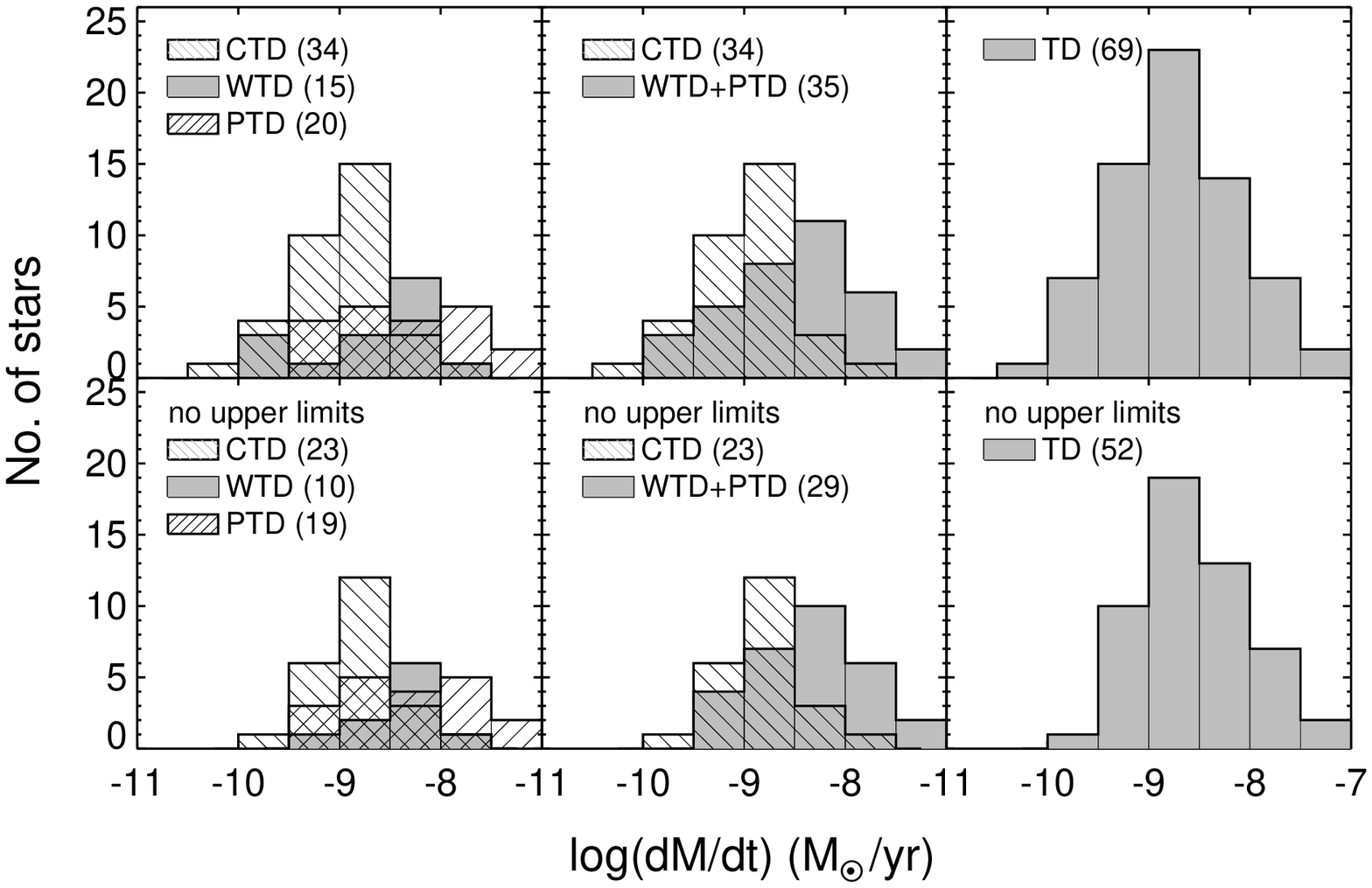}
\caption{$\dot{M}$ distributions. The upper panels include upper limits of $\dot{M}$; the lower panels do not include upper limits of $\dot{M}$. The left panels show $\dot{M}$ distributions separated by three different TD types; the middle panels compare the $\dot{M}$ distribution of disks with central clearings (CTD) and to those with gaps (WTD and PTD); the right panels result from adding all types of TDs. The results of K-S tests of $\dot{M}$ distributions between CTD and WTD$+$PTD in the middle panels: $D=0.46$, $p=0.001$ (upper middle); $D=0.5$, $p=0.002$ (lower middle). \label{fig-histMdot}}
\end{figure}
\clearpage

\clearpage
\begin{figure}[t]
\epsscale{0.5}
\plotone{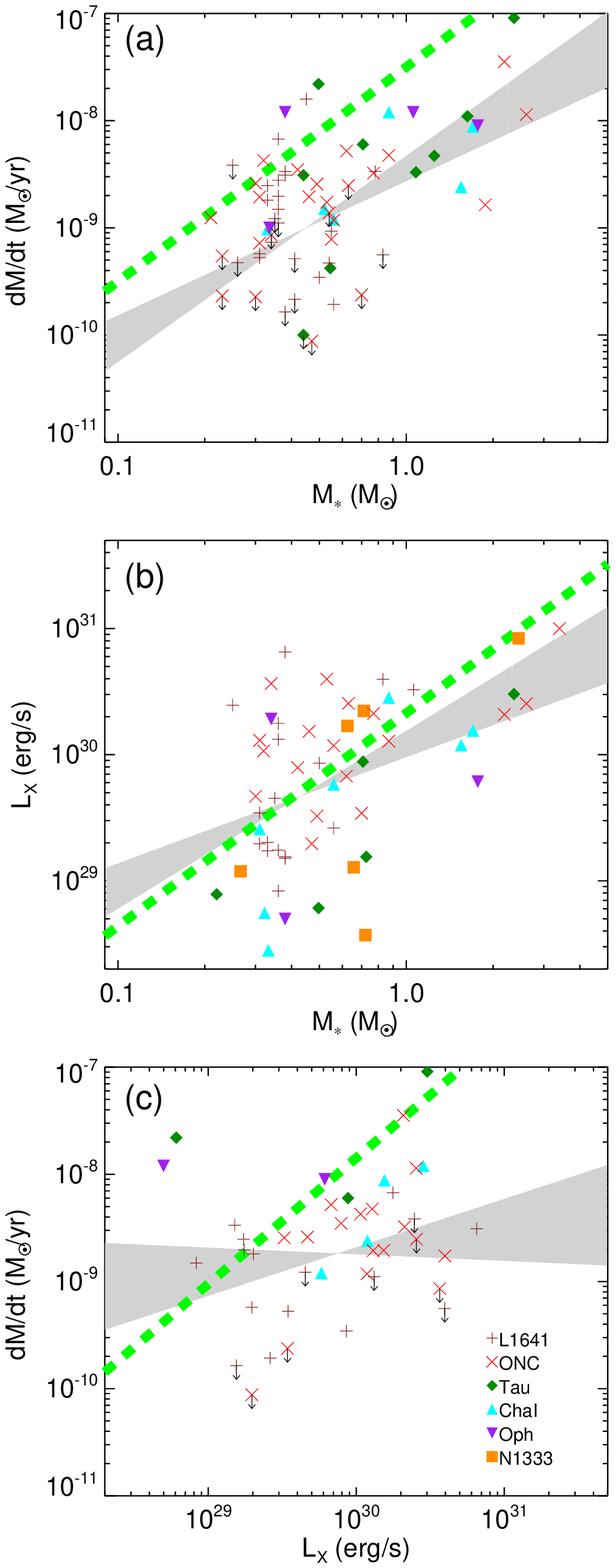}
\caption{Trends among $M_\star$, $\dot{M}$ and $L_{X}$. In each plot, different symbols indicate TDs in different star-forming regions (plus: L1641; cross: ONC; solid diamond: Tau; solid triangle: ChaI; solid inverse-triangle: Oph; solid square: N1333). In each plot, the shaded area indicates the 1$\sigma$ uncertainty of the linear regression (of the logarithms). If a shaded area is narrow with high slope (e.g., $\dot{M}$-$M_\star$), one can tell two properties in a panel is tightly correlated. If a shaded area is broad with very low slope (e.g., $\dot{M}$-$L_X$), two properties in a panel is not correlated. The thick dashed line indicates a correlation expected/observed among T Tau disks in Tau: $\dot{M}$~$\propto$~${M_\star}^{2}$ from \citet{muzerolle03mdot} in the upper panel; $L_{X}$~$\propto$~${M_\star}^{1.69}$ from \citet{Telleschi07} in the middle panel; $\dot{M}$~$\propto$~${L_{X}}^{1.2}$ in the lower panel from above two relations. The down arrows indicate $\dot{M}$ upper limits. (A color version of this figure is available in the online journal.) \label{fig-trend-MstarMdotLx}}
\end{figure}
\clearpage

\clearpage
\begin{figure}[t]
\epsscale{0.5}
\plotone{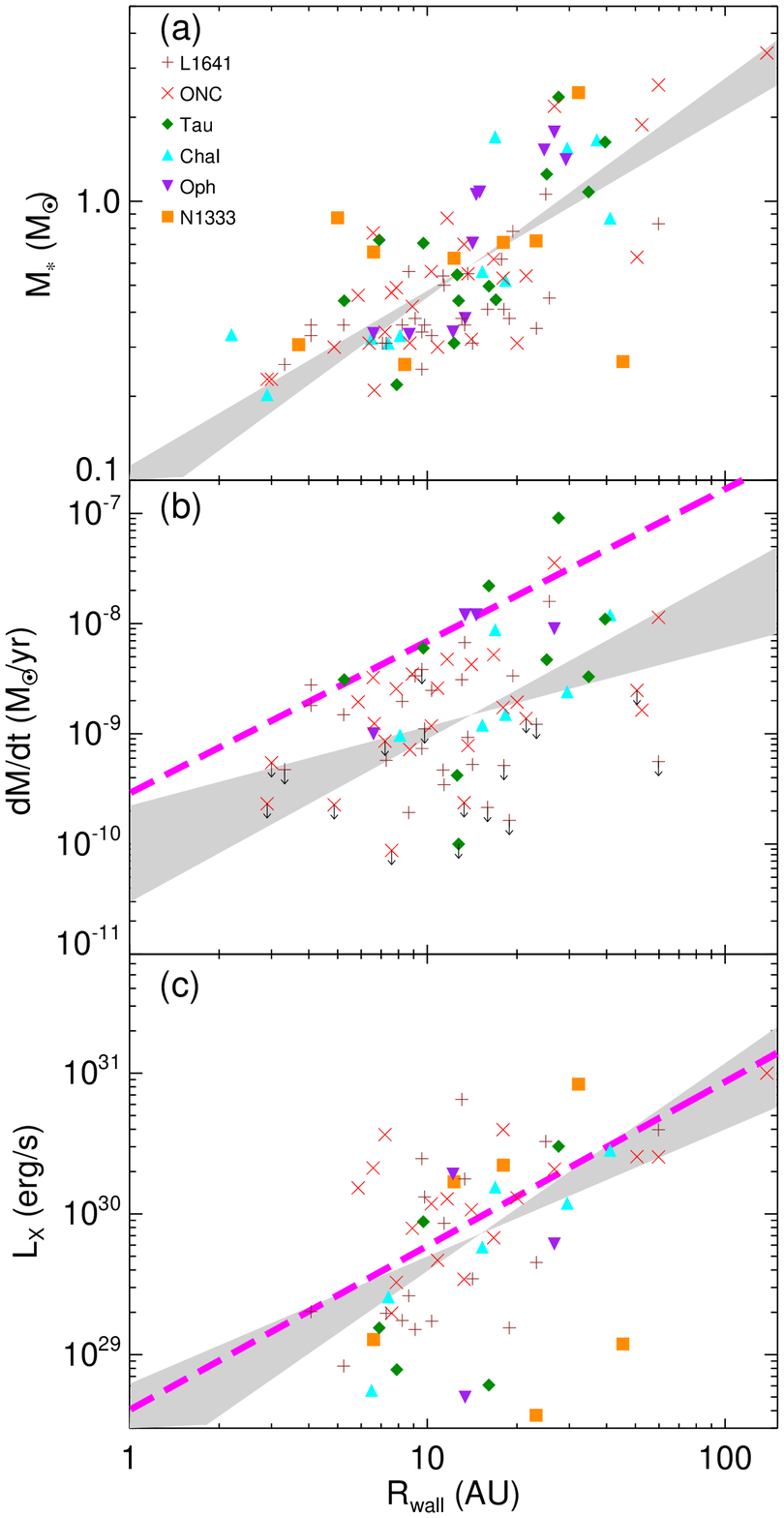}
\caption{Trends related to $R_{wall}$ with other stellar properties. Lines, shadow, and symbols have same meaning as in Figure~\ref{fig-trend-MstarMdotLx}. The thick long-dashed line indicates an expected trend derived from the relationship of a property in y-axis with $M_{\star}$ shown in Figure~\ref{fig-trend-MstarMdotLx} and the strong $M_\star$-$R_{wall}$ correlation shown in the panel (a): $\dot{M}$~$\propto$~${R_{wall}}^{1.4}$ in the panel (b); $L_{X}$~$\propto$~${R_{wall}}^{1.2}$ the panel (c). (A color version of this figure is available in the online journal.) \label{fig-trend-Rw}}
\end{figure}
\clearpage

\clearpage
\begin{figure}[t]
\epsscale{1}
\plotone{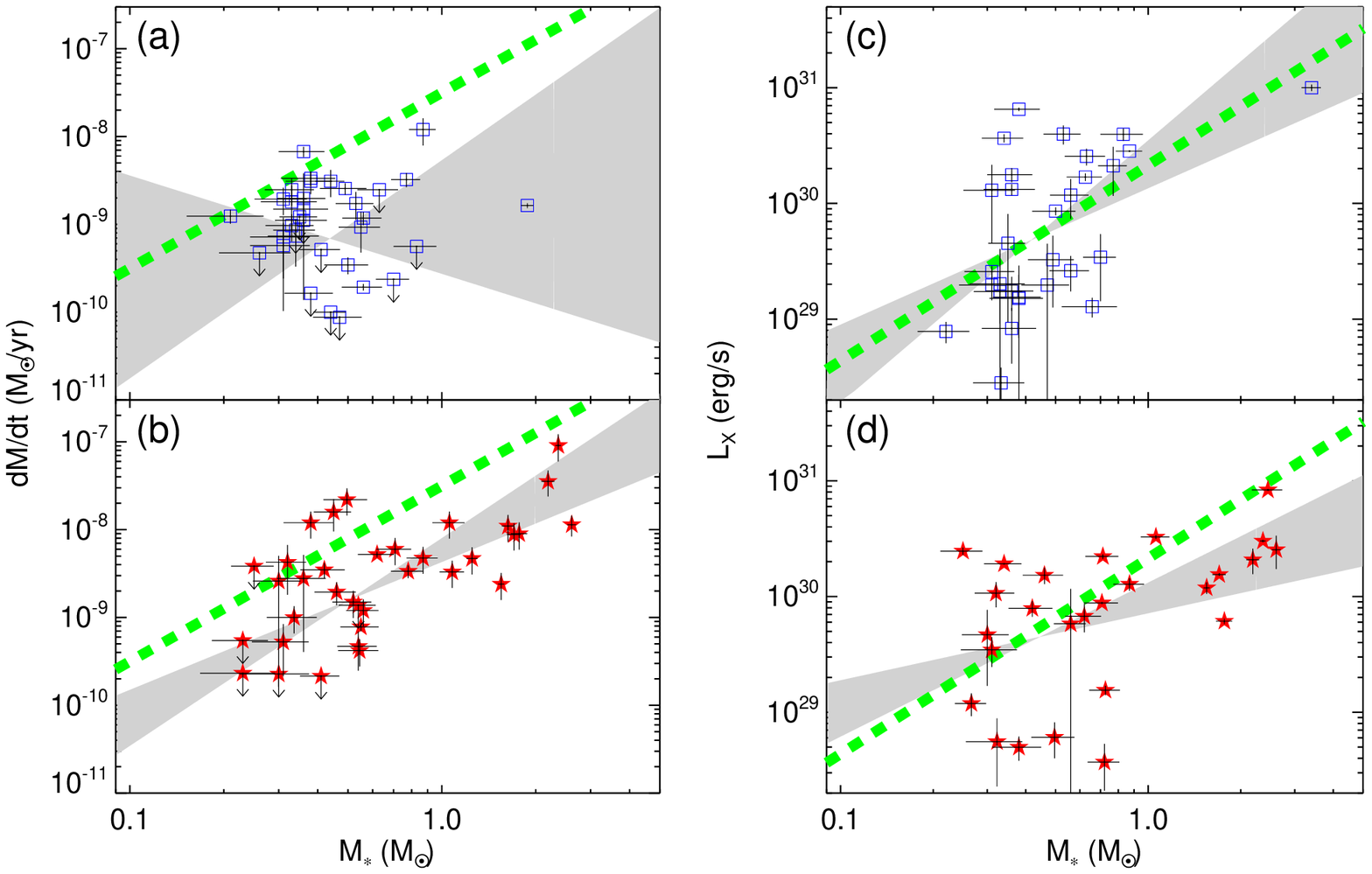}
\caption{Detailed Trends of $\dot{M}$-$M_{\star}$ (the left panels) and $L_{X}$-$M_{\star}$ (the right panels) separated by subtypes of TDs. The upper panels (a and c) show the correlation of CTDs (open square). The lower panels (b and d) show the correlation of WTDs+PTDs (solid star). The meanings of lines and shade are same as defined in the caption of Figure~\ref{fig-trend-MstarMdotLx}. (A color version of this figure is available in the online journal.) \label{fig-subtrend_Mdot-Lstar}}
\end{figure}
\clearpage

\clearpage
\begin{figure}[t]
\epsscale{1}
\plotone{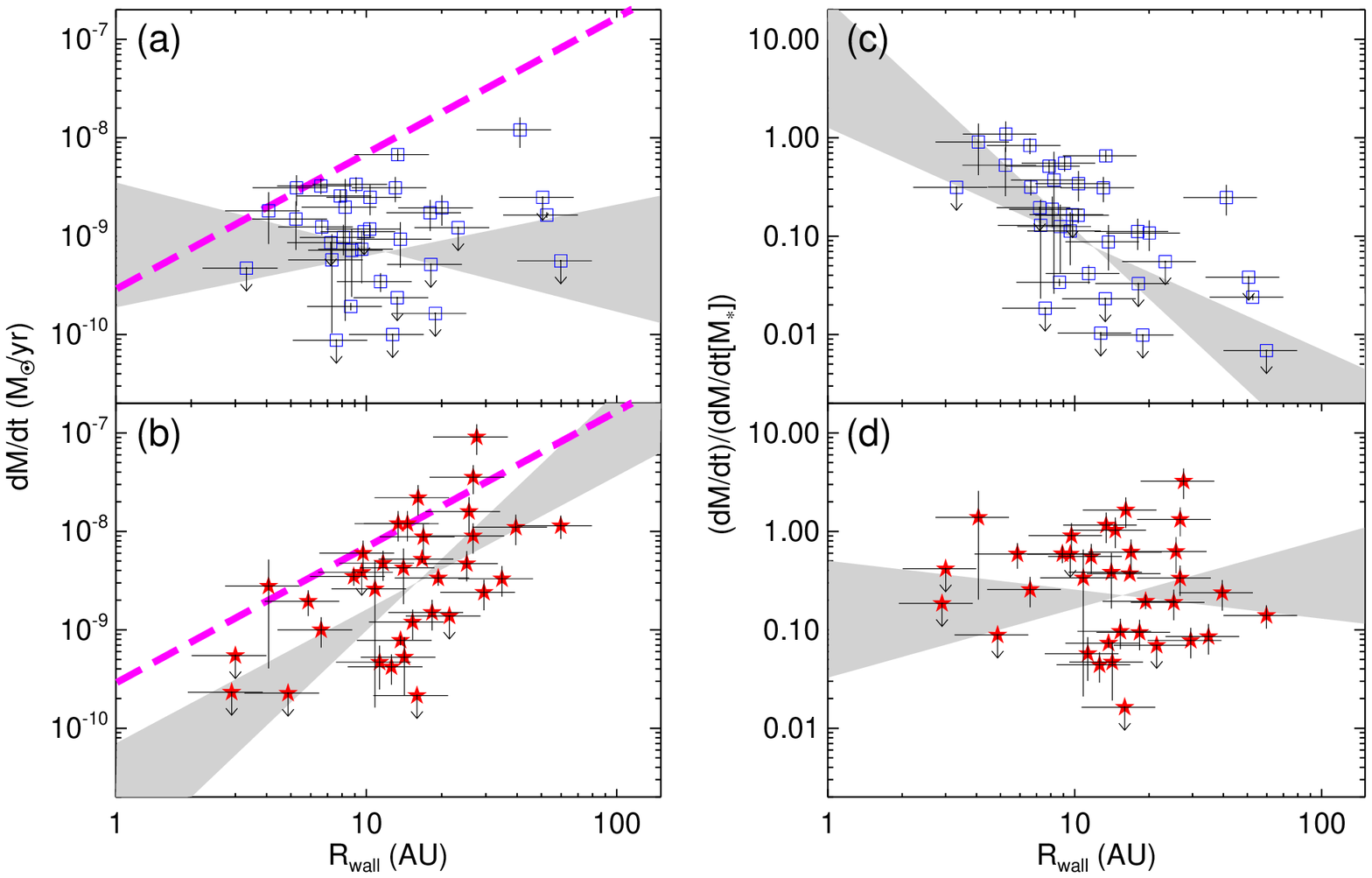}
\caption{Detailed trends of $\dot{M}$-$R_{wall}$: correlations separated in two sub-groups, CTDs (open square) and WTDs+PTDs (solid star). The symbol in each panel is same as Figure~\ref{fig-subtrend_Mdot-Lstar}. The thick long-dashed line in the panel (a) and (b) is same in Figure~\ref{fig-trend-Rw} (b). The right panels show the trend at no $M_{\star}$ dependence by presenting the deviation of $\dot{M}$ from the thick long-dashed line of $\dot{M}(M_\star)$-$R_{wall}$. (A color version of this figure is available in the online journal.) \label{fig-subtrend_Mdot-Rwall}}
\end{figure}
\clearpage

\clearpage
\begin{figure}[t]
\epsscale{1}
\plotone{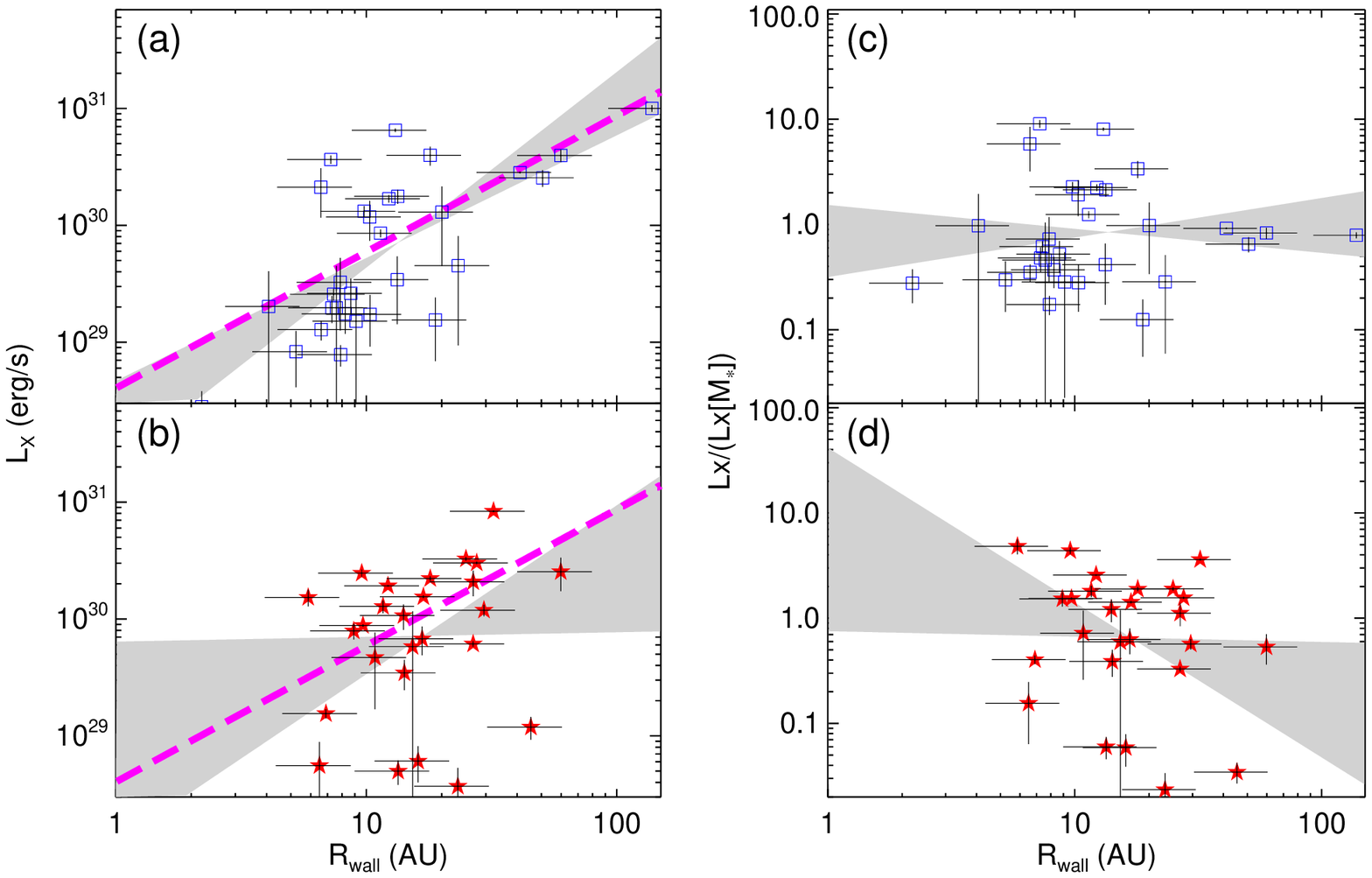}
\caption{Detailed trends of $L_{X}$-$R_{wall}$: correlations separated in two sub-groups, CTDs (open square) and WTDs+PTDs (solid star). The symbol in each panel is same as in Figure~\ref{fig-subtrend_Mdot-Lstar}. The thick long-dashed line in the panel (a) and (b) is same as in Figure~\ref{fig-trend-Rw} (c). The right panels show the trend at no $M_{\star}$ dependence by presenting the deviation of $L_{X}$ from the thick long-dashed line of $L_{X}(M_{\star})$-$R_{wall}$. (A color version of this figure is available in the online journal.) \label{fig-subtrend_Lx-Rwall}}
\end{figure}
\clearpage

\clearpage
\begin{figure}[t]
\epsscale{1}
\plotone{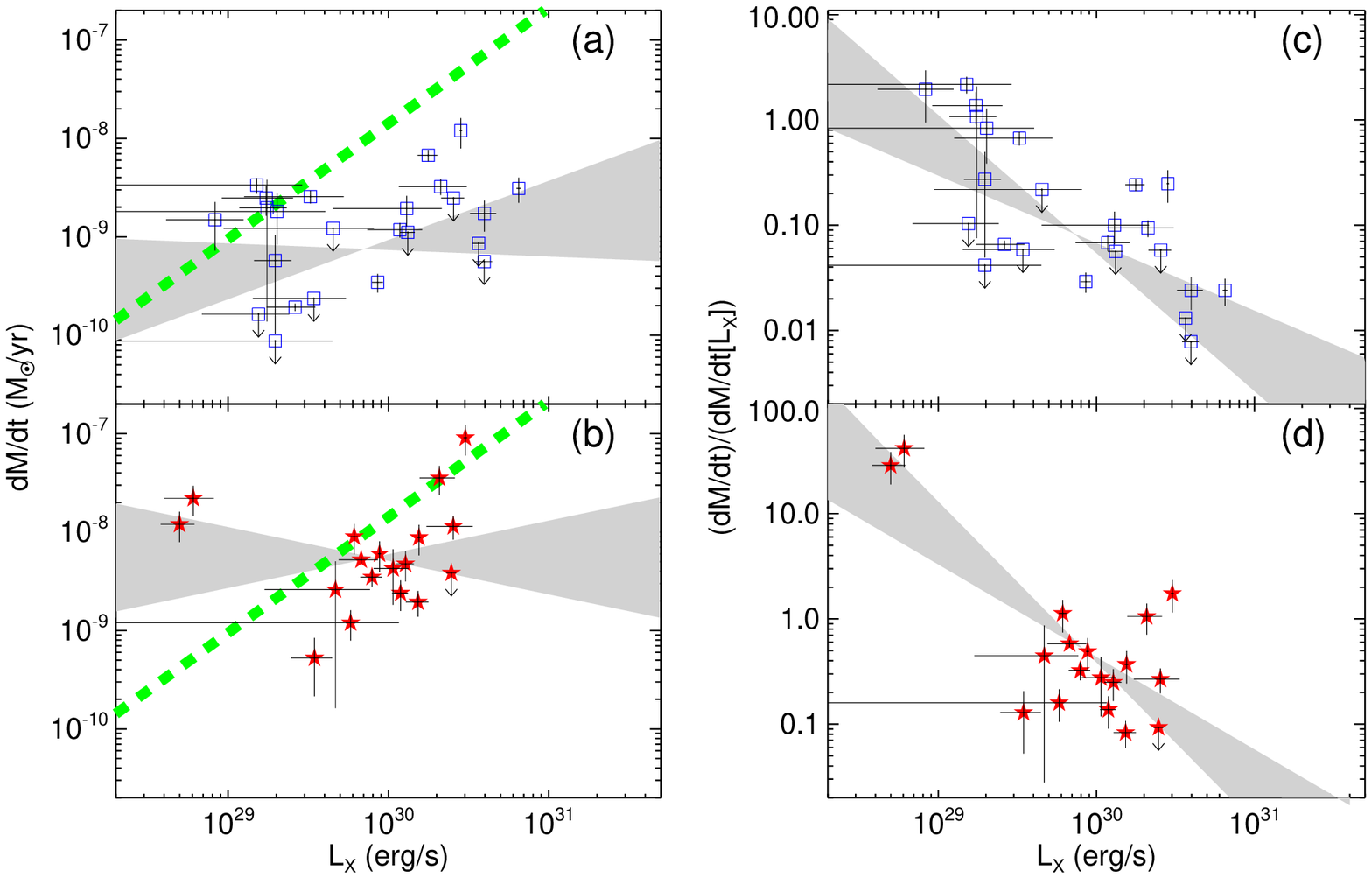}
\caption{Detailed trends of $\dot{M}$-$L_{X}$: correlations separated in two sub-groups, CTDs (open square) and WTDs+PTDs (solid star). The symbol in each panel is same as in Figure~\ref{fig-subtrend_Mdot-Lstar}. The thick dashed line represent $\dot{M}$-$L_{X}$ relation of T Tauri star in Tau as explained in \S 6.2.1 and Figure~\ref{fig-trend-MstarMdotLx}. (A color version of this figure is available in the online journal.) \label{fig-subtrend_Mdot-Lx}}
\end{figure}
\clearpage

\clearpage
\begin{figure}[t]
\epsscale{1}
\plotone{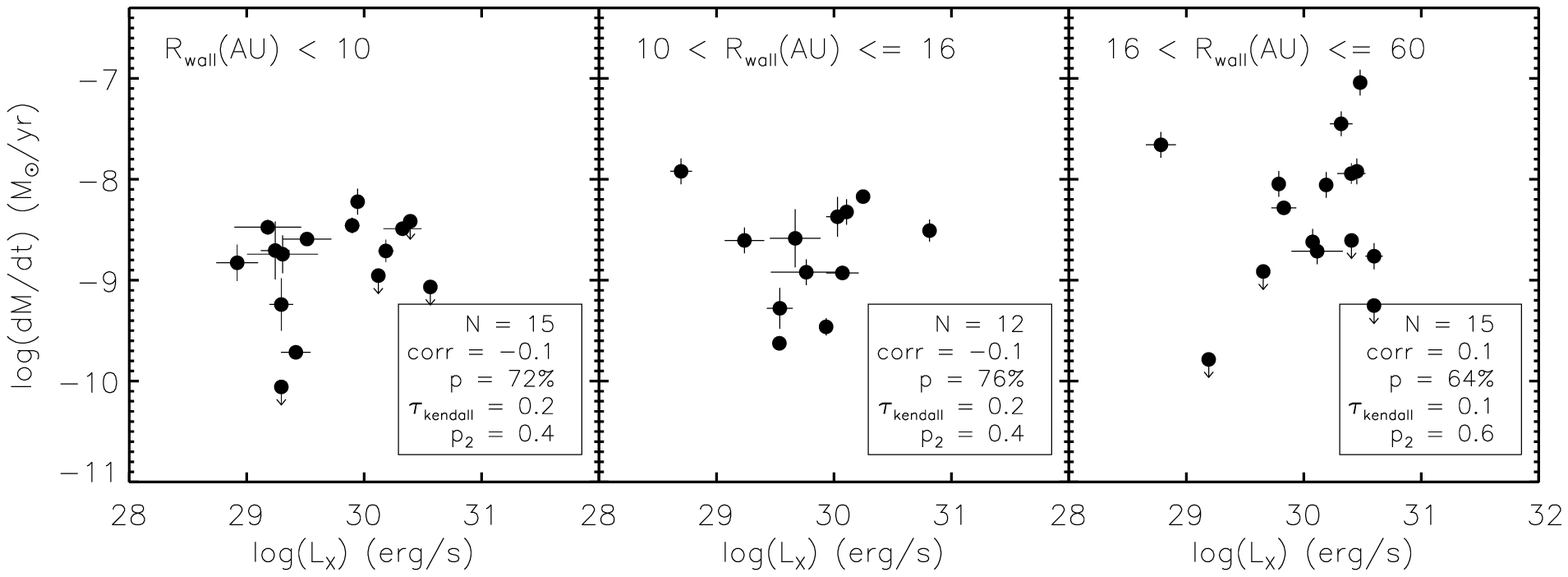}
\caption{The tests to find the relationship between $\dot{M}$ and $R_{wall}$ under a constant $R_{wall}$ condition. We present the sub-groups separated by three $R_{wall}$ bins. The coverage of $R_{wall}$ of each bin is indicated on each panel with the number of sub-sample (N), a linear correlation coefficient (corr) between $log \dot{M}$ and $logL_{X}$, and a probability (p) of getting corr from random distribution. $\tau_{kentall}$ is Kendall's tau which indicates the degree of correlation between two variables; closer to 1, tighter correlation. p${_2}$ indicate the two-sided p value of $\tau_{kentall}$; if p${_2}$=1, the probability of no correlation is 100$\%$. \label{fig-Mdot-Lx-constRwall}}
\end{figure}
\clearpage

\clearpage
\begin{figure}[t]
\epsscale{0.7}
\plotone{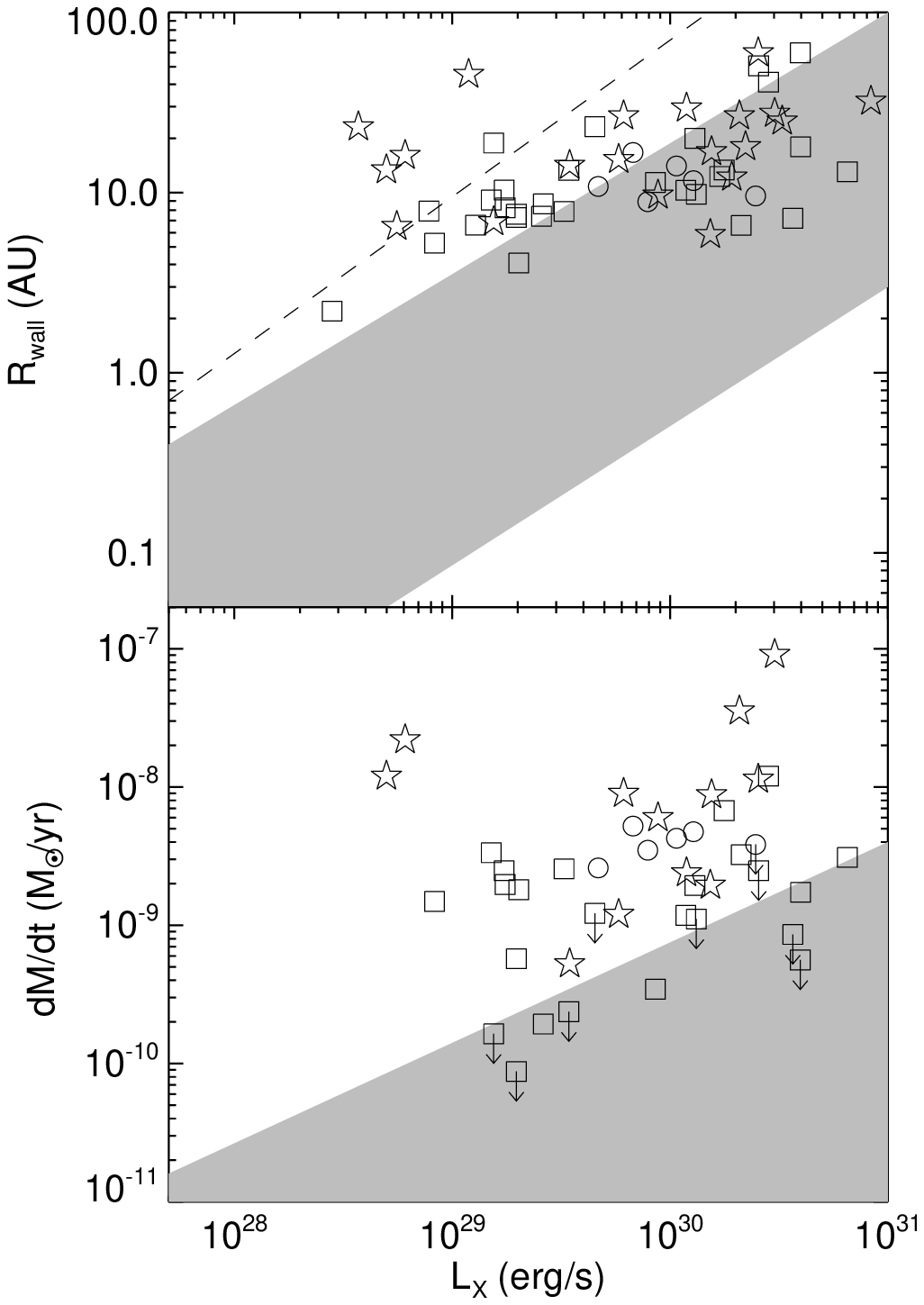}
\caption{Comparison of the predictions by X-ray photoevaporation (\citet{Owen_photoevaporation2012}, Fig 18 in their paper) to the properties of observed TDs. The gray shadows are the domains in which X-ray photoevaporation is dominant. The dashed line adopted from Fig 18. in \citet{Owen_photoevaporation2012} represents the maximum radius a TD may reach before thermal sweeping sets in and a disk dissipates beyond TD stage. The model domains are derived with $R_{g}$~$\propto$~$M_\star$ and the positive correlation between $M_\star$ and $L_{X}$ \citep{Owen_photoevaporation2012}, similar to our finding, so the photoevaporation prediction and the observed TDs properties generally show positive correlations. \label{fig-owen-XPE-fig18}}
\end{figure}
\clearpage

\clearpage
\begin{figure}[t]
\epsscale{0.9}
\plotone{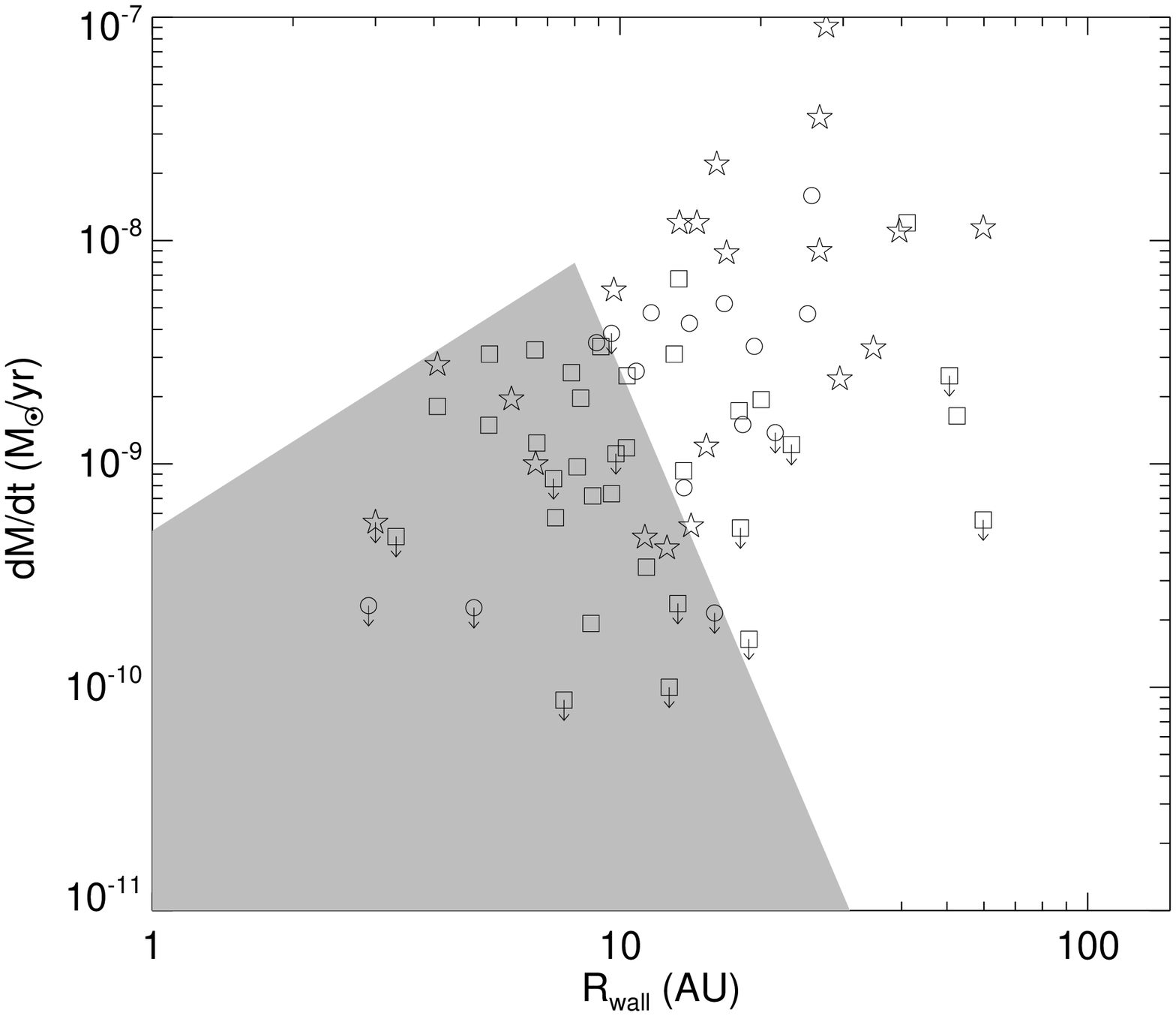}
\caption{Comparison of $\dot{M}$ vs. $R_{wall}$ to the predictions by X-ray photoevaporation (gray shadow region: \citet{Owen_photoevaporation2012}). Each symbol indicate different subtype of TDs: square (CTD); circle (WTD); star (PTD). \label{fig-mdot-Rw-model}}
\end{figure}
\clearpage

\clearpage
\begin{figure}[t]
\epsscale{0.9}
\plotone{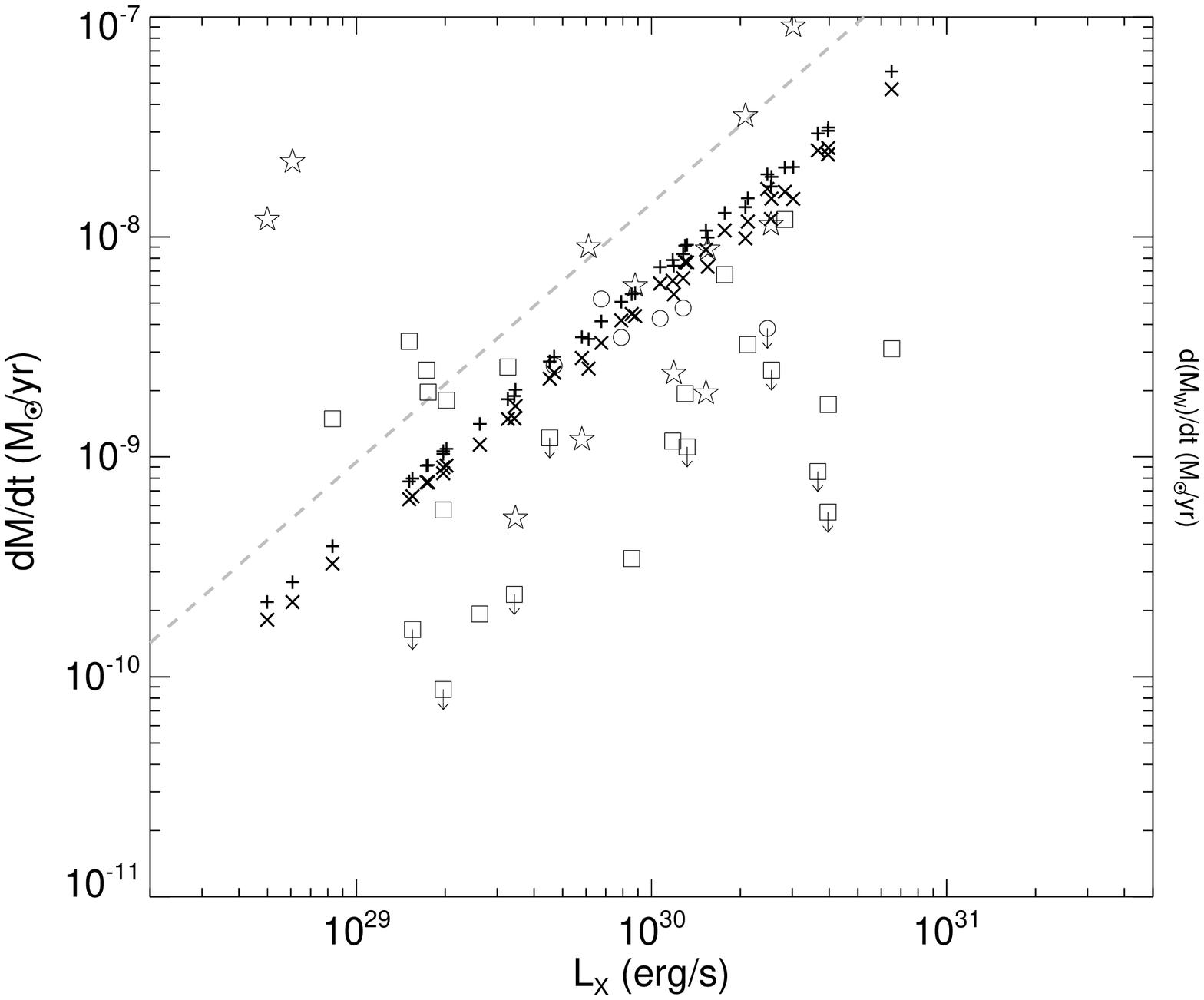}
\caption{Comparison of $\dot{M}$ vs. $L_X$ of transitional disks to $M_{W}$ vs. $L_X$ expected by the X-ray photoevaporation model \citep{Owen_photoevaporation2012}. The plus signs are for the expected X-ray photoevaporation wind rates in case of a radially continuous primordial disk with TDs $M_\star$ and $L_{X}$, and the cross signs are for that in case of transitional disks with inner hole. The squares, circles, and stars are CTDs, WTDs, and PTDs of our sample. The gray dashed line indicate the expected $\dot{M}$-$L_X$ of CTTS, which is derived from $\dot{M}$-$M_\star$ and $L_\star$-$M_\star$ of CTTS in Figure~\ref{fig-trend-MstarMdotLx}. Detail discussion is in Appendix B. \label{fig-mdot-Lx-owen}}
\end{figure}
\clearpage

\clearpage
\begin{deluxetable}{cccccccccc}
\rotate
\tablecaption{Observation and reduction log of IRS spectra \label{table:IRSlog}}
\tablehead{
\colhead{Num.}& \colhead{IRS name} & \colhead{2MASS name} & \colhead{RA (J2000)} & \colhead{DEC (J2000)} & \colhead{AORID} & \colhead{IRS Camp.} & \colhead{Date observed} & \colhead{region} & \colhead{reduction} \\
\colhead{(1)}& \colhead{(2)} & \colhead{(3)} & \colhead{(4)} & \colhead{(5)} & \colhead{(6)} & \colhead{(7)} & \colhead{(8)} & \colhead{(9)} & \colhead{(10)}
}
\tablecolumns{10}
\tablewidth{0pt}
\tabletypesize{\scriptsize}
\startdata
1	&	8336884-5290	&	05332852-0517262	&	05 33 28.52	&	-05 17 26.2	&	18804224	 &	 39	 &	 3/12/2007	 &	ONC	&	 AdOpt	 \\
5	&	8343944-5609	&	05334545-0536323	&	05 33 45.47	&	-05 36 32.5	&	18833152	 &	 39	 &	 3/12/2007	 &	ONC	&	 opse/AdOpt\tablenotemark{a}	 \\
8	&	8346141-5010	&	05335074-0500394	&	05 33 50.74	&	-05 00 39.5	&	18781952	 &	 39	 &	 3/22/2007	 &	ONC	&	 AdOpt	 \\
11	&	8347711-5533	&	05335450-0532003	&	05 33 54.51	&	-05 32 00.4	&	18844416	 &	 39	 &	 3/12/2007	 &	ONC	&	 AdOpt	 \\
13	&	8350150-5595	&	05340034-0535434	&	05 34 00.36	&	-05 35 43.5	&	18778880	 &	 39	 &	 3/12/2007	 &	ONC	&	 man	 \\
16	&	8353051-5229	&	05340731-0513454	&	05 34 07.32	&	-05 13 45.1	&	18801408	 &	 39	 &	 3/22/2007	 &	ONC	&	 AdOpt	 \\
18	&	8355083-4835	&	05341221-0450072	&	05 34 12.20	&	-04 50 07.1	&	18810624	 &	 39	 &	 3/24/2007	 &	ONC	&	 AdOpt	 \\
19	&	8355371-5480	&	05341289-0528482	&	05 34 12.89	&	-05 28 48.3	&	18779904	 &	 44	 &	 10/5/2007	 &	ONC	&	 AdOpt/opse\tablenotemark{b}	 \\
\enddata
\tablecomments{\footnotesize Column (1) Num.: the numbers are from the number sequence of 303 Class II objects observed in the IRS program number 30706. We use these numbers to identify objects easily. This number sequence will be used consistently in other future papers dealing with the objects in the PID 30706.\\
Column (2) IRS name: sometimes IRS name indicated as in the column used to identify objects. \\
Column (10) reduction: it indicate the methods of source extraction to get the SEDs in Figure 2. auto: an automated tapered column extraction in SMART with off-nod or off-order sky subtraction; man: a manual tapered column extraction in SMART with a polynomial sky subtraction; AdOpt: an optimal source extraction using an empirical point response function (PRF) in SMART; opse: an optimal source extraction using an analytical PRF.}
\tablenotetext{a} {\footnotesize SL $\&$ LL1: opsew7p; LL2: AdOpt.}
\tablenotetext{b} {\footnotesize SL: AdOpt; LL: opse.}
\tablenotetext{c} {\footnotesize SL: opse, LL:AdOpt.}
\tablenotetext{d} {\footnotesize SL: AdOpt; LL:auto onss.}
\end{deluxetable}

\clearpage

\clearpage
\begin{deluxetable}{ccccccc}
\tablecaption{Observation log of SpeX spectra \label{table:SpeXlog}}
\tablehead{
\colhead{Num.}& \colhead{IRS name} & \colhead{obs.} & \colhead{obs. date} & \colhead{slit width} & \colhead{total int. time} & \colhead{comment}\\
\colhead{}& \colhead{} & \colhead{semester} & \colhead{(UT)} & \colhead{(")} & \colhead{(sec)} & \colhead{}
}
\tablecolumns{7}
\tablewidth{0pt}
\tabletypesize{\scriptsize}
\startdata
1	&	8336884-5290	&	2011B	&	11/6/2011	&	0.5	&	480	&		\\
5	&	8343944-5609	&	2011A	&	2/28/2011	&	0.8	&	400	&		\\
11	&	8347711-5533	&	2010A	&	2/18/2010	&	0.8	&	480	&		\\
13	&	8350150-5595	&	2010A	&	2/17/2010	&	0.8	&	480	&		\\
16	&	8353051-5229	&	2010A	&	2/18/2010	&	0.8	&	240	&		\\
19	&	8355371-5480	&	2010A	&	2/17/2010	&	0.8	&	720	&	a	\\
23	&	8357031-5072	&	2011B	&	11/7/2011	&	0.8	&	240	&		\\
24	&	8358147-5505	&	2011A	&	2/27/2011	&	0.8	&	240	&		\\
\enddata
\tablecomments{\footnotesize (a) The source signal on order 8 is very faint, so that order 8 source extraction failed.; (b) It is a binary system with two sources resolved from SpeX observations. The possible primary source giving the adopted $\dot{M}$ in Table~\ref{table:property} is presented in the black line from 0.8-2.4 \micron\ in Figure~\ref{fig-SED-2}. The possible secondary is presented in the gray line in the same plot.; (c) The position angle of slit was 27$^\circ$.; (d) Resolved two sources on the guider image are dominantly displaced in the north-south direction with slight separation in east-west direction.; (e) 4 among 6$\times$120 sec exposures are combined in the spectrum.; (f) The two sources are aligned to N-S direction.; (g) The source signals in orders 6,7, and 8 is very faint so that source extraction for order 6-8 failed.; (h) The position angle of slit was -20$^\circ$.}
\end{deluxetable}
\clearpage

\clearpage
\begin{deluxetable}{cccccc}
\tablecaption{Spectral Type and Extinction \label{table:SpTypeAv}}
\tablehead{
\colhead{Num.} & \colhead{IRS name} & \colhead{Spectral type} & \colhead{SpT reference} & \colhead{$A_{V}$} & \colhead{$A_{V}$ method} \\
\colhead{(1)} & \colhead{(2)} & \colhead{(3)} & \colhead{(4)} & \colhead{(5)} & \colhead{(6)}
}
\tablecolumns{6}
\tablewidth{0pt}
\tabletypesize{\scriptsize}
\startdata
1	&	8336884-5290	&	M0.0	&	f	&	0.61	&	I-J	\\
5	&	8343944-5609	&	M5.5	&	c,f	&	0.87	&	I-J	\\
8	&	8346141-5010	&	M3.5	&	f	&	0.33	&	I-J	\\
11	&	8347711-5533	&	K7.0	&	SpeX	&	8.27	&	H-K	\\
13	&	8350150-5595	&	M3.5	&	c	&	0	&	I-J	\\
16	&	8353051-5229	&	K1.0	&	SpeX	&	3.01	&	J-H	\\
18	&	8355083-4835	&	$\cdots$	&	$\cdots$	&	1	&	CTTS J-H	\\
19	&	8355371-5480	&	K7.0	&	b	&	0.15	&	H-K	\\
\enddata
\tablecomments{\footnotesize Spectral type (column 3) and the methods or literature (column 4) from which we adopted the spectral type}
\tablenotetext{a} {\footnotesize \citet{Allen_SpT_pc}, private communication; spectral types are measured from HECTOSPEC spectra}
\tablenotetext{a'} {\footnotesize \citet{Allen_SpT_pc}, private communication; spectral types are measured from HECTOSPEC spectra, spectral types are highly uncertain}
\tablenotetext{b} {\footnotesize \citet{H97b}}
\tablenotetext{c} {\footnotesize \citet{RHS2000}}
\tablenotetext{d} {\footnotesize \citet{Allen_thesis_L1641}}
\tablenotetext{e} {\footnotesize \citet{Hernandez_SpT_MDM_pc}, private communication; spectral types measured from MDM spectroscopic data}
\tablenotetext{f} {\footnotesize \citet{Hernandez_SpT_hecfast_pc}, private communication; spectral types are measured from HECTOSPEC spectra}
\tablenotetext{f'} {\footnotesize \citet{Hernandez_SpT_hecfast_pc}; spectral types are measured from FAST spectra.}
\tablenotetext{g} {\footnotesize \citet{Dario2009_ubvri}}
\tablenotetext{h} {\footnotesize \citet{Fang2009}}
\tablenotetext{i} {\footnotesize \citet{Parihar2009_ONC_variables_SpT4V1498Ori}}
\tablenotetext{i'} {\footnotesize It is M3.6 from \citet{Parihar2009_ONC_variables_SpT4V1498Ori}, but for our analysis convenience we take it as M3.5}
\tablenotetext{SpeX} {\footnotesize Spectral typing with SpeX spectra in this work}
\tablecomments{\footnotesize Column (6) $A_{V}$ method: the method which derived $A_{V}$ in the column (5). See the text for details.}
\end{deluxetable}
\clearpage

\clearpage
\begin{deluxetable}{cccccccccc}
\tablecaption{Stellar and Disk properties \label{table:property}}
\tablehead{
\colhead{Num}	&	\colhead{IRS name}	&	\colhead{$T_{eff}$}	&	 \colhead{$L_{\star}$}	 &	 \colhead{$M_{\star}$}	 &	 \colhead{$R_{wall}$} 	 &	 \colhead{TD type}	 &	 \colhead{$\dot{M}$}	 &	 \colhead{$L_{X}$} & \colhead{region}	 \\
\colhead{ }	&	\colhead{ }	&	\colhead{($K$)}	&	\colhead{($L_{\odot}$)}	&	 \colhead{($M_{\odot}$)}	 &	 \colhead{(AU)} &	 \colhead{ }	 &	 \colhead{($M_{\odot}/yr$)}		&	 \colhead{($erg/s$)} & \colhead{ }
}
\tablecolumns{10}
\tablewidth{0pt}
\tabletypesize{\scriptsize}
\startdata
1	&	8336884-5290	&	3850	&	0.64	&	0.55	&	13.7	&	WTD	 &	 (7.8$\pm$0.9)E-10		&	 $\cdots$		 &	ONC	 \\
5	&	8343944-5609	&	3145	&	0.54	&	0.23	&	3.0	& PTD	 &	$<$5.5E-10	 &	 $\cdots$		&	 ONC	\\
8	&	8346141-5010	&	3420	&	0.17	&	0.31	&	6.4	&	CTD	 &	$\cdots$		 &	$\cdots$	 &	 ONC	\\
11	&	8347711-5533	&	4060	&	1.80	&	0.70	&	13.3	&	CTD	 &	$<$2.37E-10	 &	 (3.4$\pm$2.0)E+29		 &	ONC	 \\
13	&	8350150-5595	&	3420	&	0.45	&	0.31	&	20.0	&	 CTD	&	 (1.9$\pm$0.7)E-09		&	 (1.3$\pm$0.9)E+30	 &	 ONC	 \\
16	&	8353051-5229	&	5080	&	3.65	&	1.88	&	52.6	&	 CTD	&	 (1.6$\pm$0.1)E-09		&	 $\cdots$		&	 ONC	 \\
18	&	8355083-4835	&	3850	&	0.18	&	0.58	&	3.0	&	PTD	 &	$\cdots$		 &	$\cdots$	 &	 ONC	\\
19	&	8355371-5480	&	4060	&	0.50	&	0.77	&	6.6	&	CTD	 &	 (3.2$\pm$0.6)E-09	 &	 (2.1$\pm$1.0)E+30			&	 ONC	 \\
\enddata
\tablecomments{\footnotesize The calculation of $R_{wall}$ is explained in Section 4.2, and the values listed in the column have an uncertainty of about 33 $\%$. $\dot{M}$ of ONC and L1641 is from this work. $\dot{M}$ for other regions are from the literatures. $L_{X}$ are all from XMM-Newton Serendipitous Source Catalog ~\citep{XMMSSC-catalogue}. The energy band is total energy band of XMM-Newton (0.2-12 keV). }
\tablenotetext{a} {\footnotesize \citet{najita07}}
\tablenotetext{b} {\footnotesize \citet{Espaillat2011TDvar}}
\tablenotetext{c} {\footnotesize \citet{Guedl_XEST_tauXray}}
\tablenotetext{d} {\footnotesize \citet{Natta_2006A&A...452..245N}}
\end{deluxetable}
\clearpage

\clearpage
\begin{deluxetable}{cccccccc}
\tablecaption{Sub-classification of TDs \label{table:TDtype}}
\tablehead{
\colhead{ID}	&	\colhead{name}	&	\colhead{region}	&	\colhead{$n_{K-6}$}	&	 \colhead{$n_{13-31}$}	 &	 \colhead{$EW(10\micron)$}	&	 \colhead{subtype} 	 &	 \colhead{\textit{IDEF}}\\
\colhead{(1)}	&	\colhead{(2)}	&	\colhead{(3)}	&	\colhead{(4)}	&	 \colhead{(5)}	&	 \colhead{(6)}	 &	 \colhead{(7)}	 &	 \colhead{(8)} 	
}
\tablecolumns{8}
\tablewidth{0pt}
\tabletypesize{\scriptsize}
\startdata
1	&	8336884-5290	&	ONC	&	N	&	N	&	Y	&		WTD	&	0.43	\\
5	&	8343944-5609	&	ONC	&	N	&	N	&	Y	&		PTD	&	 0.52\tablenotemark{f}	\\
8	&	8346141-5010	&	ONC	&	Y	&	Y	&	N	&		CTD	&	 0.27\tablenotemark{c}	\\
11	&	8347711-5533	&	ONC	&	Y	&	Y	&	N	&		CTD	&	0.00	\\
13	&	8350150-5595	&	ONC	&	Y	&	Y	&	N	&			CTD	&	0.06	\\
16	&	8353051-5229	&	ONC	&	Y	&	Y	&	N	&			CTD	&	0.07	\\
18	&	8355083-4835	&	ONC	&	Y	&	N	&	N	&			PTD	&	0.58	\\
19	&	8355371-5480	&	ONC	&	Y	&	Y	&	N	&			CTD	&	0.05	\\
\enddata
\end{deluxetable}
\clearpage

\clearpage
\begin{deluxetable}{ccccccc}
\tablecaption{number of transitional disks and age trend \label{table:agefreq}}
\tablehead{
\colhead{region}	&	\colhead{Age (Myr)}	&	\colhead{CTD}	&	\colhead{WTD}	 &	 \colhead{PTD}	 &	 \colhead{total TD} 	 &	 \colhead{disks}	
}
\tablecolumns{7}
\tablewidth{0pt}
\tabletypesize{\scriptsize}
\startdata
NGC 1333 & $ < 1$ & 2 & 0 & 7 & 9 & 53\tablenotemark{a} \\
ONC & $ < 1$ & 15 & 10 & 7 & 32 & 127\tablenotemark{b} \\
L1641 & 1 & 19 & 5 & 6 & 30 & 114\tablenotemark{b} \\
Tau & 1.5 & 4 & 1 & 8 & 13 & 154\tablenotemark{c} \\
ChaI & 2 & 6 & 1 & 4 & 11 & 68\tablenotemark{d} \\
Oph & 0.3-2.1 & 1& 0 & 9 & 10 & 90\tablenotemark{e}
\enddata
\tablecomments{Number of transitional disks and the median age of each star-forming region. The median age of each star-forming region is taken from the literature listed below and references therein. The last column: n(disks) is the number of objects in our sample, which is complete for Class II objects and transitional disks with host stars of type M4 or earlier (see \S2).}
\tablenotetext{a} {\footnotesize We keep 53 objects after taking 4 Class I objects out from the 57 disk dominant objects reported in \citet{Arnold2012}.}
\tablenotetext{b} {\footnotesize from this work.}
\tablenotetext{c} {\footnotesize This number comes from IRS survey of Taurus by \citet{Furlan2011Taurus}.}
\tablenotetext{d} {\footnotesize This number comes from IRS survey of Chamaeleon I by \citet{manoj2011}.}
\tablenotetext{e} {\footnotesize This number is the number of disk dominant objects identified from IRS survey of Ophiuchus by \citet{McClure2010}.}
\end{deluxetable}
\clearpage

\clearpage
\begin{deluxetable}{ccccccc}
\tablecaption{Correlation summary I\label{table:correlation}}
\tablehead{
\colhead{Correlation}	&	\colhead{N}	&	\colhead{$\alpha$}	&	 \colhead{$\beta$}	 &	 \colhead{\textit{corr}\tablenotemark{b}}	 &	 \colhead{\textit{P}\tablenotemark{c} ($\%$)} 	&	 \colhead{\textit{$P_{range}$}\tablenotemark{d} ($\%$)}	
}
\tablecolumns{7}
\tablewidth{0pt}
\tabletypesize{\scriptsize}
\startdata
$\dot{M}$ vs. $M_{\star}/M_{\odot}$	&	69	&	-8.5$\pm$0.1	&	1.6$\pm$0.3	 &	 0.5$\pm$0.1	 &	 0.001	&	 $\ll$0.001-0.1	\\
$L_{X}$ vs. $M_{\star}/M_{\odot}$	&	56	&	30.1$\pm$0.1	&	1.1$\pm$0.3	 &	 0.5$\pm$0.1	 &	 0.01	 &	 $<$0.001-0.2	 \\
	&	55	&	30.1$\pm$0.1	&	1.1$\pm$0.3	&	0.4$\pm$0.1	&	 0.3	&	 0.01-2.6	 \\
$\dot{M}$ vs. $L_{X}$	 &	42	&	-14.6$\pm$7.7	&	0.2$\pm$0.3	 &	0.2$\pm$0.2	 &	 20.4	&	 0.9-100	\\
\tableline																		
$M_{\star}/M_{\odot}$ vs. $R_{wall}$	&	98\tablenotemark{a}	&	-1.0$\pm$0.1	&	 0.7$\pm$0.1	 &	 0.8$\pm$0.1	 &	 $\ll$0.001	 &	$\ll$0.001	\\
                                    	&	97	&	-1.0$\pm$0.1	&	0.7$\pm$0.1	&	 0.7$\pm$0.1	&	 $\ll$0.001	 &	$\ll$0.001	 \\
$\dot{M}$ vs. $R_{wall}$	&	69	&	-10.1$\pm$0.4	&	1.1$\pm$0.4	&	 0.4$\pm$0.1	&	 0.07	&	 0.001-1.2	 \\
$L_{X}$ vs. $R_{wall}$	&	56\tablenotemark{a}	&	28.5$\pm$0.3	&	1.2$\pm$0.3	&	 0.6$\pm$0.1	&	 $\ll$0.001	 &	 $\ll$0.001-0.01	\\
\enddata
\tablecomments{Summary of the results from the Bayesian linear regression between X and Y: Y vs. X; $logY=\alpha + \beta logX$. Objects has a known spectral type are only included for the correlations.}
\tablenotetext{a} {\footnotesize The results when we include OriA-88, which has a large value of $R_{wall}$, $M_{\star}$, and $L_{\star}$, and therefore is an outlier.}
\tablenotetext{b} {\footnotesize \textit{corr} is the linear correlation coefficient between $logY$ and $logX$.}
\tablenotetext{c} {\footnotesize \textit{P} is the probability of getting \textit{corr} in random distribution.}
\tablenotetext{d} {\footnotesize \textit{$P_{range}$} is the range of \textit{P} to get \textit{corr} in $1\sigma$ uncertainty.}
\end{deluxetable}
\clearpage

\clearpage
\begin{deluxetable}{ccccccccccc}
\tablecaption{Correlation summary II: details by subtypes of TDs \label{table:correlation_subTDtype}}
\tablehead{
\colhead{Correlation} & \colhead{TD type}	&	\colhead{N}	&	\colhead{$\alpha$}	&	 \colhead{$\beta$}	 &	 \colhead{\textit{corr}\tablenotemark{b}}	 &	 \colhead{\textit{P}\tablenotemark{c} ($\%$)} 	&	 \colhead{\textit{$P_{range}$}\tablenotemark{d} ($\%$)} & \colhead{MRI\tablenotemark{e}} & \colhead{XPE\tablenotemark{f}} & \colhead{GPF\tablenotemark{g}}	
}
\rotate
\tablecolumns{11}
\tablewidth{0pt}
\tabletypesize{\scriptsize}
\startdata
$\dot{M}$ vs. $M_{\star}/M_{\odot}$ & CTD	&	34	&	-8.9$\pm$0.6	&	0.7$\pm$1.8	 &	 0.1$\pm$0.3	 &	 57	 &	 1.9-26	 & & & Y\\
 & WTD+PTD	&	35	&	-8.2	$\pm$	0.1	&	1.9$\pm$0.5	&	0.7$\pm$0.1	&	 $<$0.001	 &	 $\ll$0.001-0.01	 & & & Y \\
$L_{X}$ vs. $M_{\star}/M_{\odot}$ & CTD\tablenotemark{a}	&	30	&	30.3$\pm$0.2	&	 1.8$\pm$0.6	&	 0.6$\pm$0.2	&	 0.05	 &	 $\ll$0.001-2.9	 & & & Y \\
 & CTD	&	29	&	30.5$\pm$0.3	&	2.2$\pm$0.9	&	0.5$\pm$0.2	 &	0.6	 &	 0.002-11	 & & & Y\\
 & WTD+PTD	&	26	&	30.0$\pm$0.1	&	1.0$\pm$0.4	&	0.5$\pm$0.2	&	0.9	 &	 0.007-14	 & & & Y\\
\hline																
$\dot{M}$ vs. $R_{wall}$ & CTD	&	34	&	-9.1$\pm$0.6	&	-0.1$\pm$0.6	&	 -0.0$\pm$0.2	&	 100	 &	 26-100	 & N & N$?$ & Y\\
 & WTD+PTD	&	35	&	-10.8$\pm$0.6	&	1.9$\pm$0.5	&	0.7$\pm$0.1	&	$<$0.001	 &	 $\ll$0.001-0.01	 & Y & Y$?$ & Y \\
$\dot{M}/\dot{M}(M_{\star})$ vs. $R_{wall}$ & CTD	&	34	&	0.8$\pm$0.7	&	 -1.7$\pm$0.6	 &	 -0.6$\pm$0.2	 &	 0.02	&	 $\ll$0.001-1.9	 & N & Y$?$ & Y\\
\enddata
\tablecomments{Summary of the results from the Bayesian linear regression between X and Y: Y vs. X; $logY=\alpha + \beta logX$. \\
The index $^{a, b, c,}$ and $^{d}$ have the same meanings as in Table~\ref{table:correlation}.
}
\tablenotetext{e} {\footnotesize \textit{MRI}: The inside-out disk evolution driven by Magnetic Rotational Instability.}
\tablenotetext{f} {\footnotesize \textit{XPE}: The disk dispersion by photoevaporation, specially X-ray photoevaporation}
\tablenotetext{g} {\footnotesize \textit{GPF}: Dynamical effects by infant giant planet formation.}
\tablecomments{N or Y checks indicate if a mechanism in the last three columns can explain the corresponding trends in a row.}
\end{deluxetable}
\clearpage

\clearpage
\begin{deluxetable}{ccccccc}
\tablecaption{Correlation summary III: TDs in Orion A \label{appendix-table:correlation-OriA}}
\tablehead{
\colhead{Correlation}	&	\colhead{N}	&	\colhead{$\alpha$}	&	 \colhead{$\beta$}	 &	 \colhead{\textit{corr}\tablenotemark{b}}	 &	 \colhead{\textit{P}\tablenotemark{c} ($\%$)} 	&	 \colhead{\textit{$P_{range}$}\tablenotemark{d} ($\%$)}	
}
\tablecolumns{7}
\tablewidth{0pt}
\tabletypesize{\scriptsize}
\startdata
$\dot{M}$ vs. $M_{\star}/M_{\odot}$	&	50	&	-8.6$\pm$0.2	&	1.3$\pm$0.5	 &	 0.4$\pm$0.2	 &	0.4	 &	 $<$0.001-16	\\
$L_{X}$ vs. $M_{\star}/M_{\odot}$	&	35\tablenotemark{a}	&	30.2$\pm$0.1	&	 1.0$\pm$0.3	 &	 0.5$\pm$0.2	 &	0.2	&	 $<$0.001-8 	\\
	&	34	&	30.2$\pm$0.2	&	0.9$\pm$0.4	&	0.4$\pm$0.2	&	1.9 	&	 0.02-26	 \\
$\dot{M}$ vs. $L_{X}$	 &	33	&	-20.3$\pm$8.7	&	0.4$\pm$0.3	 &	0.3$\pm$0.2	 &	 9	&	 0.3-58 	 \\
\tableline																		
$M_{\star}/M_{\odot}$ vs. $R_{wall}$	&	55\tablenotemark{a}	&	-1.1$\pm$0.1	&	 0.7$\pm$0.1	 &	 0.8$\pm$0.1	 &	 $\ll$0.001	 &	 $\ll$0.001	\\
                                    	&	54	&	-1.0$\pm$0.1	&	0.6$\pm$0.1	&	 0.8$\pm$0.1	&	 $\ll$0.001	 &	$\ll$0.001	 \\
$\dot{M}$ vs. $R_{wall}$	& 50	&	-9.8$\pm$0.5	&	0.7$\pm$0.4	&	 0.3$\pm$0.2	&	3.4 	&	 0.2-49 	 \\
$L_{X}$ vs. $R_{wall}$	&	35\tablenotemark{a}	&	28.8$\pm$0.3	&	1.0$\pm$0.3	&	 0.6$\pm$0.1	&	 0.01	 &	 $<$0.001-0.2 	 \\
\enddata
\tablecomments{Summary of the results from the Bayesian linear regression between X and Y: Y vs. X; $logY=\alpha + \beta logX$. Objects has a known spectral type and in Orion A are only included for the correlations.\\
The index $^{a, b, c,}$ and $^{d}$ have the same meanings in Table~\ref{table:correlation}.}
\end{deluxetable}
\clearpage

\clearpage
\begin{deluxetable}{ccccccccccc}
\tablecaption{Correlation summary IV: details by subtypes of TDs in Orion A \label{appendix-table:correlation_subTDtype-OriA}}
\tablehead{
\colhead{Correlation} & \colhead{TD type}	&	\colhead{N}	&	\colhead{$\alpha$}	&	 \colhead{$\beta$}	 &	 \colhead{\textit{corr}\tablenotemark{b}}	 &	 \colhead{\textit{P}\tablenotemark{c} ($\%$)} 	&	 \colhead{\textit{$P_{range}$}\tablenotemark{d} ($\%$)}
}
\rotate
\tablecolumns{8}
\tablewidth{0pt}
\tabletypesize{\scriptsize}
\startdata
$\dot{M}$ vs. $M_{\star}/M_{\odot}$ & CTD	&	30	&	-9.2$\pm$0.5	&	-0.2$\pm$1.3	 &	 0.0$\pm$0.3	 &	 100	 &	 11-100	 \\
 & WTD+PTD	&	20	&	-8.3	$\pm$	0.3	&	2.1$\pm$0.7	&	0.6$\pm$0.2	&	 0.5	 &	 0.002-8.1	 \\
$L_{X}$ vs. $M_{\star}/M_{\odot}$ & CTD\tablenotemark{a}	&	24	&	30.3$\pm$0.2	&	 1.4$\pm$0.5	&	 0.5$\pm$0.2	&	1.3	 &	 0.01-15	 \\
 & CTD	&	23	&	30.5$\pm$0.4	&	1.9$\pm$1.2	&	0.4$\pm$0.2	 &	5.7	 &	 0.2-36	 \\
 & WTD+PTD	&	11	&	30.2$\pm$0.1	&	0.5$\pm$0.4	&	0.5$\pm$0.3	&	12	 &	0.3-56 	 \\
\hline																
$\dot{M}$ vs. $R_{wall}$ & CTD	&	30	&	-8.8$\pm$0.6	&	-0.3$\pm$0.6	&	 -0.1$\pm$0.2	&	60 	 &	 11-100 	 \\
 & WTD+PTD	&	20	&	-11.1$\pm$0.9	&	2.0$\pm$0.8	&	0.7$\pm$0.2	&	0.06	 &	$\ll$0.001-2.5 	 \\
$\dot{M}/\dot{M}(M_{\star})$ vs. $R_{wall}$ & CTD	&	30	&	1.0$\pm$0.6	&	 -2.0$\pm$0.6	 &	 -0.7$\pm$0.2	 &	 0.002	&	 $\ll$0.001-0.5 	 \\
\enddata
\tablecomments{Summary of the results from the Bayesian linear regression between X and Y: Y vs. X; $logY=\alpha + \beta logX$. Objects has a known spectral type and in Orion A are only included for the correlations.\\
The index $^{a, b, c,}$ and $^{d}$ have the same meanings in Table~\ref{table:correlation}.
}
\end{deluxetable}
\clearpage

\end{document}